\PassOptionsToPackage{table}{xcolor}
\documentclass[]{lightspeed}

\usepackage{amsmath,amsfonts,bm}

\def\eqref#1{equation~\ref{#1}}

\def\1{\bm{1}}

\DeclareMathAlphabet{\mathsfit}{\encodingdefault}{\sfdefault}{m}{sl}
\SetMathAlphabet{\mathsfit}{bold}{\encodingdefault}{\sfdefault}{bx}{n}

\usepackage{amssymb}
\usepackage{url}
\usepackage{wrapfig}
\usepackage{array}
\usepackage{algorithm}
\usepackage{algpseudocode}
\usepackage{listings}
\usepackage{siunitx}
\usepackage{pifont}
\usepackage{tikz}
\usepackage{enumitem}

\microtypesetup{expansion=false}
\setcitestyle{square,comma,numbers,sort&compress}
\newcommand{\cmark}{\textcolor{green!45!black}{\ding{51}}}
\newcommand{\xmark}{\textcolor{red!75!black}{\ding{55}}}
\newcommand{\pmark}{\textcolor{green!45!black}{\ding{51}}\textsuperscript{$*$}}
\newcommand{\nmark}{--}

\definecolor{famBrowser}{HTML}{DD9B34}
\definecolor{famNative}{HTML}{2A9D8F}
\definecolor{famUnity}{HTML}{1F3C5F}
\definecolor{famTools}{HTML}{9A5F8C}

\newcommand{\floatsize}{\fontsize{8pt}{9.5pt}\selectfont}
\newlength{\ratewidth}
\newlength{\ratepctwidth}
\newcommand{\rate}[2]{%
  \begingroup\renewcommand{\arraystretch}{0.9}%
  \settowidth{\ratewidth}{000/000}\settowidth{\ratepctwidth}{(00.0\%)}%
  \ifdim\ratepctwidth>\ratewidth\setlength{\ratewidth}{\ratepctwidth}\fi
  \begin{tabular}[t]{@{}>{\raggedleft\arraybackslash}p{\ratewidth}@{}}#1\\(#2)\end{tabular}%
  \endgroup}

\newtcolorbox{rqanswer}[1]{
  enhanced, colback=white, colframe=black!70, boxrule=0.6pt,
  arc=3pt, boxsep=0pt, left=6pt, right=6pt, top=4pt, bottom=4pt,
  before skip=8pt, after skip=8pt, fontupper=\itshape,
  before upper={\textnormal{\bfseries Answer to RQ#1:}\ }
}

\lstdefinestyle{tscode}{
  basicstyle=\ttfamily\floatsize,
  keywordstyle=\bfseries\color{blue!55!black},
  commentstyle=\itshape\color{green!35!black},
  stringstyle=\color{purple!70!black},
  showstringspaces=false,
  breaklines=true,
  breakatwhitespace=true,
  columns=fullflexible,
  keepspaces=true,
  numbers=none,
  frame=none,
  morekeywords={const,await,async,function,return,test,expect,let,if,new,as,of,in,void}
}

\newcommand{\method}{StatePortal}
\newcommand{\benchmark}{DeepStateBench}

\newcommand{\confgui}{Agent S3}
\newcommand{\confguicp}{Agent S3 w/ Checkpoint}
\newcommand{\confdelegated}{\method{}-Delegated}
\newcommand{\confcua}{CUA Agent}
\newcommand{\confcuacp}{CUA Agent w/ Checkpoint}

\newcommand{\perfstudy}{performance study}
\newcommand{\replaystudy}{replay study}
\newcommand{\latentstudy}{latent-defect study}
\newcommand{\fuzzstudy}{fuzzing study}

\newcommand{\repourl}{https://github.com/pzy2000/State_Portal}
\newcommand{\repokind}{public}
\newcommand{\withheldreason}{to keep this preprint aligned with the anonymous submission}

\makeatletter
\patchcmd{\ttl@select}{\if@inlabel\else}
  {\if@inlabel\else\ifx\everypar\WF@everypar\let\everypar\WF@@everypar\fi}
  {}{\PackageError{preprint}{titlesec patch for wrapfig failed}{}}
\makeatother

\newenvironment{taskprofilesfigure}
  {\wrapfigure{R}{2.53in}}
  {\endwrapfigure}

\title{StatePortal: Agentic Checkpoint Synthesis for Testing Deep States in Interactive Software}

\author[1,*]{Zhiyuan Peng}
\author[2,*]{Wei Tao}
\author[1,\dagger]{Chenhao Ying}
\author[1,\dagger]{Yuan Luo}
\author[3,\dagger]{Yiwen Guo}

\affiliation[1]{Shanghai Jiao Tong University}
\affiliation[2]{LIGHTSPEED}
\affiliation[3]{Independent Researcher}

\renewcommand\emaillist{%
  \emailformat{pzy2000@sjtu.edu.cn}, \emailformat{wtao@ieee.org},
  \emailformat{yingchenhao@sjtu.edu.cn},\\
  \emailformat{yuanluo@sjtu.edu.cn}, \emailformat{guoyiwen89@gmail.com}}

\abstract{Large language models have reduced the effort to generate code for interactive software with graphical user interfaces (GUIs). For such software, a key testing bottleneck is reaching a \emph{deep state}, one that requires a long interaction history. The cost of reaching it recurs across commits, and the GUI automation that could amortize it assumes a DOM or accessibility tree that self-rendered engines do not expose.
We introduce \method{}, an agent that synthesizes a \emph{checkpoint}, an artifact that restores that state and can be versioned and replayed, from a natural-language functional test case and read-only access to the target's source code. The checkpoint is restored through a per-application \emph{hook} that an onboarding agent binds or synthesizes once without hand-written hook code. 
Because a checkpoint can satisfy state checks while rendering incorrectly, checkpoints are scored by \emph{hybrid programmatic-state and visual evaluation}: acceptance requires both a programmatic-state oracle and a visual oracle to pass. We calibrate the visual oracle against a blinded cross-VLM reference. 
We release \benchmark{}, 61 deep-state tasks across 10 interactive applications that span five language families and six rendering stacks, with source sizes from 874 to 7.2M lines. 
Across three registered models, a GUI agent seeded with synthesized checkpoints completes 10.4\% of trials, compared with 0\% without seeding under the same pixel action space, while \method{}, which also delivers the target action through the hook, completes 27.5\%. 
Over 1{,}361 replays of 112 checkpoints, 98.90\% reproduce the original verdict with zero model calls, 
and the hybrid oracle detects 96.25\% (77/80) of controlled visual faults invisible to programmatic-state checks alone.
}

\headercontent{%
  \begin{tabular}{c}
    \url{\repourl}
  \end{tabular}%
}

\date{September 2026}

\begin{document}
\thispagestyle{firstheader}
\maketitle
{\renewcommand{\thefootnote}{}%
 \footnotetext{$^{*}$Equal contribution. \quad $^{\dagger}$Corresponding authors.}}

\begin{flushright}
{\itshape To understand a world, first make its depths reachable.}\\
---\,Lucius
\end{flushright}

\section{Introduction}
\label{sec:intro}

Large language models have reduced the effort required to generate code for software \citep{kimi_k3_2026}, increasing the importance of effective verification \citep{playcoder2026}. In interactive software, testing a target behavior may require a long sequence of prerequisite interactions \citep{fax_icse2020}, repeated across commits. Automated tests for such software often start from hand-built fixtures, leaving states beyond them hard to verify \citep{jimenez2024swe,reproagent2026,li2026advances}. 
The shortfall is sharpest in games, where developers report substantially less test automation than in other domains \citep{murphyhill_icse2014} and only 7.33\% of 491 open-source projects contain automated tests \citep{ossgames_gas2026}. The cost of covering these states depends on how much of the running application a test harness can read and write, and the self-rendered engines widely used in game development expose little of it (Appendix~\ref{sec:appendix-motivating-example}).

\begin{figure}[t]
\centering
\includegraphics[width=0.9\linewidth]{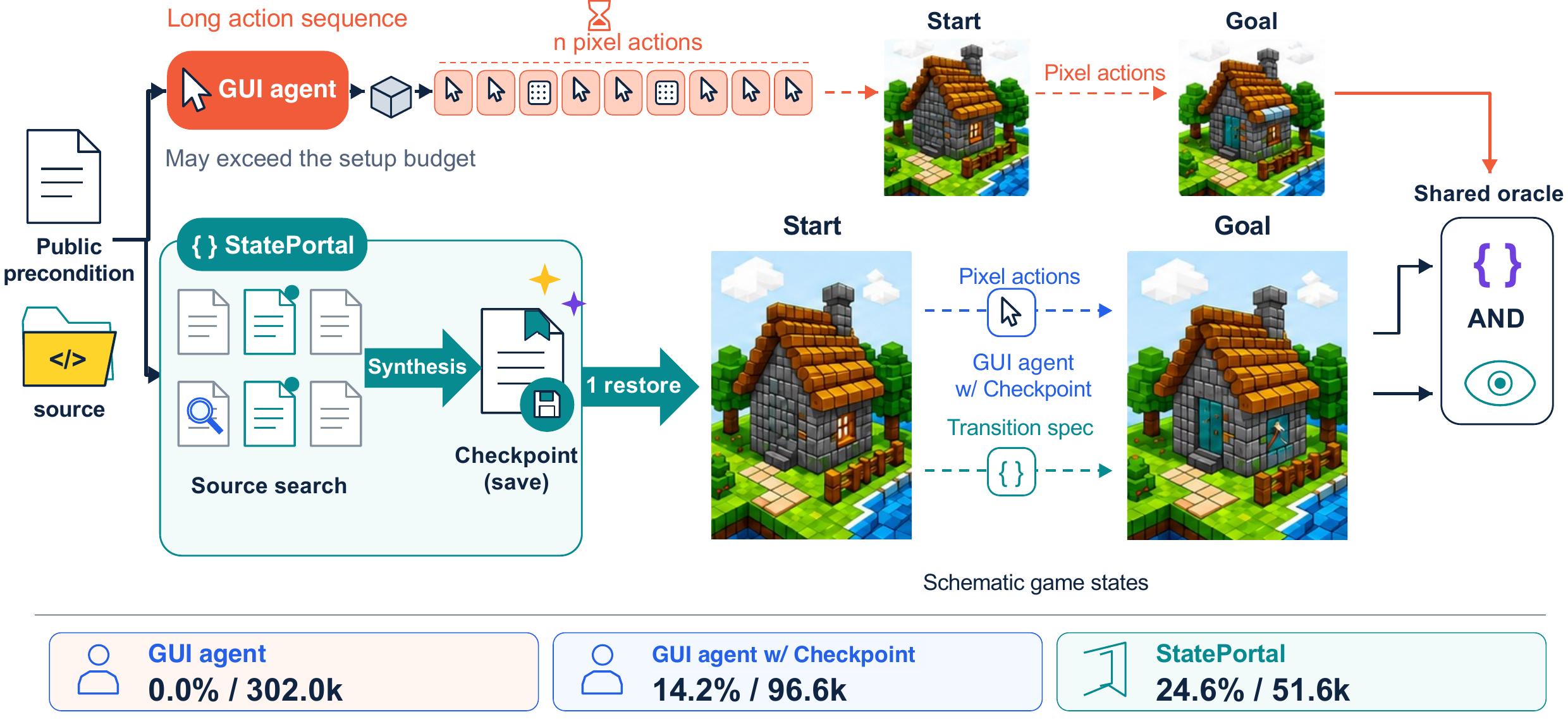}
\caption{A GUI agent reaches the deep state through a long pixel-action sequence, whereas \method{} synthesizes a checkpoint and restores the target state directly. The two pixel routes represent GUI agents, and their E2E rates and mean token costs are measured with Agent S3 \citep{gonzalezpumariega2025agents3} on \texttt{gpt-5.6-terra} over the 61 tasks.}
\label{fig:stateportal-overview}
\vspace{-6pt}
\end{figure}

Unit tests alone are insufficient: GUI-specific failures (e.g., a softlock, where the software keeps running but the player cannot progress \citep{stuckinthemiddle2025}) may emerge only during a full interaction sequence \citep{playcoder2026}, and industrial reports rank progress-blocking failures among the most common defects \citep{titan_gametesting2025}. GUI automation based on reusable DOM selectors \citep{actionengine2026} does not transfer to self-rendered engines, which expose no element tree. A pixel agent replaying an $n$-step path instead requires $O(n)$ model decisions, whose errors can compound over the sequence \citep{goexplore2021} and whose inference cost recurs on every regression run.

We study \emph{deep-state GUI testing}: testing behavior that cannot be reached from the initial screen with a small number of actions \citep{fax_icse2020}. Prior works rely on platform-specific state-inspection mechanisms that do not carry over to self-rendered engines, or generate test code without constructing the state it requires
(Section~\ref{sec:related}). From a functional test case, we instead synthesize a \emph{checkpoint}, a reusable artifact that restores a target state (the \emph{Start}) through a callable state-transition interface (the \emph{hook}) that an onboarding agent binds or synthesizes from the application's source. Our agent \method{} (Figure~\ref{fig:stateportal-overview}) explores the repository to identify the relevant state representation, persistence path, and transition logic. It jointly generates the Start, restoration payload, bounded \emph{transition spec}, and predicted assertions as one executable checkpoint. Restoration inside a running application (a \emph{warm session}) replaces the $n$-step prefix. Whether the Start is restored or reached through pixel actions, every trial is checked against the same hidden targets on runtime state and screenshots.

Existing GUI-agent benchmarks do not evaluate checkpoint synthesis for a new target from a test case and read-only source (Table~\ref{tab:benchmark-comparison}). We therefore build \benchmark{}: 61 tasks over ten applications (seven games and three general-purpose GUI applications), each scored by a \emph{conjunctive oracle} under which a programmatic-state oracle and a visual oracle must both pass, since state checks alone can miss failures visible only in rendered output \citep{osworld_arxiv24,osharm2025}. The suite spans five language families and six rendering stacks, all restored through one shared hook contract. 
Across three registered decision models, checkpoint seeding increases GUI-agent completion from 0\% to 10.4\% under the same pixel action space. 
\method{}, which also executes the target action through the hook, completes 27.5\%. 
Checkpoints support regression testing without further synthesis (98.90\% verdict reproduction over 1{,}361 replays), the hybrid oracle detects 96.25\% of controlled visual faults that leave registered state fields unchanged. On real defects that the repository's own tests miss, \method{}'s checkpoints separate buggy from fixed builds on 7 of 8 comparable tasks. 

In summary, we make the following four contributions:
\begin{itemize}[leftmargin=*]
\itemsep0em
  \item A \textbf{task formulation}: from a functional test case and read-only source, synthesize a checkpoint that restores the Start and runs the transition under test (Section~\ref{sec:task}).
  \item \textbf{\method{}}, an agent that synthesizes the Start and the transition jointly from one source exploration, with hooks onboarded for all ten applications without human intervention (Section~\ref{sec:method}).
  \item A \textbf{conjunctive evaluation protocol} in which a programmatic-state oracle and a VLM judge must both pass, so the judge can overturn a programmatic pass but never a failure (Section~\ref{sec:hybrid-eval}).
  \item \textbf{\benchmark{}}, 61 public tasks over ten applications of up to 7.2M lines. Unlike other GUI-agent benchmarks, it provides the source and leaves the Start to the method (Section~\ref{sec:benchmark}).
\end{itemize}

\section{Task Definition}
\label{sec:task}
We study checkpoint synthesis for functional testing of stateful applications.
A method receives a pair $(u, R)$: a natural-language \emph{functional test case} $u$ derived from a bug report or feature request and read-only access to the target's source repository $R$. For each test, the method needs to output a single configuration file, the \emph{checkpoint}, declaring the semantic state to restore (the \emph{Start}), the strategy and payload that construct it, a finite \emph{transition spec} (the post-Start action sequence under test), and assertions over the resulting state. 
The declared Start is constructed through a per-application \emph{hook}, a callable state-transition interface that an \emph{onboarding agent} binds or synthesizes from source once per application. The same agent writes the hook's \emph{capability contract} $\Sigma = h(R, B)$, a declarative file specifying the hook kind, its activation gate, its calling convention, and the readable and writable state fields, with $B$ the public onboarding brief of Section~\ref{sec:synthesis-validation-reuse}. $\Sigma$ contains neither per-test state assignments nor hidden grading criteria. A checkpoint is synthesized as $\hat c = g(u, G, \Sigma)$, where $G$ denotes the source evidence retrieved during exploration of the sandboxed repository $R$. Validity is checked in two gated stages. $\mathrm{Valid}_{\mathrm{Start}}$ requires that the restored state match the declared Start, the runtime stay healthy, the frame be ready, and, for a save-restorable checkpoint, the persisted state round-trip through a save/reload cycle. After Start validity and the hidden Start oracles pass, $\mathrm{Valid}_{\mathrm{Trans}}$ checks that the transition spec executes without error. Section~\ref{sec:hybrid-eval} scores whether the restored and post-transition states match the grader's respective hidden Start and post-transition criteria. The grader's reference specifications and oracle remain hidden throughout synthesis: $u$ states the scenario as a human tester would without exposing the grader's internal reference specification, and a leakage guard rejects trials that access oracle-reserved paths (Algorithm~\ref{alg:synthesis-agent}, Table~\ref{tab:task-examples}).

\section{Method: Agentic Checkpoint Synthesis}
\label{sec:method}
\label{sec:testing-state-management}
\label{sec:synthesis-overview}

\begin{figure}[t]
\centering
\includegraphics[width=0.9\linewidth]{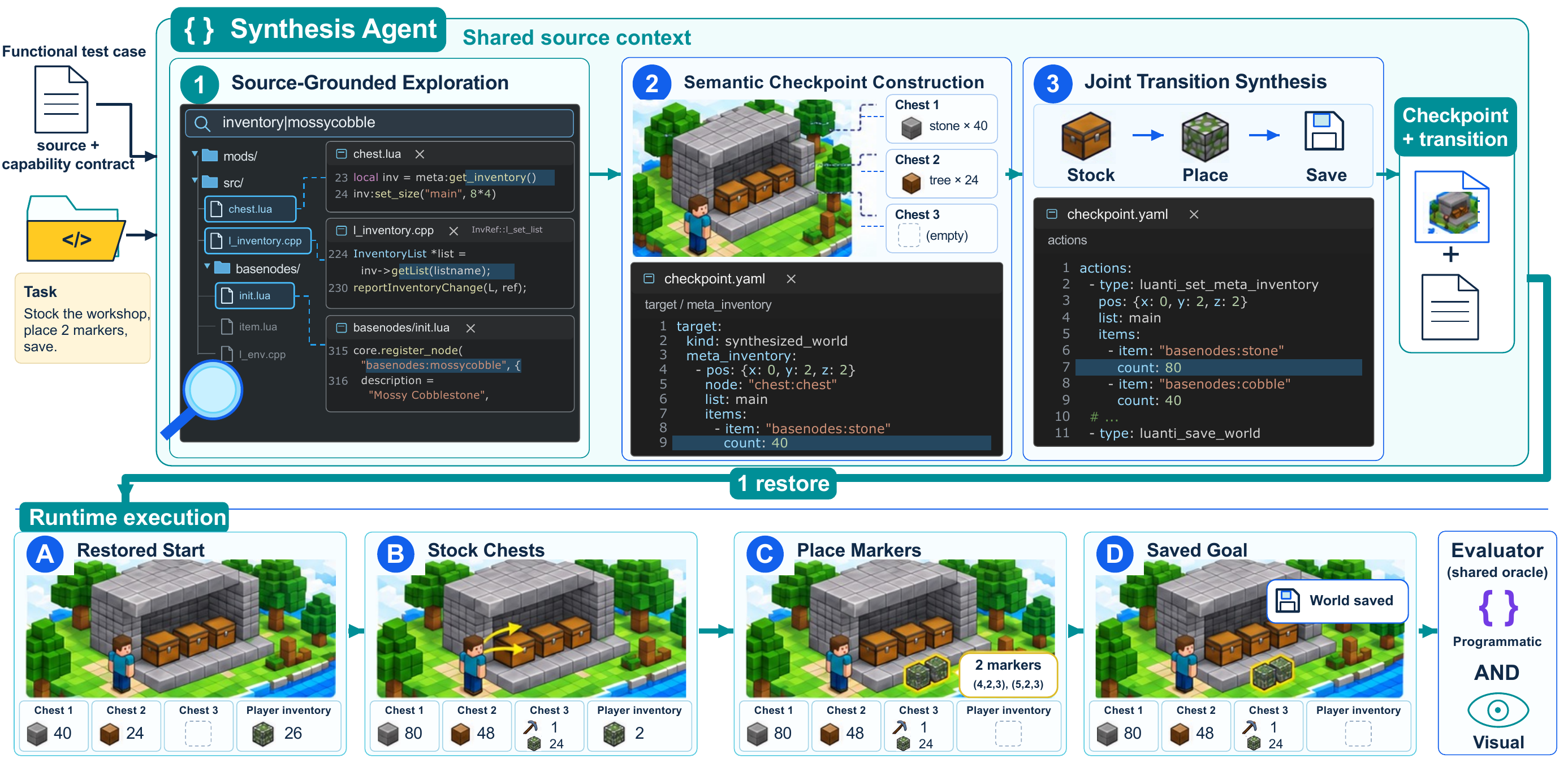}
\caption{\method{} on a Luanti workshop task. Shared source exploration supports joint synthesis of the Start checkpoint and transition, with state changes shown below the scenes.}
\label{fig:synthesis-agent}
\vspace{-6pt}
\end{figure}

\method{} translates a functional test case into an executable checkpoint through \emph{source-grounded exploration}, \emph{semantic checkpoint construction}, and \emph{joint transition synthesis}. Given $u$, a bounded map of $R$, and the onboarding contract $\Sigma$ (Section~\ref{sec:synthesis-validation-reuse}), one model emits the checkpoint and its transition in a single final response from one shared exploration history (Figure~\ref{fig:synthesis-agent}).

\subsection{Source-Grounded Exploration}
\label{sec:source-grounded-exploration}

The agent identifies the objects, identifiers, and control flow relevant to the task scenario through a bounded ReAct loop \citep{yao2023react} over \texttt{list\_dir}, \texttt{grep}, and \texttt{read\_file}. Writing $H_t$ for the conversation history and $q_t$ for a repository query, an exploration step is
\begin{equation}
q_t= g_\theta(H_t),\qquad o_t=\mathrm{Tool}_R(q_t),\qquad
H_{t+1}=H_t\mathbin{\|}(q_t,o_t),
\label{eq:synthesis-exploration}
\end{equation}
and the accumulated observations constitute $G$. A fixed public prompt $\mathcal P$ conditions $g_\theta$ to construct the Start from $u$, reserve the tested operation for the transition, and cite supporting source files. Exploration allows up to 16 turns with at most one repository query per turn. The first final response after a query terminates exploration, and citations must refer to accessed files (Algorithm~\ref{alg:synthesis-agent}).

\subsection{Semantic Checkpoint Construction}
\label{sec:semantic-checkpoint-construction}

The agent maps the public preconditions to a semantic Start target $\hat z$ and a payload $\eta$ under a restoration strategy $\tau$ supported by $\Sigma$. Each performance-study task admits a single $\tau$. The agent must satisfy the public preconditions without executing the tested operation in advance, which hidden Start checks verify through registered observations. A family-level adapter serializes the payload into native save bytes or dispatches it through the software's own API. The validity class (\emph{save-restorable}, \emph{replay-reachable}, \emph{injection-only}, or \emph{unknown}) determines the $\mathrm{Valid}_{\mathrm{Start}}$ checks.

\subsection{Joint Transition Synthesis}
\label{sec:joint-transition-synthesis}

Using the same exploration context, the agent maps the tested operation to registered action types, resolves arguments from source, orders the actions by their dependencies, and predicts the after-state assertions. The combined checkpoint, emitted in one response, is
\begin{equation}
\hat c=(\tau,\eta,\hat z,V,\alpha,A),\qquad
\hat c=g_{\theta,\mathcal P}(u,G,\Sigma)\in\mathcal C_{\mathrm{pub}}(u,G,\Sigma),
\label{eq:joint-checkpoint-synthesis}
\end{equation}
where $V$ is validity metadata, $\alpha$ the finite action sequence, and $A$ the predicted assertions. 
The public admissible set $\mathcal C_{\mathrm{pub}}$ requires schema compliance, interface compatibility, and evidence of source access, but membership does not establish runtime validity or semantic correctness (Table~\ref{tab:synthesis-check-boundaries}). The predicted assertions $A$ do not affect acceptance. Task-specific Start and Goal verdicts are
determined by the hidden programmatic and visual oracles (Algorithm~\ref{alg:checkpoint-synthesis}, example in Appendix~\ref{sec:appendix-running-example}).

\subsection{Onboarding, Validation, and Reuse}
\label{sec:synthesis-validation-reuse}

Before task construction, a source-guided onboarding agent reads the clean upstream repository and, from a capability brief and runtime notes, generates the hook, a driver, $\Sigma$, and save/reload self-tests, validated on five cases per application (two held out). With \texttt{gpt-5.6-sol}, onboarding succeeds on all ten applications without human intervention (Table~\ref{tab:onboarding-results}: three via existing APIs, three via extension mechanisms, four via test-gated entry points leaving product source otherwise untouched), using 5.49M tokens and 2{,}838~s. An independent replication also succeeds with 6.07M tokens, about \$7 (Appendix~\ref{sec:appendix-hook-onboarding}). Each synthesis trial then evaluates a single candidate without runtime or hidden-oracle feedback, with at most one repair of its output structure from public feedback. Execution restores the proposed Start, checks its validity, applies the programmatic and visual Start gate, and runs the transition once within the task budget (Section~\ref{sec:hybrid-eval}), after which the frozen checkpoint replays without further synthesis or model calls (Section~\ref{sec:replay}).

\section{Benchmark Suite and Evaluation Protocol}
\label{sec:benchmark}

Evaluating the method requires software with accessible source code, target states that require long interaction sequences, and evaluation oracles hidden from the agent. The eleven benchmarks in Table~\ref{tab:benchmark-comparison} do not jointly satisfy these requirements. We therefore introduce \benchmark{}: 10 applications and 61 public tasks derived mechanically from source evidence. In \benchmark{}, constructing the Start is part of the task: the synthesized checkpoint must restore a valid state in the running software and support the target action. Six applications render to a single GPU surface without a DOM or accessibility tree. Evaluation therefore combines programmatic runtime-state checks with screenshot-based checks, and a trial passes only when both do.

\begin{table}[t]
\centering
\caption{\benchmark{} compared with representative digital-agent benchmark environments.}
\label{tab:benchmark-comparison}
\begingroup
\fontfamily{lmr}\selectfont
\floatsize
\setlength{\tabcolsep}{2.4pt}
\renewcommand{\arraystretch}{1.1}
\begin{tabular*}{\linewidth}{@{\extracolsep{\fill}}lrrcccc@{}}
\toprule
Benchmark & \#\,Tasks & \#\,Apps & Source & Deep start & Method-built & Hybrid oracle \\
\midrule
GUITester \citep{guitester2026} & 143 & 12 & \xmark & \nmark & \xmark & \xmark \\
FlashAdventure \citep{flashadventure2025} & 34 & 34 & \xmark & \xmark & \xmark & \xmark \\
OSWorld \citep{osworld_arxiv24} & 369 & \nmark & \xmark & \cmark & \xmark & \xmark \\
WindowsWorld \citep{windowsworld2026} & 181 & 17 & \xmark & \cmark & \xmark & \xmark \\
ProSoftArena \citep{prosoftarena2026} & 456 & 13 & \xmark & \cmark & \xmark & \xmark \\
PPT-Eval \citep{ppteval2026} & 120 & 1 & \xmark & \cmark & \xmark & \pmark \\
MobileWorld \citep{mobileworld2026} & 201 & 20 & \xmark & \cmark & \xmark & \xmark \\
VenusBench-Mobile \citep{venusbench2026} & 149 & 27 & \xmark & \cmark & \xmark & \xmark \\
ColorBench \citep{colorbench2026} & 175 & \nmark & \xmark & \xmark & \xmark & \xmark \\
GTA \citep{gta2026} & 5{,}600 & 50+ & \xmark & \nmark & \xmark & \xmark \\
MMBench-GUI \citep{mmbenchgui2026} & 8{,}123 & \nmark & \xmark & \nmark & \xmark & \nmark \\
\midrule
\rowcolor{black!8}
\textbf{\benchmark{} (ours)} & \textbf{61} & \textbf{10} & \cmark & \cmark & \cmark & \cmark \\
\bottomrule
\end{tabular*}
\par\vspace{3pt}
\begin{minipage}{\linewidth}
\floatsize
\raggedright
\hypersetup{hidelinks}
\textit{Notes.} \cmark{}/\xmark{} record each paper's own claim, and ``--'' marks a value not stated comparably. \emph{Source}: the method receives the target's repository. \emph{Deep start}: a non-default, work-in-progress state. \emph{Method-built}: written by the method, not the benchmark. \emph{Hybrid oracle}: a programmatic and a visual check, both required. \pmark{} PPT-Eval merges both channels into one score.
\end{minipage}
\endgroup
\end{table}

\subsection{Benchmark Composition}
\label{sec:coverage}

\begin{figure}[t]
\centering
\newlength{\galgap}\setlength{\galgap}{2pt}
\newlength{\galw}\setlength{\galw}{\dimexpr(\linewidth-4\galgap)/5\relax}
\newlength{\galh}\setlength{\galh}{0.625\galw}
\newcommand{\galtile}[5]{%
  \node[inner sep=0pt, anchor=north west] (tile)
    at ({#1*(\galw+\galgap)}, {-#2*(\galh+\galgap)})
    {\includegraphics[width=\galw, height=\galh]{figures/showcase/gal-#3.png}};
  \draw[black!25, line width=0.3pt] (tile.south west) rectangle (tile.north east);
  \node[anchor=north west, fill=#5, text=white, rounded corners=0.8pt,
    inner xsep=1.6pt, inner ysep=1pt,
    font=\sffamily\bfseries\fontsize{6pt}{6.5pt}\selectfont]
    at ([shift={(1.5pt,-1.5pt)}]tile.north west) {#4};}
\makebox[\linewidth]{\begin{tikzpicture}
\galtile{0}{0}{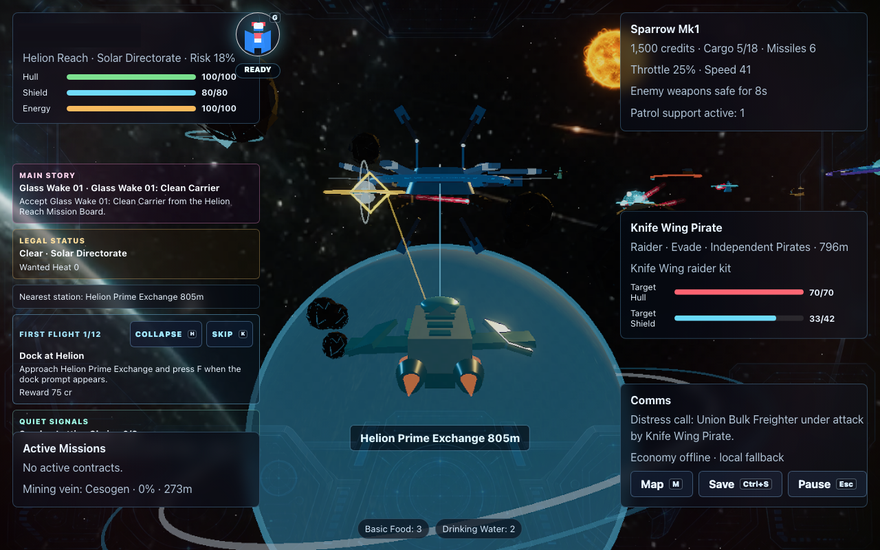}{SpaceRPG}{famBrowser}
\galtile{1}{0}{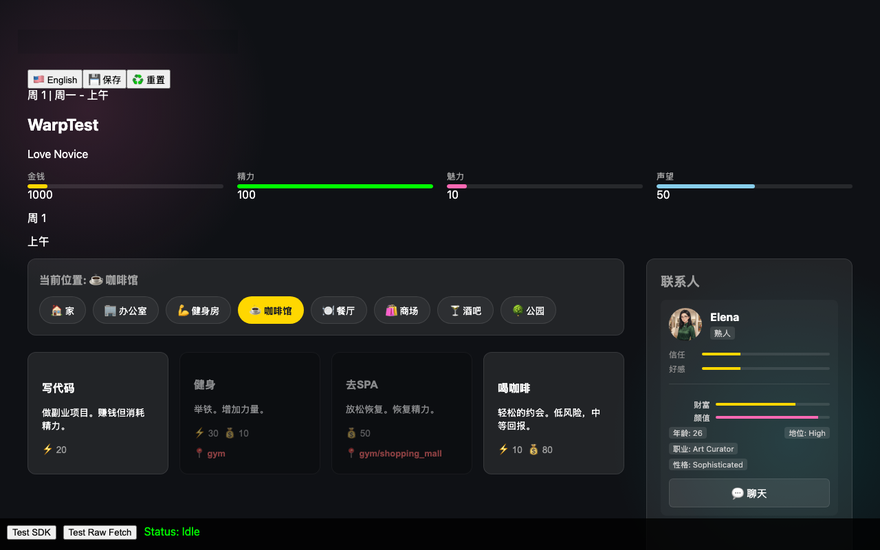}{DatingSim}{famBrowser}
\galtile{2}{0}{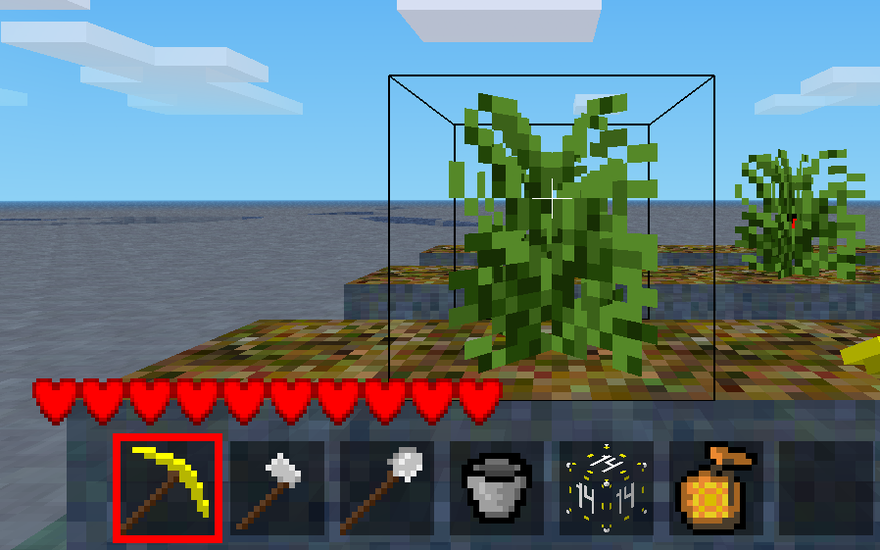}{Luanti}{famNative}
\galtile{3}{0}{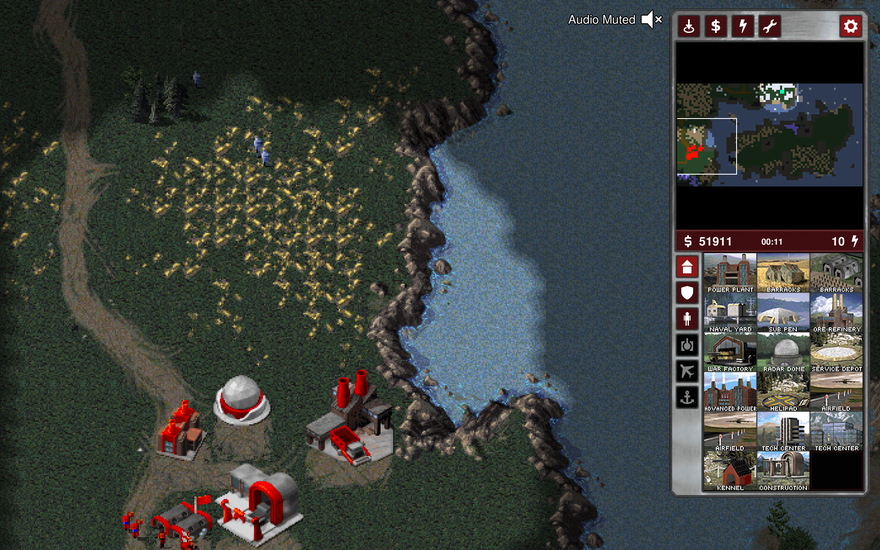}{OpenRA}{famNative}
\galtile{4}{0}{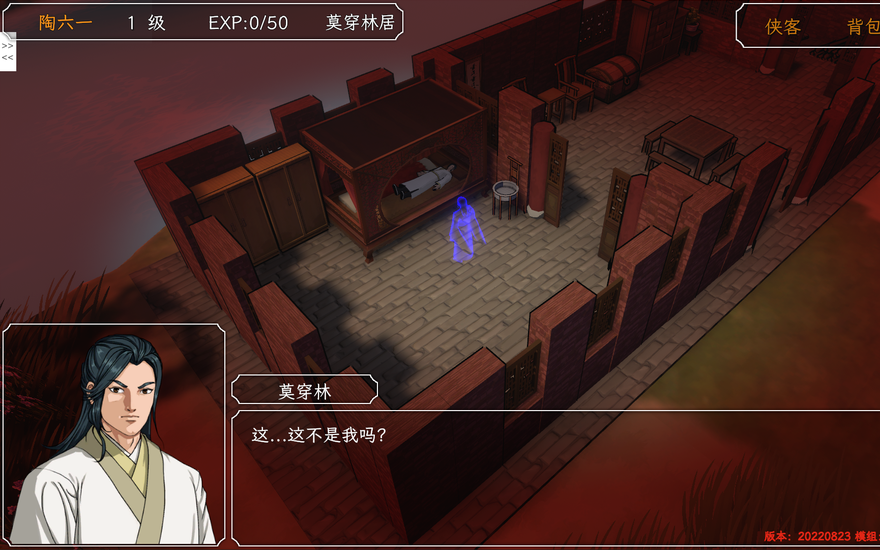}{jynew}{famUnity}
\galtile{0}{1}{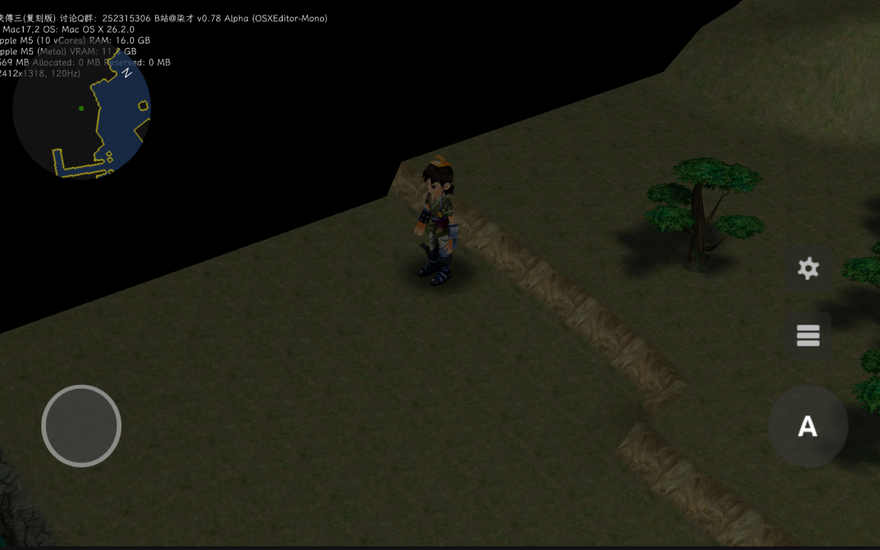}{PAL3}{famUnity}
\galtile{1}{1}{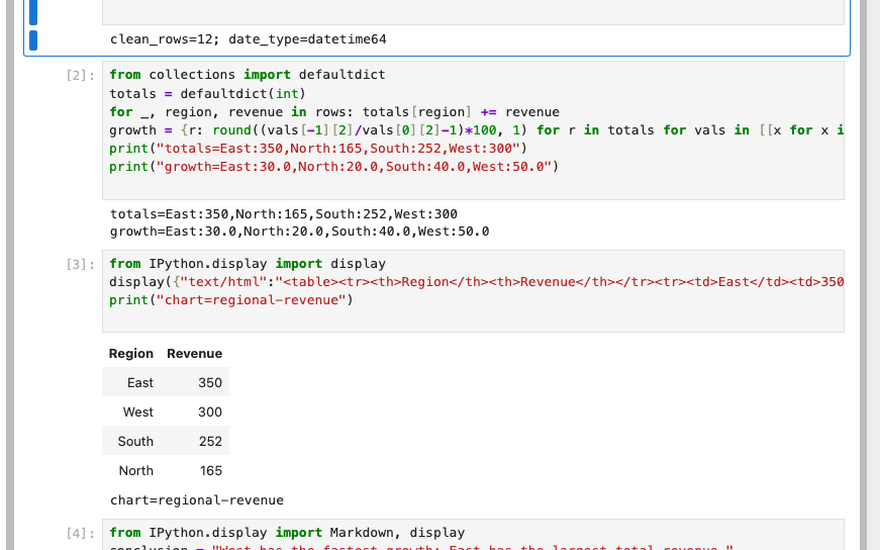}{JupyterLab}{famTools}
\galtile{2}{1}{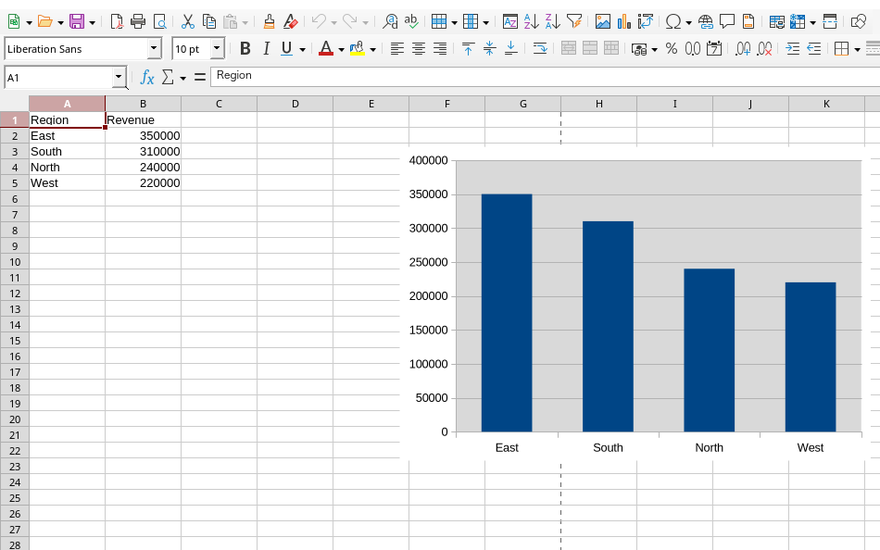}{LibreOffice}{famTools}
\galtile{3}{1}{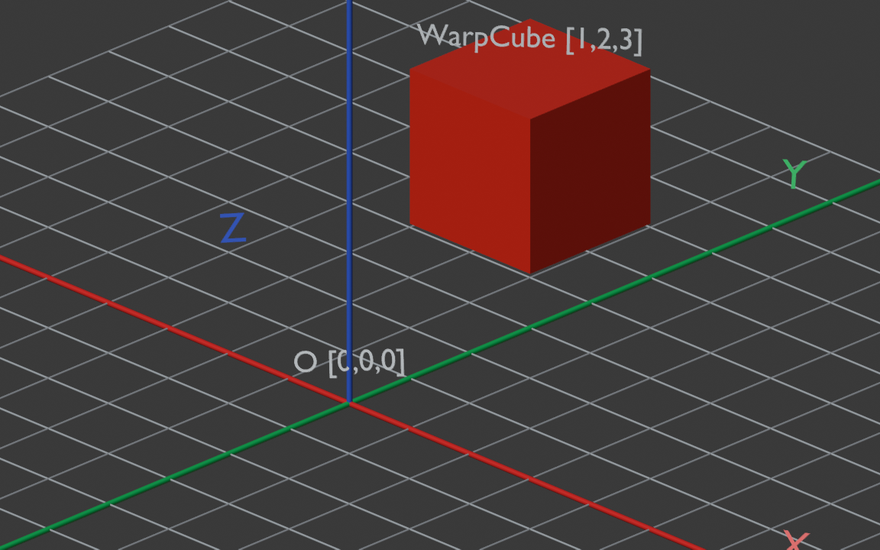}{Blender}{famTools}
\galtile{4}{1}{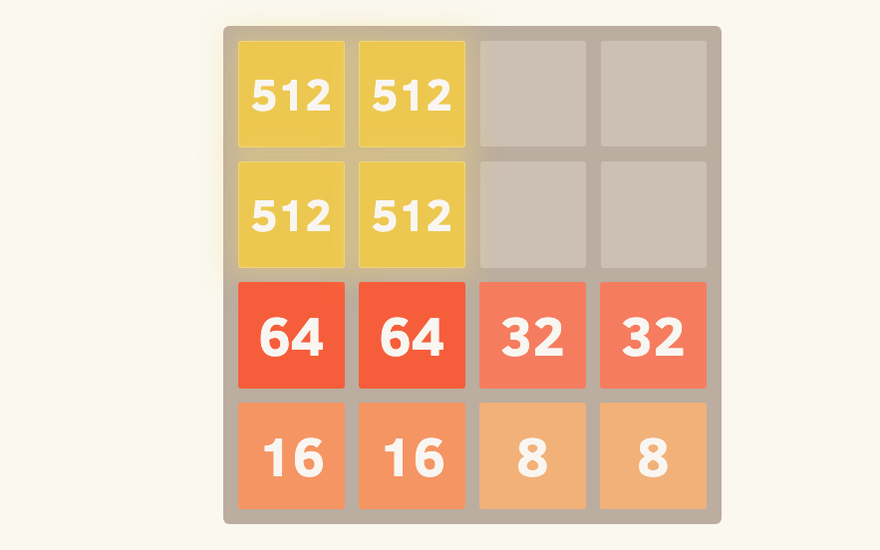}{2048}{famBrowser}
\end{tikzpicture}}%
\caption{Validated deep states of all ten applications, tagged by domain family (Figure~\ref{fig:task-profiles}).}
\label{fig:gallery}
\end{figure}

\begin{taskprofilesfigure}
\centering
\includegraphics[width=\linewidth, trim=0 0 0 9bp, clip]{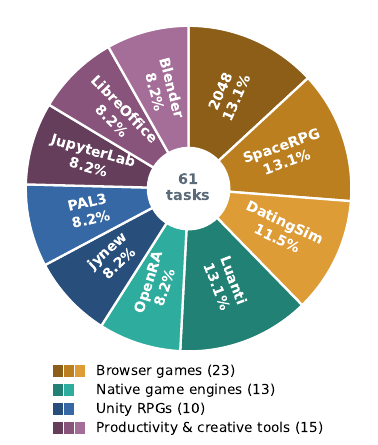}%
\vspace{-1em}
\caption{The 61 tasks by profile, colored by domain family (full form, Figure~\ref{fig:task-distribution}).}
\label{fig:task-profiles}
\end{taskprofilesfigure}
The suite spans two Unity RPGs, three browser games, Luanti, OpenRA, JupyterLab, LibreOffice, and Blender (Figures~\ref{fig:gallery} and~\ref{fig:task-profiles}, Table~\ref{tab:coverage}). These applications comprise 10.9M lines of source across five language families and six rendering stacks, ranging from 874 lines for 2048 to 2.8M for Blender and 7.2M for LibreOffice. All share the hook contract, checkpoint schema, and evaluation protocol. Seven application profiles (46 tasks) are games, including four self-rendered games that form our primary target domain, and the remaining three (15 tasks) are general-purpose applications. Table~\ref{tab:coverage} also records the channel through which each task's registered transition spec delivers the target action. The channel is a property of the task: \confgui{} and \confguicp{} always act through pixels, and every trial is scored against the same hidden targets.

\subsection{Metric-Grounded Task Provenance}
\label{sec:task-provenance}
\label{sec:task-selection}
\label{sec:ground-truth}
\label{sec:provenance-gates}

Each task is produced by a three-stage mechanical pipeline driven by fixed public inputs (Appendix~\ref{sec:appendix-provenance}). \emph{Selection} converts per-function complexity, nesting, fan-in/fan-out, and git-churn signals \citep{mccabe1976complexity,henry1981software,nagappan2005churn} into percentile ranks combined under weights frozen before ranking, then keeps the regions whose state is observable and controllable through a discovered semantic surface, as execution-based benchmarks in the OSWorld lineage \citep{osworld_arxiv24} also require. The criterion admits only deterministic transitions expressible through that interface. \emph{Ground truth} is recorded from one controlled real execution of branch-covering seed states through the hook. These seeds and their follow-on actions are the pipeline's only authored input, and assertions are restricted to the \emph{replay-stable} projection of the after-state. \emph{Discrimination} confines language-agnostic single-point mutants \citep{demillo1978hints,jia2011analysis} to the region's line span and excludes regions with low mutation kill rates. A final \emph{phrasing} step paraphrases the specification into a bug-report-style issue under a guard requiring equivalence of the specification tokens. Four gates verify the pipeline (Table~\ref{tab:provenance-gates}): re-running selection reproduces the ranking, and on the 2048 pilot region, assertions hold on an unpinned run (3/3), the leakage guard passes on both phrasing variants (6/6), and the oracle kills 7 of 8 mutants. Catalog-wide, all 61 oracle-built references pass both programmatic qualification runs (122/122, Table~\ref{tab:reference-coverage}).

\subsection{Hybrid Programmatic-State and Visual Evaluation}
\label{sec:hybrid-eval}
\label{sec:execution-based-eval}
\label{sec:conjunctive-oracle}

A checkpoint may satisfy registered state checks while producing an incorrect rendered frame. We therefore evaluate both semantic state and visual output. A trial runs in two phases, \emph{Phase~A} establishing the Start and \emph{Phase~B} executing the bounded target action in the live software. A registered gate between them runs the programmatic Start check first, then the visual check on a frame captured with the runtime suspended, and both channels are checked again for the Goal verdict. Acceptance is the conjunction $O_{\text{prog}} \wedge O_{\text{vis}}$ of a \emph{Programmatic State Oracle}, which reads semantic state fields, persistence round-trips, and runtime health, and a \emph{VLM Oracle}, which grades each gate's frame against that stage's golden reference and a rubric of what must, must not, and may appear. Visual rubrics tolerate benign non-determinism and leave value-level checks to the programmatic oracle. Conjunction imposes a monotonicity constraint: $O_{\text{vis}}$ can downgrade a trial that passes the programmatic oracle to a failure but cannot accept a trial that fails it, so skipping visual evaluation when $O_{\text{prog}} = \textsc{Fail}$ saves a VLM call and leaves the verdict unchanged (Appendix~\ref{sec:visual-audit-cascade}).

\paragraph{Cross-VLM calibration.} \label{sec:cross-vlm-calibration} \label{sec:hybrid-eval-results} The evaluation judge is a single fixed model, \texttt{qwen3-vl-flash}. Two blinded VLM annotators (\texttt{glm-5.3-flash} and \texttt{doubao-seed-2-1-turbo}) independently label every sample, and \texttt{claude-sonnet-5} adjudicates only disagreements and \emph{uncertain} labels. The 200 samples are independent of every study in Section~\ref{sec:experiments}: 40 base states drawn under a frozen hash, each expanded into a clean before-frame, a clean after-frame, a same-profile wrong state, a blank render, and a distributed striped occlusion. All four pre-registered gates pass (Table~\ref{tab:calibration}). GLM and Doubao agree on 199 of 200 samples ($\kappa = 0.9896$). Against this consensus, Qwen's strict accuracy is 96.0\% ($\kappa = 0.9173$), with false-accept rate 3.39\% and false-reject rate 4.88\%. The 80 fault samples change no programmatic field and so pass the programmatic oracle by construction. The visual channel, and hence the hybrid oracle, rejects 77 of them (96.25\%, 95\% task-clustered bootstrap CI $[91.25\%, 100\%]$).

\section{Experiments}
\label{sec:experiments}

\subsection{RQ1: Can checkpoint synthesis improve deep-state testing efficiency?}
\label{sec:performance}

\subsubsection{Checkpoint Synthesis: Cost and Accuracy}
\label{sec:reproducible-protocol}

The \emph{\perfstudy{}} examines two factors: the source of the Start state and the interface used to execute the target action. \confgui{} \citep{gonzalezpumariega2025agents3}\label{sec:agent-s} runs from the initial screen with pixel execution. \confguicp{} restores a \method{}-synthesized Start, then executes only the target action with the same policy and pixel action space, isolating state construction. \method{} restores the same kind of Start but delivers the target action as the transition spec from the exploration that produced the checkpoint, isolating the execution interface. All three run on all 61 tasks under three decision models (\texttt{gpt-5.6-terra}, \texttt{gemini-3.5-flash}, and \texttt{kimi-k2.6}) and seeds 0--2: 1,647 registered trials in a Latin-square rotation blocked by model and seed, 549 per configuration. Table~\ref{tab:performance-main} adds a second pixel-policy family, \confcua{} and \confcuacp{}, which run ScaleCUA \citep{liu2026scalecua} cold and seeded, and a fourth decision model, \texttt{gpt-6-astra} \citep{openai2026astra}. These extension rows ($n=183$ each) are never pooled with the registered aggregates.

\begin{table}[t]
\caption{Performance-study results. Ckpt.\ marks rows that start from a restored synthesized checkpoint; the marked \confgui{} and \confcua{} rows are \confguicp{} and \confcuacp{} in the text. Start and E2E are success rates, and ``Goal $\mid$ Start'' is conditioned on trials that reached Start. Time is per-trial wall clock including synthesis. Mean token usage, cost, and time include terminal failures. TiPS is total time per E2E success, and ``\textemdash{}'' denotes none.}
\label{tab:performance-main}
\centering
\fontfamily{lmr}\selectfont
\floatsize
\setlength{\tabcolsep}{4.5pt}
\renewcommand{\arraystretch}{1.0}
\resizebox{\linewidth}{!}{\begin{tabular}{@{}>{\columncolor{white}[0pt][\tabcolsep]}llcS[table-format=2.1]S[table-format=2.1]S[table-format=2.1]rrr>{\columncolor{white}[\tabcolsep][0pt]}r@{}}
\toprule
Model & Agent & Ckpt. & \multicolumn{3}{c}{Success (\%) $\uparrow$} & \multicolumn{3}{c}{Mean $\downarrow$} & TiPS $\downarrow$ \\
\cmidrule(lr){4-6}\cmidrule(lr){7-9}
 & & & {Start} & {Goal $\mid$ Start} & {E2E} & tokens (k) & cost (\$) & time (s) & (s) \\
\midrule
 &  & \xmark{} & 8.7 & 0.0 & 0.0 & 302.0 & 1.148 & 234 & \textemdash \\
 & \multirow{-2}{*}{\confgui{}} & \cmark{} & 30.1 & 47.3 & 14.2 & 96.6 & 0.367 & 81 & 569.0 \\
\cmidrule(l){2-10}
 &  & \xmark{} & 10.9 & 0.0 & 0.0 & 193.2 & 0.734 & 126 & \textemdash \\
 & \multirow{-2}{*}{\confcua{}} & \cmark{} & 33.9 & 37.1 & 12.6 & 56.1 & 0.213 & 38 & 303.4 \\
\cmidrule(l){2-10}
\rowcolor{blue!4}
\multirow{-5}{*}{gpt-5.6-terra} & \textbf{\method{}} & \cmark{} & 30.6 & 80.4 & \bfseries 24.6 & 51.6 & 0.196 & 41 & 166.6 \\
\midrule
 &  & \xmark{} & 7.1 & 0.0 & 0.0 & 554.3 & 1.580 & 323 & \textemdash \\
 & \multirow{-2}{*}{\confgui{}} & \cmark{} & 47.0 & 24.4 & 11.5 & 181.4 & 0.517 & 159 & 1386.9 \\
\cmidrule(l){2-10}
 &  & \xmark{} & 12.6 & 13.0 & 1.6 & 444.0 & 1.265 & 241 & 14694.4 \\
 & \multirow{-2}{*}{\confcua{}} & \cmark{} & 51.9 & 21.1 & 10.9 & 119.6 & 0.341 & 100 & 917.6 \\
\cmidrule(l){2-10}
\rowcolor{blue!4}
\multirow{-5}{*}{gemini-3.5-flash} & \textbf{\method{}} & \cmark{} & 50.3 & 70.7 & \bfseries 35.5 & 115.1 & 0.328 & 84 & 236.2 \\
\midrule
 &  & \xmark{} & 7.7 & 0.0 & 0.0 & 368.6 & 0.552 & 294 & \textemdash \\
 & \multirow{-2}{*}{\confgui{}} & \cmark{} & 32.8 & 16.7 & 5.5 & 213.4 & 0.320 & 115 & 2111.2 \\
\cmidrule(l){2-10}
 &  & \xmark{} & 8.7 & 18.8 & 1.6 & 404.6 & 0.607 & 285 & 17357.2 \\
 & \multirow{-2}{*}{\confcua{}} & \cmark{} & 33.9 & 16.1 & 5.5 & 175.2 & 0.263 & 88 & 1613.8 \\
\cmidrule(l){2-10}
\rowcolor{blue!4}
\multirow{-5}{*}{kimi-k2.6} & \textbf{\method{}} & \cmark{} & 34.4 & 65.1 & \bfseries 22.4 & 153.5 & 0.230 & 68 & 305.4 \\
\midrule
 &  & \xmark{} & 21.3 & 0.0 & 0.0 & 248.9 & 4.280 & 205 & \textemdash \\
 & \multirow{-2}{*}{\confgui{}} & \cmark{} & 48.1 & 29.5 & 14.2 & 162.6 & 2.797 & 128 & 903.4 \\
\cmidrule(l){2-10}
 &  & \xmark{} & 18.0 & 21.2 & 3.8 & 233.7 & 4.020 & 164 & 4296.4 \\
 & \multirow{-2}{*}{\confcua{}} & \cmark{} & 46.4 & 30.6 & 14.2 & 146.7 & 2.524 & 94 & 660.0 \\
\cmidrule(l){2-10}
\rowcolor{blue!4}
\multirow{-5}{*}{gpt-6-astra} & \textbf{\method{}} & \cmark{} & 48.1 & 63.6 & \bfseries 30.6 & 135.7 & 2.333 & 87 & 285.8 \\
\bottomrule
\end{tabular}
}
\vspace{-6pt}
\end{table}

\textbf{State construction under a fixed pixel policy.} Pooled across models and seeds, \confgui{} reaches Start on 7.8\% of trials and 0\% E2E success, using 408.3k tokens on average, whereas \confguicp{} reaches Start on 36.6\% and records 10.4\% E2E (task-clustered bootstrap 95\% CI $[4.9,16.6]$) at 163.8k mean tokens. Per-model exact McNemar tests are significant after Holm correction ($p\leq0.00195$). At the request level (Figure~\ref{fig:request-decay}), checkpoint seeding cuts the 42.4 requests per completed trial of \confgui{} to 16.2, and \method{} to 7.4. A restore takes only 1--3~s.

\textbf{Execution interface after checkpoint restoration.} \method{} reaches Start on 38.4\% of trials and records 27.5\% E2E (95\% CI $[19.3,36.1]$) at 106.7k mean tokens. Because comparison with \confgui{} changes both factors, the informative contrast is \confguicp{}, which restores a Start from the same synthesis agent at a similar rate (36.6\%). The 17.1-point E2E gap between the two supports the benefit of delivering the target action through the generated transition spec.

\textbf{Joint synthesis versus delegated planning.} In the diagnostic \confdelegated{}, a separate bounded planner generates the target action from the test-case text, outside the exploration that built the Start. It records 24.2\% E2E against 27.5\% (Table~\ref{tab:delegated-diagnostic}). It reaches Start more often (43.9\% versus 38.4\%) but converts fewer Starts into Goal successes (55.2\% versus 71.6\%, a 16.4-point difference with CI $[7.0,25.9]$), consistent with a jointly synthesized transition fitting its checkpoint better.

\textbf{Additional pixel policies and decision models.} On the matched population (Table~\ref{tab:performance-matched-seed0}, $n=183$ per configuration), cold \confcua{} reaches Start on 8.7--12.6\% of trials and records at most 1.6\% E2E, against 7.1--8.7\% Start and no E2E success for \confgui{} under the three registered models. Seeded, \confcuacp{} and \confguicp{} both record 5.5--14.2\% E2E against 22.4--35.5\% for \method{}. The same ordering holds for \texttt{gpt-6-astra}, whose two cold configurations reach at most 3.8\% E2E: at similar Start rates (46.4--48.1\%), both seeded pixel configurations reach 14.2\% E2E and \method{} 30.6\%, at 135.7k mean tokens against 248.9k for cold \confgui{}.

\subsubsection{Checkpoint Reuse: Reliability and Regression Cost}
\label{sec:replay}

The \emph{\replaystudy{}} re-executes frozen checkpoints through the execution and grading path of the \perfstudy{}, with all model calls disabled. We replay 112 checkpoints from 8 profiles 5--13 times each across two environments, for 1{,}361 graded replays. Of these, 98.90\% reproduce the original verdict (bootstrap 95\% CI $[96.8\%, 100\%]$, clustered by checkpoint), including all 328 replays of the 29 checkpoints accepted in the \perfstudy{}. The median replay takes 422~ms. One checkpoint is inconsistent across replays, and one other is consistent but differs from its original verdict (Appendix~\ref{sec:ablations}). Because a replay adds no model calls, \method{}'s cumulative token cost stays below both \confgui{} configurations from the first run on, for every model (\texttt{gpt-5.6-terra}, first run: 51.6k against 302.0k for cold \confgui{} and 96.6k for \confguicp{}).

\begin{rqanswer}{1}
\method{} improves test completion and enables reuse without model calls. Most of the cost of deep-state testing lies in reaching the state rather than in the tested action, so a checkpoint synthesized once amortizes that cost over model-free, sub-second replays.
\end{rqanswer}

\subsection{RQ2: Can checkpoint synthesis expose real defects the registered tests miss?}
\label{sec:latent-defect-results}
\label{sec:latent-defect}

The \emph{\latentstudy{}} asks whether synthesized checkpoints expose visual defects that the registered tests miss while the software keeps running \citep{playcoder2026}. Its population (Table~\ref{tab:latent-defect-tasks}) is 11 real issues with merged fixes (6 JupyterLab, 3 Luanti, 2 OpenRA) that have no fail-to-pass (F2P) test: every registered pass-to-pass (P2P) test passes on \emph{both} the fixed version and the buggy version, which we reconstruct by reversing the fix's production-code changes. F2P absence is an admission requirement, because issue-to-test generation succeeds on about one third of real issues or fewer \citep{jimenez2024swe,swtbench2024,otter2025} and defects confined to rendered output require visual evaluation \citep{visiondroid2025}. Both versions must reach visually equivalent Start states before their post-transition outputs are compared. \confgui{} and \confcua{} reach Start from the initial screen through pixel actions, whereas \confguicp{}, \confcuacp{}, and \method{} restore a shared synthesized checkpoint, after which the first two act through pixels and \method{} through generated semantic actions. The comparison covers four decision models, two versions, and five configurations: 440 cells, 88 each. We analyze 430 after dropping one ten-cell group that exceeded its wall-clock budget (Table~\ref{tab:latent-defect-results}).

\begin{table}[t]
\caption{Latent-defect-study results. Columns follow Table~\ref{tab:performance-main}, over versioned cells rather than trials. The three means are the method-attributed synthesis, restore, and execution totals per analyzed cell. Table~\ref{tab:latent-defect-cost} provides the cost-attribution rule and complete breakdown.}
\label{tab:latent-defect-results}
\centering
\fontfamily{lmr}\selectfont
\floatsize
\setlength{\tabcolsep}{4.5pt}
\renewcommand{\arraystretch}{1.0}
\resizebox{\linewidth}{!}{\begin{tabular}{@{}>{\columncolor{white}[0pt][\tabcolsep]}llcS[table-format=2.1]S[table-format=2.1]S[table-format=2.1]rrr>{\columncolor{white}[\tabcolsep][0pt]}r@{}}
\toprule
Model & Agent & Ckpt. & \multicolumn{3}{c}{Cell-level success (\%) $\uparrow$} & \multicolumn{3}{c}{Mean $\downarrow$} & TiPS $\downarrow$ \\
\cmidrule(lr){4-6}\cmidrule(lr){7-9}
& & & {Start} & {Goal $\mid$ Start} & {E2E} & tokens (k) & cost (\$) & time (s) & (s) \\
\midrule
 &  & \xmark{} & 36.4 & 12.5 & 4.5 & 343.5 & 0.608 & 266 & 5853.5 \\
 & \multirow{-2}{*}{\confgui{}} & \cmark{} & 63.6 & 7.1 & 4.5 & 80.7 & 0.123 & 77 & 1695.1 \\
\cmidrule(l){2-10}
 &  & \xmark{} & 31.8 & 14.3 & 4.5 & 156.8 & 0.342 & 174 & 3946.2 \\
 & \multirow{-2}{*}{\confcua{}} & \cmark{} & 68.2 & 0.0 & 0.0 & 26.7 & 0.053 & 44 & \textemdash \\
\cmidrule(l){2-10}
\rowcolor{blue!4}
\multirow{-5}{*}{gpt-5.6-terra} & \textbf{\method{}} & \cmark{} & 59.1 & 38.5 & \bfseries 22.7 & 12.4 & 0.020 & 10 & 42.1 \\
\midrule
 &  & \xmark{} & 9.1 & 50.0 & 4.5 & 238.6 & 0.391 & 351 & 7018.7 \\
 & \multirow{-2}{*}{\confgui{}} & \cmark{} & 72.7 & 6.2 & 4.5 & 120.2 & 0.190 & 115 & 2303.5 \\
\cmidrule(l){2-10}
 &  & \xmark{} & 9.1 & 0.0 & 0.0 & 133.4 & 0.218 & 196 & \textemdash \\
 & \multirow{-2}{*}{\confcua{}} & \cmark{} & 68.2 & 0.0 & 0.0 & 76.0 & 0.117 & 48 & \textemdash \\
\cmidrule(l){2-10}
\rowcolor{blue!4}
\multirow{-5}{*}{gemini-3.5-flash} & \textbf{\method{}} & \cmark{} & 63.6 & 64.3 & \bfseries 40.9 & 72.5 & 0.111 & 35 & 78.2 \\
\midrule
 &  & \xmark{} & 9.1 & 0.0 & 0.0 & 86.5 & 0.064 & 225 & \textemdash \\
 & \multirow{-2}{*}{\confgui{}} & \cmark{} & 54.5 & 0.0 & 0.0 & 147.8 & 0.068 & 186 & \textemdash \\
\cmidrule(l){2-10}
 &  & \xmark{} & 9.1 & 0.0 & 0.0 & 114.7 & 0.074 & 118 & \textemdash \\
 & \multirow{-2}{*}{\confcua{}} & \cmark{} & 63.6 & 0.0 & 0.0 & 91.3 & 0.030 & 36 & \textemdash \\
\cmidrule(l){2-10}
\rowcolor{blue!4}
\multirow{-5}{*}{kimi-k2.6} & \textbf{\method{}} & \cmark{} & 59.1 & 23.1 & \bfseries 13.6 & 89.6 & 0.029 & 29 & 211.7 \\
\midrule
 &  & \xmark{} & 31.8 & 14.3 & 4.5 & 115.6 & 0.776 & 214 & 4699.7 \\
 & \multirow{-2}{*}{\confgui{}} & \cmark{} & 95.5 & 23.8 & 22.7 & 103.7 & 0.433 & 117 & 515.3 \\
\cmidrule(l){2-10}
 &  & \xmark{} & 31.8 & 14.3 & 4.5 & 91.3 & 0.287 & 119 & 2628.4 \\
 & \multirow{-2}{*}{\confcua{}} & \cmark{} & 86.4 & 15.8 & 13.6 & 62.9 & 0.181 & 49 & 356.5 \\
\cmidrule(l){2-10}
\rowcolor{blue!4}
\multirow{-5}{*}{gpt-6-astra} & \textbf{\method{}} & \cmark{} & 90.9 & 50.0 & \bfseries 45.5 & 60.2 & 0.154 & 37 & 82.1 \\
\bottomrule
\end{tabular}
}
\vspace{-6pt}
\end{table}

\textbf{Start reachability and post-restoration success.}
Pooled over models, the two cold pixel configurations reach Start on 21.6\% (\confgui{}) and 20.5\% (\confcua{}) of cells, against 71.6\% for \confguicp{} and \confcuacp{} and 68.2\% for \method{}. At similar Start rates, E2E still differs: 30.7\% for \method{}, 8.0\% for \confguicp{}, and 3.4\% for \confcuacp{}, against 2.3\% for cold \confcua{}. With \texttt{gpt-6-astra}, \method{} records 45.5\% E2E on 22 cells against 22.7\% for \confguicp{} and 13.6\% for \confcuacp{} at similar Start rates (90.9\%, 95.5\%, 86.4\%), while both unseeded configurations record 4.5\%.

\textbf{Visual discrimination between buggy and fixed versions.}
For pixels that differ between the fixed and buggy post-transition references, we assign each observed frame to the closer reference without model inference. This separates the two versions on 7 of the 8 tasks where \method{} has a comparable pair, against 4/8 and 2/8 for the two checkpoint-seeded pixel configurations and 2/3 and 1/2 for the cold ones. It separates none of the 85 before-frame pairs and 4 of 26 same-version pairs, an observed 15.4\% false-separation rate (Appendix~\ref{sec:appendix-differential}). All 129 fixed-version cells of the three restoring configurations reuse the buggy-side checkpoint without further synthesis (Table~\ref{tab:latent-defect-cost}).

\begin{rqanswer}{2}
Checkpoint synthesis can expose real defects the registered tests miss. Because one checkpoint places the buggy and fixed versions in the same deep state, comparing their rendered outputs without a model gives a differential oracle for defects no fail-to-pass test encodes.
\end{rqanswer}

\section{Related Work}
\label{sec:related}

\textbf{GUI and computer-use agents.} Pixel-driven GUI agents such as the Agent S line \citep{agashe2025agents,agashe2025agents2,gonzalezpumariega2025agents3}, ScaleCUA \citep{liu2026scalecua}, and UI-TARS \citep{qin2025ui} map screenshot observations to coordinate-based actions. OSWorld and OSWorld-2 established execution-grounded, programmatic-state scoring for computer-use evaluation \citep{osworld_arxiv24,osworld2_2026}, and OS-Harm shows that such an oracle can miss failures observable only in rendered output \citep{osharm2025}. Our hybrid oracle keeps the programmatic channel as the necessary condition and adds a visual channel under an explicit monotonicity constraint. None of these evaluations asks the agent to synthesize the deep starting state itself.

\textbf{Benchmarks for GUI agents and game testing.} Agent benchmarks score navigation or task completion in benchmark-prepared environments, on desktop and professional software \citep{osworld_arxiv24,windowsworld2026,prosoftarena2026,ppteval2026}, on mobile, web, and multi-platform GUIs \citep{mobileworld2026,venusbench2026,colorbench2026,gta2026,mmbenchgui2026}, on full story arcs in adventure games \citep{flashadventure2025}, and on exploratory defect discovery \citep{guitester2026} (Table~\ref{tab:benchmark-comparison}). LLM agents play-test games, where progress-blocking failures rank among the most common defects \citep{titan_gametesting2025} and softlocks are a recognized level-design hazard \citep{stuckinthemiddle2025}. PlayCoder checks whether LLM-generated GUI code is playable \citep{playcoder2026}, and VisionDroid finds non-crash functional bugs from the screen \citep{visiondroid2025}.

\textbf{State construction in software testing.} Android GUI testing reaches deep states without full replay by constructing launching contexts \citep{fax_icse2020}, snapshotting and rolling back VM state \citep{timemachine_icse2020}, and recomposing mined use-case fragments \citep{combodroid_icse2020,epidroid2026}. These approaches use Android-specific state-inspection mechanisms, whereas \method{} generates a hook and capability contract for each application (Section~\ref{sec:synthesis-validation-reuse}). Hand-written fixtures and debug save states \citep{hamster_icse2026,gametestsurvey_ast2021} provide reusable starting states but require developers to encode them. A further line generates executable tests from natural language, by reproducing a described bug \citep{libro2023,swtbench2024,reproagent2026} or by synthesizing end-to-end test scripts from a specification \citep{genia_e2etest2025,xtestgen2025,autoe2e2025}. Our task additionally requires valid application-native checkpoints.

\section{Conclusion}
\label{sec:conclusion}

We propose \method{}, an agent that synthesizes deep-state checkpoints from a functional test case and read-only source, and \benchmark{}, 61 tasks over ten applications scored by a conjunctive oracle. Our results support checkpoint synthesis as a complementary mechanism for deep-state GUI testing. Under the same pixel action space and conjunctive oracle, seeding \confgui{} with synthesized checkpoints raises completion from 0\% to 10.4\%, and \method{}, which also executes the transition through the hook, completes 27.5\% (Section~\ref{sec:performance}). Replays reproduce 98.90\% of 1{,}361 verdicts without model calls, including every replay of an accepted checkpoint. On real repairs missed by every registered test, \method{}'s checkpoints separate the buggy and fixed versions on 7 of 8 comparable tasks in this diagnostic study. Extending it to more real repairs and seeds would measure how reliably synthesized tests separate regressions from fixes. We open-source our code and benchmark at \url{\repourl}.

\subsection*{AI use statement}
In this work, we used generative AI tools to help polish portions of the manuscript text (including translation and language editing). We also used generative AI tools to help write data-conversion scripts. We did not use generative AI tools to design the research methodology, formulate the paper's claims, or interpret the reported results. The LLM-based agents evaluated in this paper (e.g., \confgui{}, \method{}, and \confdelegated{}) are research components whose experimental roles are described in the relevant method and experiment sections. Their experimental use is distinguished here from the AI assistance in preparing the manuscript and implementing the research software disclosed above. All AI-assisted text and code were reviewed and verified by the authors before inclusion, and we take responsibility for the final content of this work.

\subsection*{Ethics statement}
This work involves no human-subjects data collection: the only annotation used
anywhere in the paper is blinded, model-only cross-VLM labeling
(Section~\ref{sec:cross-vlm-calibration}), with no crowdworker or other human
annotator data. All ten benchmark applications (Table~\ref{tab:coverage})
are open-source software exercised locally under their licenses; no
production service or third-party user data is accessed.

\subsection*{Reproducibility statement}
The experimental protocol (models, seeds, etc.) is
specified in Section~\ref{sec:reproducible-protocol}, with the manifest/content hashes, and per-stage gates in the appendix sections, and the benchmark's mechanical
task-provenance pipeline (selection, ground truth, discrimination, and the four
gates of Table~\ref{tab:provenance-gates}) is detailed in
Section~\ref{sec:task-provenance} and Appendix~\ref{sec:appendix-provenance}.
The appendix further gives the checkpoint-synthesis algorithm
(\ref{sec:appendix-algorithm}), the hook-onboarding protocol and its independent
replication (\ref{sec:appendix-hook-onboarding}), the pixel action space of the
\confgui{} and \confcua{} baselines (\ref{sec:appendix-action-space}), the
harness configuration keys (\ref{sec:appendix-config-keys}), the runtime-hold and
Start-parity protocols (\ref{sec:appendix-runtime-hold} and~\ref{sec:appendix-start-parity}),
the delegated-transition diagnostic (\ref{sec:appendix-e-prime}), the
execution-channel strata and the \confcua{} failure audit
(\ref{sec:appendix-channel-strata} and~\ref{sec:appendix-cua-failures}), the
evaluation-efficiency and reference-control analyses
(\ref{sec:evaluation-efficiency} and~\ref{sec:ablations}), and the
complete per-model, per-seed, and per-task result tables underlying every number
in Section~\ref{sec:experiments}
(\ref{sec:appendix-performance-full}--\ref{sec:appendix-latent-defect}). Every reported statistic is computed from archived run records by
the result-aggregation and evidence-collection procedures, which verify task,
specification, and report hashes before generating any table. The supplementary
evidence includes aggregate results and complete trial-level records. Source code for the \method{} harness,
benchmark adapters, and evidence-collection scripts is available at the
\repokind{} repository \url{\repourl}.

\bibliographystyle{assets/plainnat}
\bibliography{iclr2027_conference}

\clearpage
\appendix
\renewcommand{\topfraction}{0.92}
\renewcommand{\bottomfraction}{0.75}
\renewcommand{\textfraction}{0.06}
\renewcommand{\floatpagefraction}{0.80}
\setcounter{topnumber}{3}
\setcounter{bottomnumber}{2}
\setcounter{totalnumber}{5}
\section{Motivation}
\label{sec:appendix-motivation}

\subsection{A Motivating Example}
\label{sec:appendix-motivating-example}

The cost of deep-state testing is not uniform across interactive software: it depends almost entirely on what the runtime lets a test harness reach into. We make this concrete by contrasting one end of the spectrum, a browser game whose entire state is addressable from the test process, with the other end, engine-backed software, whose state the harness cannot reach.

\paragraph{The cheap end: a browser game.} SpaceRPG ships a Playwright suite in which a single test walks the first two chapters of the main storyline: an opening cinematic, mission acceptance, an inter-system jump, docking, two boss kills, salvage recovery, and a debriefing at a second station. Listing~\ref{lst:spacerpg-smoke} abridges that 140-line TypeScript test to its two load-bearing steps.

\begin{lstlisting}[style=tscode,caption={Abridged SpaceRPG smoke test with story names neutralized. Comments are ours.},label={lst:spacerpg-smoke}]
test("smokes the first two story chapters", async ({ page }) => {
  await resetApp(page);
  await startNewGame(page, { keepIntro: true });

  // (1) The store is callable from the test process through a window handle, so the
  // dialogue can be dismissed by name rather than by locating and clicking a control.
  await page.evaluate(() => {
    (window.__SPACERPG_E2E__!.getState() as { closeDialogue: () => void }).closeDialogue();
  });
  // ... accept the first mission, then jump systems and dock via the store's own actions ...

  // (2) Kill the first boss deterministically instead of aiming and firing in real time:
  // co-locate the player and the target at the origin, push every other enemy far away
  // so it cannot interfere, inject one lethal projectile owned by the player, and then
  // advance the simulation by exactly one fixed 50 ms step.
  await page.evaluate(() => {
    const e2e = window.__SPACERPG_E2E__!;
    const state = e2e.getState() as { player: Record<string, unknown>; runtime: {
      enemies: Array<Record<string, unknown>>; projectiles: Array<Record<string, unknown>>; } };
    e2e.setState({
      targetId: "scout-drone",
      player: { ...state.player, position: [0, 0, 0] },
      runtime: { ...state.runtime,
        enemies: state.runtime.enemies.map((ship) => ship.id === "scout-drone"
          ? { ...ship, position: [0, 0, 0] } : { ...ship, position: [900, 0, 900] }),
        projectiles: [{ id: "e2e-drone-shot", owner: "player", kind: "laser",
          position: [0, 0, 0], direction: [0, 0, 0], speed: 0, damage: 999, life: 1,
          targetId: "scout-drone" }] }
    });
    (e2e.getState() as { tick: (delta: number) => void }).tick(0.05);
  });
  // The kill must be observable through the player-facing surface, not only in the store.
  await expect(page.getByTestId("story-notification")).toContainText("Boss Defeated");
  await expect(page.getByTestId("story-tracker")).toContainText("Drone Prime");

  // ... a second boss by the same recipe, then salvage recovery and the debrief ...
});
\end{lstlisting}

Three properties of the runtime, not of the test, are what make this test writable in 140 lines and runnable in a commit-triggered pipeline. First, the game's state store is exposed to the test process as \texttt{window.\_\_SPACERPG\_E2E\_\_}, so the test both reads state (\texttt{getState()}) and writes it (\texttt{setState()}) in the same address space the game runs in. Second, the simulation step is itself callable: \texttt{tick(0.05)} advances physics and combat by one fixed 50~ms increment, which converts ``destroy a boss'' from a real-time skill problem into a deterministic state edit (place the target at the origin, exile the other enemies to \texttt{[900, 0, 900]} so they cannot interfere, inject a projectile with \texttt{damage: 999}, and step once). Third, the rendered result stays queryable: because the view is a DOM, the test can insist that the kill be visible on the player-facing surface (\texttt{story-notification}, \texttt{story-tracker}) rather than only in the store, which is what keeps the shortcut a shortcut in how the state is reached and not a weakening of the oracle. Outside the boss-kill shortcut, which edits positions directly, the writes are not arbitrary state injection: \texttt{closeDialogue} and the elided travel, docking, and mission-completion steps call the same store actions the UI dispatches, so the state the assertions see is production state a player could also have reached.

\paragraph{The cheap end, at scale.} SpaceRPG demonstrates the strongest case, but test counts alone do not establish GUI coverage. Table~\ref{tab:motivation-testing-gap} therefore separates repository scale and static complexity from test depth and runtime affordances. The five browser projects are larger (50.6--299.7 kLOC versus 2.8--57.6 kLOC), while both groups have P90 cyclomatic complexity between 4 and 8 and overlapping densities of high-complexity functions. All five browser projects run behavioral tests in CI; none of the nine engine projects does. The browser projects' 10{,}850 tests are nevertheless primarily unit, reducer, headless-simulation, jsdom, and API-integration tests rather than GUI E2E tests. Only OpenFrontIO includes a real-browser, interaction-reaching harness, and that harness sits outside its default CI suite, so no commit-triggered run exercises it.

\begin{table}[!htbp]
\centering
\caption{Scale, tests, and external runtime affordances across five browser and nine engine-backed games. Static complexity overlaps, while automation-ready state, view, and transition surfaces concentrate in browser games; no engine project has a full-runtime interaction test.}
\label{tab:motivation-testing-gap}
\begingroup
\fontfamily{lmr}\selectfont
\floatsize
\setlength{\tabcolsep}{2.5pt}
\renewcommand{\arraystretch}{1.08}
\begin{tabular*}{\linewidth}{@{\extracolsep{\fill}}lrrrrrccc@{}}
\toprule
 & Repository & \multicolumn{3}{c}{Production source} & Test suite & & Test depth & Runtime surface \\
\cmidrule(lr){2-2}\cmidrule(lr){3-5}\cmidrule(lr){6-6}\cmidrule(lr){8-8}\cmidrule(l){9-9}
Project & Stars/PRs & kLOC & CC$_{90}$ & H$_{10}$ & Tests/Cov. & CI & F/D & S/V/T \\
\midrule
\rowcolor{black!6}[0pt][0pt]
\multicolumn{9}{@{}l@{}}{\hspace{5pt}\rule[-3pt]{0pt}{13pt}\textbf{Browser games}} \\
\addlinespace[2pt]
habitica & 14{,}145/5{,}213 & 146.5 & 5 & 1.22 & 1{,}171/57.6\% & \cmark & \xmark/\xmark & \cmark/\cmark/\xmark \\
pokemon-showdown & 5{,}890/8{,}944 & 104.5 & 8 & 2.33 & 2{,}364/38.4\% & \cmark & \xmark/\xmark & \cmark/\xmark/\cmark \\
pokerogue & 5{,}837/4{,}721 & 299.7 & 5 & 0.93 & 3{,}028/62.2\% & \cmark & \xmark/\xmark & P/P/\xmark \\
OpenFrontIO & 2{,}658/2{,}801 & 166.4 & 6 & 1.92 & 4{,}026/60.8\% & \cmark & \cmark/\cmark$^\dagger$ & \cmark/P/\xmark \\
hanabi-live & 213/794 & 50.6 & 6 & 1.92 & 261/75.4\% & \cmark & \xmark/\xmark & \cmark/P/\xmark \\
\addlinespace[5pt]
\rowcolor{black!6}[0pt][0pt]
\multicolumn{9}{@{}l@{}}{\hspace{5pt}\rule[-3pt]{0pt}{13pt}\textbf{Unity games}} \\
\addlinespace[2pt]
jynew & 8{,}960/815 & 53.7 & 5 & 0.84 & 0/-- & \xmark & \xmark/\xmark & P/\xmark/\xmark \\
open-project-1 & 6{,}095/199 & 19.8 & 4 & 0.30 & 0/-- & \xmark & \xmark/\xmark & P/\xmark/\xmark \\
Pal3.Unity & 1{,}047/57 & 57.6 & 5 & 1.11 & 0/-- & \xmark & \xmark/\xmark & P/\xmark/\xmark \\
\addlinespace[5pt]
\rowcolor{black!6}[0pt][0pt]
\multicolumn{9}{@{}l@{}}{\hspace{5pt}\rule[-3pt]{0pt}{13pt}\textbf{Godot games}} \\
\addlinespace[2pt]
oh-my-git & 2{,}884/34 & 2.8 & 5 & 1.06 & 0/-- & \xmark & \xmark/\xmark & P/\xmark/\xmark \\
tabletop-club & 1{,}456/85 & 20.6 & 8 & 3.50 & 0/-- & \xmark & \xmark/\xmark & P/\xmark/\xmark \\
a-little-game-called-mario & 1{,}451/426 & 10.7 & 4 & 1.12 & 0/-- & \xmark & \xmark/\xmark & P/\xmark/\xmark \\
\addlinespace[5pt]
\rowcolor{black!6}[0pt][0pt]
\multicolumn{9}{@{}l@{}}{\hspace{5pt}\rule[-3pt]{0pt}{13pt}\textbf{Unreal Engine games}} \\
\addlinespace[2pt]
ActionRoguelike & 4{,}584/0 & 11.6 & 4 & 0.43 & 0/-- & \xmark & \xmark/\xmark & P/\xmark/\xmark \\
EpicSurvivalGame & 3{,}397/10 & 8.2 & 5 & 0.61 & 0/-- & \xmark & \xmark/\xmark & \xmark/\xmark/\xmark \\
Bomber & 381/68 & 24.6 & 5 & 1.10 & 0/-- & \xmark & \xmark/\xmark & P/\xmark/\xmark \\
\bottomrule
\end{tabular*}
\par\vspace{4pt}
\begin{minipage}{\linewidth}
\floatsize
\raggedright
\textit{Notes.} CC$_{90}$ is nearest-rank P90 per-function McCabe complexity; H$_{10}$ is functions with CC $\geq 10$ per physical kLOC. ``Tests/Cov.'' gives declared first-party cases and measured line coverage. CI means default-branch behavioral tests. ``F/D'' means full-runtime test/interaction-reaching full-runtime test. ``S/V/T'' means externally callable live state/queryable semantic view/deterministic transition surface. ``P'' denotes partial access limited to persistence, a manual console, a protocol-specific state channel, or DOM controls beside a canvas. $\dagger$ OpenFrontIO's real-browser harness is not in default CI. The reported browser test counts are predominantly non-GUI tests.
\end{minipage}
\endgroup
\end{table}

We measured coverage rather than trusting repository badges: each browser repository's own test command was run at its default branch under a coverage tool (\texttt{vitest} with the V8 provider for Pok\'eRogue and OpenFront, \texttt{jest} for Hanabi Live, \texttt{c8} over \texttt{mocha} for Pok\'emon Showdown, and \texttt{nyc} over \texttt{mocha} for Habitica), with line coverage taken over the repository's own source. Pok\'emon Showdown's 38.4\% includes large static data tables; its battle simulator alone reaches 83.5\%. Habitica's full suite requires MongoDB, so we ran its three independent suites covering shared game rules; the other repositories ran every loadable suite. The supplementary evidence records the scripts, scopes, commits, and raw counts behind every metric and classification in Table~\ref{tab:motivation-testing-gap}.

\paragraph{The expensive end: engine-backed software.} The three properties are not available as a uniform automation-ready surface in a self-rendered engine. A shipped Unity, Godot, or Unreal build presents an external harness with a native process and pixels; its internal state, scene/UI tree, and clock are not exposed as stable out-of-process read/write, query, and step APIs. Some surveyed games persist saves, provide a local console, or carry a narrow multiplayer protocol (P in Table~\ref{tab:motivation-testing-gap}), but adapting any of these requires application-specific instrumentation. The consequence is visible across vendors: all three engines ship a test framework, yet these projects leave it unused.

We surveyed 47 open-source playable games on GitHub (22 on Unity, 14 on Godot, and 11 on Unreal), excluding engines, libraries, editor tooling, and asset collections. Table~\ref{tab:motivation-testing-gap} reports three per engine, chosen to span genres, popularity, and maintenance activity and to include the two used by \benchmark{}. Across these nine projects, none declares a Unity \texttt{[Test]} or \texttt{[UnityTest]} method, a Godot GUT or gdUnit4 suite, or an Unreal \texttt{IMPLEMENT\_*\_AUTOMATION\_TEST}, \texttt{BEGIN\_DEFINE\_SPEC}, or \texttt{FAutomationTestBase} declaration. We verified each zero from a local clone and confirmed that each clone contains the game's source, ruling out both GitHub code-search limitations and missing source files in a partial checkout.

Across the wider survey the picture is the same on all three engines. Seventeen of the 22 Unity projects contain no first-party automated test. Twelve of the 14 Godot projects install no test framework; the two exceptions are Thrive (3{,}699 stars), which maintains a real gdUnit4 suite, and OpenVic (471), which ships the gdUnit4 addon over a directory holding only a placeholder file. None of the 11 Unreal projects declares an automation test in first-party source; the single \texttt{Tests/} directory the survey turned up, in WarriOrb (340), belongs to a vendored third-party dialogue plugin.

Directories named \texttt{Test/} or \texttt{tests/} rarely hold tests (Relintai/broken\_seals' \texttt{game/tests/} contains debug \emph{scenes}, and jynew's two \texttt{using NUnit.Framework} lines are unused editor-inserted imports), and four of the nine projects run 19 GitHub Actions workflows on their default branch, none of which executes a test. The projects that do carry tests make the pattern precise: engine developers test only what is reachable without the engine. \texttt{Unity-Technologies/FPSSample} (5{,}129 stars), for example, places every test in an Editor-only assembly over network serialization and delta compression, and Thrive's gdUnit4 suite follows the same rule. Across all 47 projects the survey contains exactly one test that renders at all, in \texttt{Unity-Technologies/BoatAttack} (2{,}781): a single \texttt{[UnityTest]} that loads a scene and compares one camera frame against a reference image. Even that one asserts a scene's initial render rather than any state reached by playing, so not a single test in the survey drives a game to a state that requires interaction to reach.

The contrast isolates what this paper pays for. Writing the SpaceRPG test required no infrastructure work: the browser had already granted the harness a state handle, a steppable clock, and a queryable view. Obtaining the equivalent for an engine-backed title requires building each of those affordances first, a per-application cost that no individual test amortizes, which is why it does not get paid, and why 1{,}694 merged pull requests across nine games on three engines ship with no automated coverage. \method{} therefore separates the two costs and automates both: an onboarding agent binds or synthesizes the hook and its capability contract once per application (440 and 456 added lines for PAL3 and jynew, driver and self-tests included, with no product logic rewritten; Table~\ref{tab:onboarding-replication-apps}), after which a checkpoint for a new target state is synthesized once per test.

\section{Additional Artifact Details}
\label{sec:appendix}

\subsection{Benchmark Composition}
\label{sec:appendix-coverage}

\begin{table}[!htbp]
\caption{\benchmark{} coverage: 10 applications and 61 tasks, with implementation languages, source sizes, and target-action delivery channels. Source size (kLOC) counts Git-tracked source lines at each application's frozen onboarding commit (Table~\ref{tab:onboarding-results}), restricted to its implementation directories using versioned counting rules. Delivery channels are derived from the transition specifications executed by \method{} and \confdelegated{}.}
\label{tab:coverage}
\begin{center}
\floatsize
\setlength{\tabcolsep}{3pt}
\begin{tabular*}{\linewidth}{@{\extracolsep{\fill}}lllrrrrr@{}}
\hline
Profile & Domain & Language & kLOC & Tasks & Native input & UI event & Source API \\
\hline
2048 & browser puzzle & JavaScript & 0.9 & 8 & 8 & 0 & 0 \\
SpaceRPG & browser space RPG & TypeScript & 11 & 8 & 1 & 0 & 7 \\
DatingSim & browser dating sim & TypeScript & 2.7 & 7 & 0 & 2 & 5 \\
Luanti & native voxel engine & C++/Lua & 245 & 8 & 0 & 0 & 8 \\
OpenRA & native RTS engine & C\#/Lua & 283 & 5 & 0 & 0 & 5 \\
JupyterLab & notebook IDE & TypeScript/Python & 277 & 5 & 0 & 0 & 5 \\
jynew & Unity wuxia RPG & C\#/Lua & 53 & 5 & 0 & 0 & 5 \\
PAL3 & Unity RPG remake & C\# & 59 & 5 & 0 & 0 & 5 \\
LibreOffice & office suite (UNO) & C++/Python & 7{,}190 & 5 & 0 & 0 & 5 \\
Blender & creative tool (\texttt{bpy}) & C++/Python & 2{,}766 & 5 & 0 & 0 & 5 \\
\hline
Total & & & & 61 & 9 & 2 & 50 \\
\hline
\end{tabular*}
\end{center}
\end{table}

\begin{figure}[!htbp]
\centering
\includegraphics[width=0.92\linewidth]{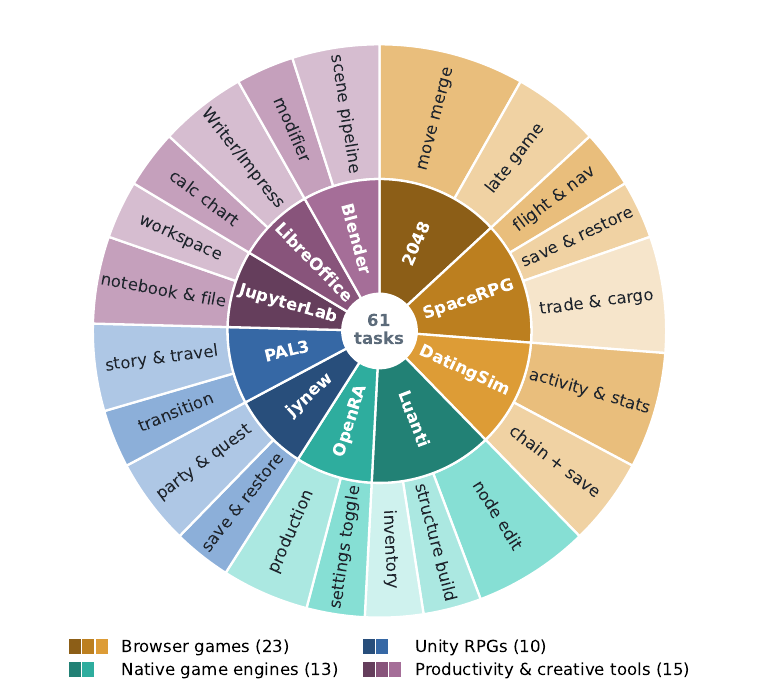}
\caption{Distribution of the \benchmark{} tasks, the full form of Figure~\ref{fig:task-profiles}. The inner ring is each task's registered profile, the outer ring the kind of operation it exercises. Colour marks the domain family (Table~\ref{tab:coverage}); shades within a family only separate its profiles and encode no ordering.}
\label{fig:task-distribution}
\end{figure}

Table~\ref{tab:coverage} gives the per-profile composition of \benchmark{} referenced throughout Section~\ref{sec:coverage}: each application's language and source size, its task count, and how many of its tasks deliver the target action by native input, a UI event, or a source-API call. Source counts are restricted to implementation directories, excluding bundled third-party code such as Blender's libraries and the Unity projects' asset-store plugins. Whole-repository counts would instead yield 657 kLOC for jynew and 3{,}947 kLOC for Blender, compared with the reported 53 and 2{,}766 kLOC. The counting rules are implemented in \texttt{scripts/collect\_app\_codebase\_stats.py}. \emph{SpaceRPG} and \emph{DatingSim} are placeholder names: both are open-source browser games whose upstream names are withheld \withheldreason{}, and both are released under their own names with the benchmark. The operation types in Figure~\ref{fig:task-distribution}'s outer ring are descriptive labels read off each task's registered \texttt{preconditions}/\texttt{reference\_steps}; they summarize task composition and are excluded from task schemas, candidate selection, checkpoint synthesis, and evaluation.

\paragraph{Delivery taxonomy.}
The three delivery columns record how each task's target action reaches the software, making the scope distinction of Section~\ref{sec:coverage} auditable rather than a prose claim. They are mechanically derived from each task's registered action type against the schema's own action-type vocabulary, whose three values are \texttt{native\_input}, \texttt{ui\_event}, and \texttt{source\_api}, so a schema change cannot silently desynchronize the column from the tasks it describes. Native input (9 tasks: 2048's 8 move tasks plus SpaceRPG's flight-capable task) delivers a real input event through the software's own input pipeline, bypassing only pixel grounding. A UI event (2 tasks, both DatingSim dialogue chains) dispatches a structured selector event to a real control's handler, bypassing hit-testing and occlusion. A source-API call (50 tasks) reaches a source-level interface (e.g., a test-gated hook, an engine API, UNO, \texttt{bpy}, a worldmod, or a registration probe), bypassing the UI input path entirely. This is a property of the \emph{transition spec}, so it applies only to configurations that execute one (\method{} and the diagnostic \confdelegated{}); \confgui{} and \confguicp{} always execute the target action by pixel input and are outside this column, whatever a task's registered spec says. All three are judged identically: every judgment reads the same real runtime state and the same real screenshot, so bypassing UI input changes only which defects can be \emph{triggered}, never which can be \emph{seen} (Appendix~\ref{sec:evaluation-efficiency} returns to this with a concrete case). The 50/61 source-API share follows from the selection stage (Section~\ref{sec:task-selection}), which admits only targets expressible as deterministic transitions over a discovered surface, rather than from the fraction of these applications' behavior that is source-drivable. Because the column describes the task rather than the configuration, and every configuration runs all 61 tasks, the outcomes of all four can still be read within each channel: Table~\ref{tab:performance-channel} splits the three-seed \perfstudy{} that way, which is what separates the contribution of the restored Start from that of the execution interface.

\subsection{Task Examples}
\label{sec:appendix-task-examples}

Table~\ref{tab:task-examples} shows, for three applications, the split that Section~\ref{sec:task} defines: the preconditions a method is given, the Start it must restore, and the hidden checks it never sees.

\begin{table}[!htbp]
\centering
\begingroup
\fontfamily{lmr}\selectfont
\renewcommand{\ttdefault}{lmtt}
\caption{Examples of \benchmark{} preconditions and simplified hidden evaluation scripts. Screenshots show the restored Start states; descriptions and checks are abridged.}
\label{tab:task-examples}
\floatsize
\setlength{\tabcolsep}{3pt}
\renewcommand{\arraystretch}{1.04}
\lstset{
  language=Python,
  basicstyle=\ttfamily\floatsize,
  keywordstyle=\color{black},
  commentstyle=\color{black},
  stringstyle=\color{black},
  identifierstyle=\color{black},
  showstringspaces=false,
  columns=fullflexible,
  keepspaces=true,
  breaklines=true,
  breakatwhitespace=true,
  numbers=none,
  frame=none,
  aboveskip=0pt,
  belowskip=0pt,
  xleftmargin=0pt,
  xrightmargin=0pt
}
\begin{tabular}{@{}>{\raggedright\arraybackslash}p{\dimexpr.26\linewidth-3.12pt\relax}>{\raggedright\arraybackslash}p{\dimexpr.35\linewidth-4.20pt\relax}>{\raggedright\arraybackslash}p{\dimexpr.39\linewidth-4.68pt\relax}@{}}
\toprule
\textbf{Restored State} & \textbf{Preconditions} & \textbf{Evaluation Script (Simplified)} \\
\midrule
\textbf{Luanti}\par\smallskip
\includegraphics[width=\linewidth]{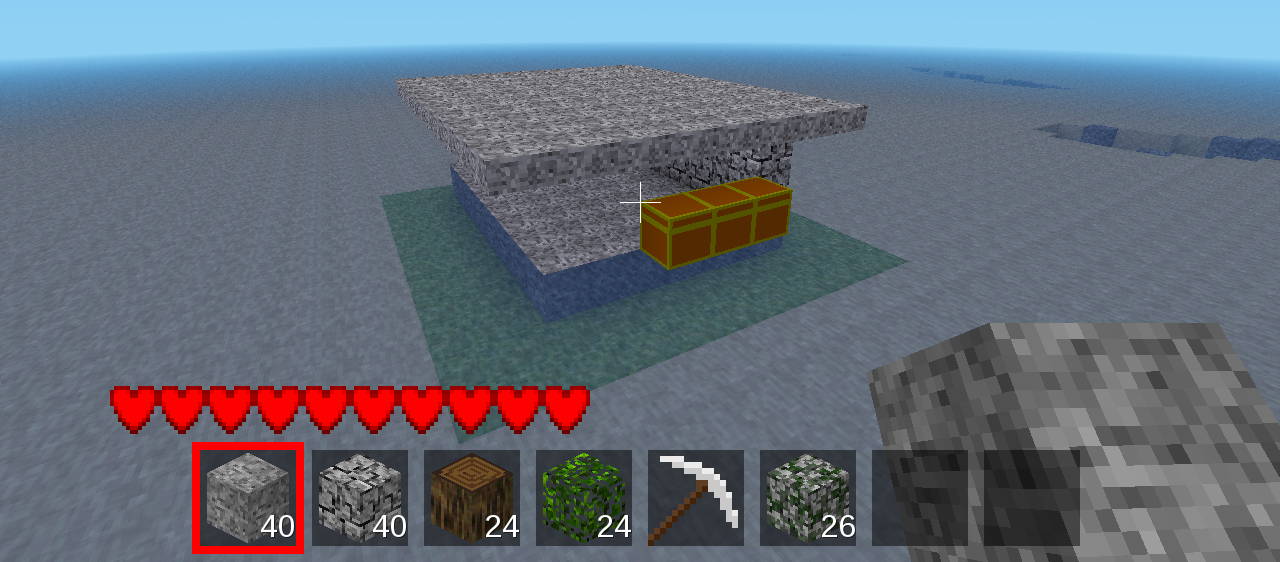}
&
A roofed workshop has three chests holding 40 stone, 24 tree nodes, and no items, respectively. The player has HP 20 and carries 40 stone, 40 cobble, 24 tree nodes, 24 leaves, one steel pick, and 26 mossy cobble.
&
\begin{minipage}[t]{\linewidth}\vspace{0pt}
\begin{lstlisting}
w = read_world(env)
for pos, item, n in [
    ((0,2,2), "stone", 80),
    ((0,2,3), "tree", 48),
    ((0,2,4), "steel_pick", 1),
    ((0,2,4), "mossycobble", 24)]:
    assert chest_count(w,pos,item) >= n
assert node_count(w,(4,2,3),(5,2,3),
                  "mossycobble") >= 2
\end{lstlisting}
\end{minipage}
\\
\addlinespace[3pt]
\midrule
\textbf{OpenRA}\par\smallskip
\includegraphics[width=\linewidth]{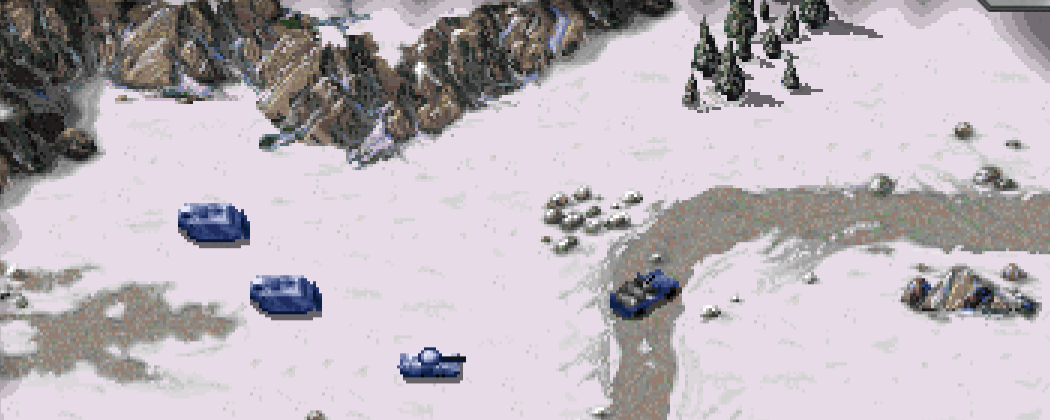}
&
Red Alert map \texttt{allies-08a} is initialized for Allied Greece with an undeployed MCV, sufficient cash, fast production enabled, and map anchor (26,64).
&
\begin{minipage}[t]{\linewidth}\vspace{0pt}
\begin{lstlisting}
g = read_gameplay_probe(env)
for a in ["powr", "proc", "tent",
          "weap", "2tnk"]:
    assert newly_built(g,"Greece",a) >= 1
assert prerequisite_respected(
    g, "Greece", "2tnk")
\end{lstlisting}
\end{minipage}
\\
\addlinespace[3pt]
\midrule
\textbf{LibreOffice Calc}\par\smallskip
\includegraphics[width=\linewidth]{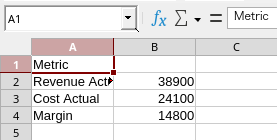}
&
A Calc workbook has January, February, March, and Summary sheets. Monthly actual revenue/cost pairs are 11,800/7,600, 13,200/8,100, and 13,900/8,400. Summary formulas compute quarterly revenue, cost, and margin.
&
\begin{minipage}[t]{\linewidth}\vspace{0pt}
\begin{lstlisting}
d = read_calc_via_uno(env)
assert cell_value(
    d, "Summary", "B2") == 40000
assert cell_value(
    d, "Summary", "B4") == 16000
assert sheet_exists(d, "March")
assert persistence_checks_pass(env)
\end{lstlisting}
\end{minipage}
\\
\bottomrule
\end{tabular}
\par\vspace{3pt}
\begin{minipage}{\linewidth}
\floatsize\raggedright
\textit{Shared protocol.} Programmatic Start check $\wedge$ visual Start check $\rightarrow$ agent transition $\rightarrow$ listed After checks $\wedge$ runtime checks $\wedge$ visual After check. A failed check stops evaluation. Code is grader-only pseudocode; visual checks use the full reference frames and rubrics, while screenshots here are cropped.
\end{minipage}
\endgroup
\end{table}

\subsection{Core Protocol Algorithm}
\label{sec:appendix-algorithm}

\textbf{Admissible outputs and decision rule.}
The public admissible set is
\begin{equation}
\mathcal C_{\mathrm{pub}}(u,G,\Sigma)=
\left\{c:\mathrm{Schema}(c)\land\mathrm{Conform}_{u,\Sigma}(c)
\land\mathrm{Evidence}(c,G)\right\}.
\label{eq:public-checkpoint-set}
\end{equation}
Here $\mathrm{Schema}$ checks the output envelope and field types;
$\mathrm{Conform}_{u,\Sigma}$ checks the task's public target kind, payload and
restoration interface, validity declaration, and registered action signatures;
$\mathrm{Evidence}$ verifies that citations accompanying $c$ refer to accessed files and contain no private
grader references. These checks establish source access but do not verify that the cited
code supports the model's interpretation. The fixed prompt $\mathcal P$ assigns the public
preconditions to \texttt{target}, the tested operation to \texttt{actions}, and
the predicted after-state to \texttt{assertions}. Its semantic instruction is
``Never place inferred after-state values into target.'' Both initial generation and the optional format repair are subject to these public constraints on checkpoint structure and source evidence.

Algorithm~\ref{alg:synthesis-agent} returns the first final candidate admitted by
these checks, after at most one repair for public format errors, or returns
failure. It performs no candidate ranking, learned verifier selection, or search
against hidden verdicts. The exploration budget is 16 response turns with at
most one repository query each; a query is \texttt{list\_dir}, \texttt{grep}, or
\texttt{read\_file}. Responses are limited to 8{,}192 tokens, and all configurations use the
same bounded retry policy for empty responses. The generated
transition is a finite list of registered actions, executed once within the
2{,}400-second task budget, without an additional global limit on action count.

Public admissibility is distinct from runtime validity. After restoration,
$s_0=J_\tau(\eta)$ must satisfy $\mathrm{Valid}_{\mathrm{Start}}(c;s_0)$ before
the hidden Start checks run. Task-level correctness additionally requires the
public preconditions and the requested operation to be represented faithfully.
The prompt instructs the agent to leave that operation for the transition;
the grader tests its registered Start and Goal observations. These observations provide only a partial assessment of semantic correctness. For a read-only or idempotent operation, a valid
Start may already satisfy Goal. Evaluation therefore depends on the coverage of the registered observations with respect to the specified operation.

\begin{table}[t]
\centering
\floatsize
\caption{Responsibility and feedback boundary of checkpoint checks. Runtime and
hidden-oracle outcomes are never returned to the synthesis agent.}
\label{tab:synthesis-check-boundaries}
\begin{tabular}{p{0.25\linewidth}p{0.42\linewidth}p{0.23\linewidth}}
\toprule
Condition & Enforcement & Synthesis feedback \\
\midrule
Output shape and interface & Public parser, task contract, registered action signatures & One format repair \\
Source-access evidence & Cited paths must have been accessed; private references rejected & No additional retry \\
Restoration and validity & Live state checks, runtime health, render readiness, and declared persistence requirements & None \\
Task-specific Start and Goal & Hidden programmatic and visual predicates, with Start gating the transition & None \\
No pre-applied transition & Prompt requirement; assessed only through registered task observations & No oracle feedback \\
Predicted assertions $A$ & Retained as auxiliary checkpoint outputs; excluded from acceptance & None \\
\bottomrule
\end{tabular}
\end{table}

Algorithm~\ref{alg:synthesis-agent} describes the agent-visible exploration and joint generation. Algorithm~\ref{alg:checkpoint-synthesis} places its output in the core runtime protocol. The grader's hidden target and oracle are passed only to the evaluation stages. Here $\mathrm{Valid}_{\mathrm{Start}}$ is the first stage of Section~\ref{sec:task}: restored semantic state, the persistence its declared validity class demands, runtime health, and render readiness. The single execution of $\alpha$ supplies $\mathrm{Valid}_{\mathrm{Trans}}$; validation does not execute the transition a second time. The agent's own assertions $A$ are retained with the generated checkpoint; the registered hidden assertions determine the verdict independently of these predictions. We write $J_\tau(\eta)$ for restoring payload $\eta$ into the running instance through the hook under strategy $\tau$, and $\mathrm{Frame}(\cdot)$ for a state's captured frame, which is the only evidence the visual oracle receives. Each programmatic check short-circuits its visual counterpart, a failed Start prevents transition execution, and Start evidence is captured and held under the named protocol that Section~\ref{sec:execution-based-eval} registers for that profile's runtime backend.

\begin{algorithm}[!htbp]
\floatsize
\caption{Synthesis Agent: Source Exploration and Joint Generation}
\label{alg:synthesis-agent}
\begin{algorithmic}[1]
\Require Functional test case $u$, sandboxed repository $R$, agent-generated contract $\Sigma$ (Section~\ref{sec:synthesis-validation-reuse}), turn budget $B=16$
\Ensure Checkpoint $\hat c$ with source evidence $G$, or synthesis failure
\State $H\gets[\mathcal P,u,\mathrm{RepoMap}(R),\Sigma]$; $G\gets\emptyset$; $r\gets\bot$
\For{$t=1,\ldots,B$}
    \State $a\gets g_\theta(H)$ \Comment{Choose a repository query or a joint final artifact}
    \If{$a$ is a permitted repository query $q$}
        \State $o\gets\mathrm{Tool}_R(q)$; $H\gets H\mathbin{\|}(a,o)$
        \State $G\gets\Call{RecordSourceAccess}{G,q,o}$
    \ElsIf{$a$ is a final artifact and at least one tool call was made}
        \State $r\gets a$; \textbf{break}
    \Else
        \State $H\gets H\mathbin{\|}(a,\mathrm{ProtocolGuidance})$
    \EndIf
    \State Append a finalization reminder to $H$ when the budget is nearly exhausted
\EndFor
\State \textbf{if} $r=\bot$ or private grader references occur in $r$ \textbf{then return} failure
\State $\hat c,e\gets\Call{ParseAndCheckContract}{r,\Sigma}$
\If{$e\neq\emptyset$}
    \State $r\gets\Call{RepairFormatOnce}{g_\theta,u,\Sigma,r,e}$ \Comment{Public information only}
    \State $\hat c,e\gets\Call{ParseAndCheckContract}{r,\Sigma}$
\EndIf
\State \textbf{if} $e\neq\emptyset$ or $\neg\Call{EvidenceAndBoundaryChecks}{\hat c,G}$ \textbf{then return} failure
\State \Return $(\hat c,G)$
\end{algorithmic}
\end{algorithm}

\begin{algorithm}[!htbp]
\floatsize
\caption{Checkpoint Synthesis and Gated Evaluation}
\label{alg:checkpoint-synthesis}
\begin{algorithmic}[1]
\Require Functional test case $u_i$, repository $R$, agent-generated capability contract $\Sigma$
\Require Grader-only Start/Goal targets and programmatic/visual oracles $O^h$
\Ensure Report with checkpoint, source evidence, runtime validity, and Start/Goal verdicts
\State $(\hat c_i,G_i)\gets\Call{SynthesisAgent}{u_i,R,\Sigma}$ \Comment{Algorithm~\ref{alg:synthesis-agent}; no hidden inputs}
\If{synthesis failed}
    \State \Return synthesis-failure report
\EndIf
\State Extract $(\tau_i,\eta_i,\hat z_i,V_i,\alpha_i,A_i)$ from $\hat c_i$
\State $s_i\gets J_{\tau_i}(\eta_i)$
\If{restoration failed or $\neg\mathrm{Valid}_{\mathrm{Start}}(\hat c_i;s_i)$}
    \State \Return restoration/validity-failure report
\EndIf
\State $p_A\gets O^h_{\mathrm{prog,Start}}(s_i)$ \Comment{First hidden-oracle check}
\If{$\neg p_A$}
    \State \Return Start-failure report
\EndIf
\State $v_A\gets O^h_{\mathrm{vis,Start}}(\mathrm{Frame}(s_i))$
\If{$\neg v_A$}
    \State \Return Start-failure report
\EndIf
\State $(y_i,e_i)\gets\Call{ExecuteOnce}{\alpha_i,s_i}$ \Comment{Execute the transition spec once}
\If{execution error $e_i$}
    \State \Return transition-failure report
\EndIf
\State Retain $A_i$ with the checkpoint \Comment{Auxiliary predictions; not acceptance checks}
\State $p_B\gets O^h_{\mathrm{prog,Goal}}(y_i)$
\If{$\neg p_B$}
    \State \Return Goal-failure report
\EndIf
\State $v_B\gets O^h_{\mathrm{vis,Goal}}(\mathrm{Frame}(y_i))$
\State \Return report $(\hat c_i,V_i,G_i,A_i,\mathrm{Start}=\mathrm{Pass},\mathrm{Goal}=p_B\wedge v_B)$
\end{algorithmic}
\end{algorithm}

\subsection{Running Example: \texttt{2048-merge-left}}
\label{sec:appendix-running-example}

The box below walks Section~\ref{sec:method}'s three reasoning functions through one task. It is a mechanism illustration derived from the functional test case and source, not a recorded model trajectory.

\begin{tcolorbox}[colback=blue!2!white,colframe=blue!35!black,boxrule=0.5pt,
  sharp corners,left=6pt,right=6pt,top=5pt,bottom=5pt]
\textbf{Input.} The top row is $[2,2,0,0]$, all other cells are empty, and the score is 0; make exactly one left move.

\textbf{Explore.} \texttt{Grid.serialize} reveals the coordinate layout; \texttt{LocalStorageManager} and \texttt{GameManager.setup} expose the save/reload path. \texttt{GameManager.move} maps left to direction 3 and adds the merged tile's value to the score.

\textbf{Construct.} Emit the pre-move board and score, restored by writing a \texttt{localStorage} payload and reloading. The adapter serializes the board into \texttt{gameState}; the nonterminal flags default to false. The restored board still contains two 2 tiles.

\textbf{Synthesize the transition.} Emit \texttt{\{type: move, direction: left\}} and two assertions: a tile of value 4 is present, and the score has increased by at least 4. They follow from the merge logic. The randomly added tile's location is not prescribed. These generated assertions are the method's own output and are distinct from the grader's hidden checks.
\end{tcolorbox}

\subsection{Task-Provenance Pipeline Details}
\label{sec:appendix-provenance}

This section expands the three-stage pipeline of Section~\ref{sec:task-provenance} and its phrasing step, giving the signals, weights, execution steps, and guard variants behind each stage and the gates of Table~\ref{tab:provenance-gates}.

\paragraph{Selection signals and weights.}
Three per-function signals are computed with a single off-the-shelf analyzer, \texttt{lizard} (covering JS/TS/C\#/C++/Python/Lua under one definition): McCabe cyclomatic complexity, nesting depth, NLOC, and token count \citep{mccabe1976complexity}; a name-based, non-call-graph textual fan-in/fan-out proxy in the spirit of Henry--Kafura information-flow coupling \citep{henry1981software}; and git churn (commits plus $0.01\times$ changed lines) following the classical churn--defect-density signal \citep{nagappan2005churn}. Each signal is converted to a within-corpus percentile rank before being combined, so that a signal with a much larger raw scale cannot dominate the composite; the composite score weights complexity 0.35, nesting 0.20, fan-in 0.10, fan-out 0.10, and churn 0.25, with trivial getters/setters and anonymous functions (complexity $<3$) removed before ranking. These weights were frozen \emph{before} any ranking was produced, under a versioned, content-hashed configuration (\texttt{v1.0-preregistered-2026-07}) that every downstream artifact records, so a report can cite the configuration that produced its ranking. Intersection granularity follows each profile's controllable-surface shape: 2048 requires candidates to land on a concrete symbol anchor, while Luanti's controllable surface is the entire registry driven by a test worldmod (which can write any registered node/meta/player state), so no anchor filter applies there. The top-ranked region for 2048 is \texttt{GameManager.prototype.move} (final score 0.812; cyclomatic complexity 4, nesting 3, fan-out 15, 122 commits), matching the region all eight 2048 tasks target; for Luanti, the top three among 1{,}587 Lua functions are \texttt{core.item\_place\_node}, \texttt{core.node\_dig}, and \texttt{falling.lua::set\_node}, the core routines that mutate world state.

\paragraph{From-scratch unreachability diagnostic.}
This is a 2048-specific, bounded-playout diagnostic that quantifies why a given target cannot be reached by playing from scratch within budget; it is not a cross-benchmark selection signal. A Monte-Carlo playout under 2048's own rules \citep{metropolis1949monte} (a greedy immediate-merge policy, 200 trials, a several-thousand-step budget) reaches a given tile value from scratch with sharply declining probability: about 0.99 for 64, 0.88 for 128, 0.43 for 256, and near 0 for 512 and above. Because a greedy policy is stronger than a random one, this estimate is conservative; we do not claim these tile values are unreachable in principle, only that a stronger 2048 heuristic could reach them. What carries weight is narrower: the near-win task fixes one specific, score-consistent board (two tiles of 1024 awaiting a winning merge) that almost never appears as the exact post-state of an ordinary from-scratch playout within budget, so that behavior is only exercisable on a constructed checkpoint. This signal enters the final task score multiplicatively rather than being averaged in ($\text{final} = \text{composite} \times (0.5 + 0.5 \times \text{unreach})$), because its natural grain is a candidate target, not a function. Luanti has neither a cheap faithful simulator nor a from-scratch generator for arbitrary constructed deep structure, so it is assigned the documented conservative default of 1.0 rather than an estimated number.

\paragraph{Ground-truth recording.}
The selected region's control-flow branches mechanically induce branch-covering seed states, one per task: 2048's three \texttt{move} branches yield its three pilot tasks (a merge branch, a win branch, and a terminated/no-moves branch). Each scenario runs five fixed steps: pin \texttt{Math.random} to a fixed value so the post-\texttt{move} new-tile placement is deterministic and the recorded ground truth is itself replayable; restore the seed board through the game's own persistence path (\texttt{localStorage}); call the game's real \texttt{move} through the \texttt{\_\_WARPTEST\_2048\_\_} test-gated hook; read back \texttt{getState()}, whose after-state is the ground truth; and record the assertion as the \emph{replay-stable} projection of that state (score delta, resulting tile value, terminal flag, plus \texttt{no\_severe\_console}). The only human input is the seed board and the bounded follow-on action, both derived from the region's branches; the post-state is computed by the game itself. Restricting assertions to the replay-stable projection is what lets the same assertion hold on an unpinned, ordinary run (Gate B). Luanti's ground truth reuses the same worldmod path; where the engine cannot host a given scenario, synthesis reports a neutral \texttt{blocked} rather than a failure.

\paragraph{Mutation operators and kill-rate accounting.}
Mutants are confined to the selected region's line span (the move-handling region in 2048), so a mutant only perturbs code the task claims to cover. Operators are language-agnostic, single-point textual mutations \citep{demillo1978hints,jia2011analysis}: relational-operator swaps, arithmetic-operator swaps, boolean flips, and small constant perturbations (e.g., $2048 \to 4096$), at most three per operator. Mutants that error out for syntactic or environmental reasons are removed from the denominator rather than counted as survivors. Kill rate is reported per task rather than pooled: the near-win task kills nearly all merge/score/win-class mutants, the near-loss task (where \texttt{move} is a no-op) kills none on its own, and the two together cover the region, so kill rate should be read over the task \emph{set}, not any single task.

\paragraph{Phrasing guard variants.}
The token-equivalence guard has two variants: \emph{shallow} requires every grader-critical token to be recoverable with no template-sounding phrasing; \emph{explore} additionally requires that the strategy name, validity class, and framework action-type identifiers be absent (left for the agent to recover from source and the capability contract), while every target value is preserved. A rendering that drops a token or leaks a mechanism token is resampled against the guard's feedback (up to 4 attempts). All 6 checks pass (3 tasks $\times$ 2 variants); the \emph{explore} variant correctly hides \texttt{localStorage\_restore\_reload}, \texttt{save-restorable}, and \texttt{move}. Rendering quality therefore affects only how readable a task description is, never the task or its assertions.

\begin{table}[!htbp]
\caption{Task-provenance verification gates. Gates B, C, and D are measured on the 2048 pilot region and its tasks; Gate A is a property of the selection stage over the whole corpus.}
\label{tab:provenance-gates}
\begin{center}
\floatsize
\begin{tabular}{lll}
\hline
Gate & Check & Status \\
\hline
A & Reproducibility (re-running selection reproduces the ranking) & Pass \\
B & Oracle validity (assertions hold on an unpinned run) & Pass (3/3) \\
C & Leakage guard (shallow \& explore) & Pass (6/6) \\
D & Discrimination (region mutation kill rate above threshold) & Pass (0.875) \\
\hline
\end{tabular}
\end{center}
\end{table}

\paragraph{Provenance pipeline scope.}
Three caveats bound the pipeline's current scope. The fan-in/fan-out signal is a name-based textual approximation, not a call graph. Luanti's unreachability diagnostic uses the documented default of 1.0, not an estimate. The weights' validity rests on having been frozen before any ranking was produced, which the version tag and content hash make auditable. A metric-vs-kill-rate correlation over a larger cross-region task corpus is left for future scaling.

\subsection{Automatic Hook Onboarding: Agent Architecture, Protocol, and Results}
\label{sec:appendix-hook-onboarding}

Every hook and capability contract $\Sigma$ that the experiments of Section~\ref{sec:experiments} restore through is produced by the onboarding step summarized in Section~\ref{sec:synthesis-validation-reuse}.  We evaluate that one-time step separately from per-test checkpoint synthesis: can a coding agent start from a clean upstream repository, discover the software's semantic and persistence surfaces, and construct a reusable hook without access to the grader-side profiles?  The study covers all ten applications in \benchmark{} and treats onboarding as an executable software-engineering task rather than a manually supplied benchmark fixture.

\paragraph{Frozen inputs and outputs.}
Each trial materializes a detached checkout of a fixed upstream commit by depth-one fetch, removes its remote before the agent starts, and exposes only a repository map, a public brief of the capabilities to target, public runtime notes, and three agent-visible development cases.  The agent must emit four artifacts: (i)~\texttt{warptest\_onboarding/contract.yaml}, declaring the hook kind, activation gate, driver, capabilities, and touched product files; (ii)~a file-based driver with the exact interface \texttt{python warptest\_onboarding/driver.py --request REQUEST --report REPORT}; (iii)~five to ten self-authored cases, with both live and save/reload round trips; and (iv)~the minimal product-side hook, extension, or public-API adapter needed by the driver.  A successful save/reload case must save through the software, restart or reconnect, read the state back from the software or native artifact, and return a nonempty artifact produced by the software.  Copying the request into the reported observation is rejected.

\paragraph{Agent and trust boundary.}
All runs use \texttt{gpt-5.6-sol} with medium reasoning through the Responses API.  This is our own bounded coding-agent loop, not an invocation of an off-the-shelf coding agent.  Table~\ref{tab:onboarding-architecture} summarizes the separation of responsibilities.  The model receives exactly seven function tools (\texttt{list\_dir}, \texttt{grep}, \texttt{read\_file}, \texttt{write\_file}, \texttt{apply\_patch}, \texttt{get\_diff}, and \texttt{run\_check}) and no arbitrary shell.  \texttt{run\_check} accepts only symbolic commands registered by the experiment manifest, so the orchestrator, rather than the model, owns process launch, timeouts, environment filtering, and report paths.  Credentials, parent directories, other checkouts, the reference implementation, and hidden cases are outside the model-visible workspace.

\begin{table}[!htbp]
\caption{Automatic hook-onboarding architecture and information boundary.  The model can author tests, but it cannot author or inspect the final grader.}
\label{tab:onboarding-architecture}
\begin{center}
\floatsize
\begin{tabular}{@{}p{0.14\linewidth}p{0.38\linewidth}p{0.40\linewidth}@{}}
\hline
Layer & Model-visible responsibility & Orchestrator-only responsibility \\
\hline
Source workspace & clean repository map and sandboxed file edits & fixed commit materialization; remote removal; path confinement \\
Coding agent & hook/adapter, contract, driver, self-tests & model and budget enforcement; usage accounting \\
Tool layer & seven typed functions; named \texttt{run\_check} calls & command registry, environment filtering, timeouts, log capture \\
Development validation & six generated self-tests in the realized runs; three public cases & harness-owned execution and semantic comparison \\
Final grading & no access & post-final reruns; two hidden cases; gate and artifact checks \\
\hline
\end{tabular}
\end{center}
\end{table}

\paragraph{Three-layer validation.}
Self-tests measure whether the agent actively probes independent values, optional-field omission, partial updates, nested or collection state, and boundary shapes, rather than stopping after one smoke case.  Because a model can encode the same misconception in its hook and its own tests, self-tests are never the final oracle.  The harness also executes three human-authored, agent-visible development cases and, only after finalization, two agent-hidden held-out cases per application.  Every source write advances a mutation epoch and invalidates all earlier check passes; finalization is accepted only if self-tests, the development suite, and the build all pass after the last edit.  The orchestrator then independently reruns the generated tests and public suite from the final tree before exposing the hook to either hidden case.  A reference implementation written by the authors is run separately to confirm that all 30 public and 20 hidden cases are feasible; it is not shown to the agent, and no experiment in this paper restores through it.

\paragraph{Population.}
Table~\ref{tab:onboarding-results} records the frozen commits and realized interface kinds.  The population spans three existing public automation surfaces (Jupyter Server, UNO, and \texttt{bpy}), four synthesized test-gated patches, and three extensions.  Product-side gates are checked in normal mode; an existing public API is recorded as such rather than being relabeled as a test-only gate.

\paragraph{Product-side footprint.}
The three interface kinds differ in what they touch.  An \emph{existing API} binding adds only an external driver and changes no product file, so its contract records \texttt{normal\_mode\_absent} as not applicable.  An \emph{extension} is loaded through the software's own mod, worldmod, or editor mechanism and likewise leaves the core source untouched.  A \emph{test-gated patch} adds an explicit entry point to product source; the prompt requires that such a hook be absent during normal startup and reachable only through a query, flag, command, editor method, or test-only extension, and the contract must declare the gate, state whether the hook is absent in normal mode, and list every product file changed.  The verifier recomputes the changed-file list from the git diff and rejects a contract that omits a file; the normal-mode declaration is recorded as the gate status, and a product-side hook that does not declare it fails.  All seven product-side hooks declare and pass it.  Across the ten applications the replication lands 3{,}973 added and 3 deleted lines (Table~\ref{tab:onboarding-replication-apps}), a total that includes the driver, contract, and self-tests as well as the hook, so no product logic is rewritten and the tested build differs from the upstream commit only by additive, gated code.

\paragraph{Human input and one-time cost.}
The human contribution to onboarding an application is the public capability brief, the runtime notes, the three development cases the agent may run, and the two held-out cases used only for grading; the repository map is generated, and no human intervenes while the agent runs.  Against that input, the one-time cost of the step is the ten-application agent run reported below (5.49M tokens and 2{,}838~s in the frozen configuration; 6.07M tokens, 2{,}794~s, and an estimated \$7.19 in the replication), after which every checkpoint for that application is synthesized per test through $\Sigma$ without revisiting the hook.  The grader-side semantic profiles of Table~\ref{tab:effort} (93 to 503 lines each) are a separate, method-invisible component of the benchmark harness: they read runtime state for the oracle and are not part of the hook.

\begin{table}[!htbp]
\caption{Per-application automatic hook-onboarding results under one frozen configuration.  One runner invocation materializes and evaluates all ten fixed commits; every row passes post-final re-execution before its hidden cases are exposed, so no row is graded on a tree the agent can still edit.  Three rows pin a revision that resolves only in our own checkout; \nmark{} marks a hash withheld \withheldreason{}, and the pinned trees ship with the artifact.}
\label{tab:onboarding-results}
\begin{center}
\floatsize
\setlength{\tabcolsep}{3pt}
\begin{tabular}{llllrrr}
\hline
Application & Upstream commit & Interface kind & Status & Self & Public & Hidden \\
\hline
2048 & \texttt{478b6ec346e3} & test-gated patch & pass & 6/6 & 3/3 & 2/2 \\
SpaceRPG & \nmark{} & test-gated patch & pass & 6/6 & 3/3 & 2/2 \\
DatingSim & \nmark{} & test-gated patch & pass & 6/6 & 3/3 & 2/2 \\
OpenRA & \texttt{f31a0b06a0c5} & extension & pass & 6/6 & 3/3 & 2/2 \\
JupyterLab & \texttt{4a19f8fff404} & existing API & pass & 6/6 & 3/3 & 2/2 \\
LibreOffice & \texttt{5f60ba8b3103} & existing API & pass & 6/6 & 3/3 & 2/2 \\
Blender & \texttt{3d018bcabdae} & existing API & pass & 6/6 & 3/3 & 2/2 \\
jynew & \texttt{773f3e76492f} & extension & pass & 6/6 & 3/3 & 2/2 \\
PAL3 & \texttt{78c3dcfe8d01} & test-gated patch & pass & 6/6 & 3/3 & 2/2 \\
Luanti & \nmark{} & extension & pass & 6/6 & 3/3 & 2/2 \\
\hline
\end{tabular}
\end{center}
\end{table}

\paragraph{Uniform protocol and result.}
The frozen configuration gives every application the same model, seven-tool architecture, self-test contract, and ceilings of 120 response rounds, 280 tool calls, and 10{,}800 seconds.  One runner invocation requests all ten applications in Table~\ref{tab:onboarding-results}; each is materialized from its listed clean commit, and no row is imported or substituted from another run.  All ten applications complete onboarding and pass their normal-mode gate.  After independent post-final execution, the generated tests pass 60/60, the agent-visible development cases pass 30/30, and the agent-hidden held-out cases pass 20/20.  All ten hooks are reusable under the benchmark contract, with zero runtime human intervention during agent execution.  The whole run costs 272 responses, 370 tool calls, and 5.49M tokens over 2{,}838~s of agent time, across all ten applications.  These results demonstrate onboarding across five language families and source sizes from 874 lines to 7.2M lines under a shared configuration (Table~\ref{tab:coverage}), without language-specific prompts or tools.

\paragraph{Interpretation boundary.}
The study uses one model and two hidden cases per application, and the ten systems are a coverage-oriented suite rather than a statistical sample of interactive software.  Self-tests are model-authored and therefore measure active validation behavior rather than supply an independent oracle; the separate public and hidden suites provide that independence.  We use the 10/10 result as evidence that automatic onboarding is feasible across all interface families represented in \benchmark{}, not as an estimate of its success rate on arbitrary software.

\paragraph{Independent replication.}
We repeat the ten-application evaluation with \texttt{gpt-5.6-sol} at medium reasoning under the same frozen commits, prompt, seven text-only tools, validation cases, and agent limits. The agent receives no screenshots or OCR. All ten applications complete onboarding; independent post-final execution passes 60/60 self-tests, 30/30 development cases, and 20/20 hidden cases. Table~\ref{tab:onboarding-replication-apps} reports the per-application results. This run costs 263 responses, 369 tool calls, and 6.07M tokens over 2{,}794~s of agent time, and lands 3{,}973 added and 3 deleted lines for an estimated \$7.19 at the provider's catalog prices, which we did not independently verify.

\begin{table}[!htbp]
\caption{Per-application results for an independent \texttt{gpt-5.6-sol} replication of automatic hook onboarding.  Every hidden entry is executed post-final and passes; patch size is added/deleted lines.}
\label{tab:onboarding-replication-apps}
\begin{center}
\floatsize
\setlength{\tabcolsep}{3.5pt}
\begin{tabular}{l l r r r r r r r}
\hline
Application & Hook kind & Self & Visible & Hidden & Rounds & Tools & Tokens & Patch $+/-$ \\
\hline
2048 & test-gated patch & 6/6 & 3/3 & 2/2 & 17 & 22 & 247{,}865 & 350/1 \\
SpaceRPG & test-gated patch & 6/6 & 3/3 & 2/2 & 18 & 28 & 415{,}180 & 360/0 \\
DatingSim & test-gated patch & 6/6 & 3/3 & 2/2 & 21 & 35 & 395{,}895 & 396/1 \\
OpenRA & extension & 6/6 & 3/3 & 2/2 & 25 & 42 & 832{,}472 & 379/1 \\
JupyterLab & existing API & 6/6 & 3/3 & 2/2 & 19 & 22 & 233{,}340 & 404/0 \\
LibreOffice & existing API & 6/6 & 3/3 & 2/2 & 31 & 31 & 553{,}289 & 413/0 \\
Blender & existing API & 6/6 & 3/3 & 2/2 & 27 & 29 & 353{,}876 & 444/0 \\
jynew & extension & 6/6 & 3/3 & 2/2 & 54 & 67 & 1{,}794{,}265 & 456/0 \\
PAL3 & test-gated patch & 6/6 & 3/3 & 2/2 & 25 & 52 & 835{,}103 & 440/0 \\
Luanti & extension & 6/6 & 3/3 & 2/2 & 26 & 41 & 405{,}443 & 331/0 \\
\hline
\end{tabular}
\end{center}
\end{table}

\subsection{Specification and Report Fields}

The current specification schema includes an artifact or benchmark name, target mapping, checkpoint strategy list, expected checkpoint validity, the bounded transition spec, assertions, optional replay-from-start data, and optional visual-oracle metadata. The core method-checkpoint schema records task id, benchmark, generated specification, source evidence, patch/hook/save payload, runtime validation, validity class, inputs read, hidden files read, and generating model. The report schema records repository metadata, semantic profile, adapter layers, detected capabilities, runtime access methods, touched storage keys, hook availability, instrumentation mode, checkpoint strategy, checkpoint validity class, validation status, post-checkpoint interaction status, oracle status, visual-oracle status, console errors, screenshots, logs, unsupported baseline reasons, and timing metrics. Synthesis reports additionally record prompt and raw-response artifact paths, parse errors, schema errors, hidden-leakage guard results, and normalization warnings.

\subsection{Configuration Keys in the Released Harness}
\label{sec:appendix-config-keys}

The paper names configurations by the policy they run and by where their Start comes from, while the released harness identifies the same configurations by the manifest keys of \texttt{benchmarks/p5/manifest.yaml}, under which every archived run record is stored. \confgui{} is \texttt{c1\_agent\_s} and \confguicp{} is \texttt{c1\_agent\_s\_synth}; \confcua{} is \texttt{c1\_agent\_scalecua} and \confcuacp{} is \texttt{c1\_agent\_scalecua\_synth}; \method{} is \texttt{c1\_state} and the delegated diagnostic \confdelegated{} is \texttt{c1\_state\_minimal\_agent}. The two-phase Start gate of Section~\ref{sec:execution-based-eval} is registered as \texttt{c1\_start\_gate\_protocol} (Appendix~\ref{sec:appendix-runtime-hold} gives its step sequence). The \texttt{c1} prefix is the harness's internal name for the \perfstudy{}; it is kept in code, manifests, and archived records so that every number reported here remains traceable to the run that produced it.

\subsection{Visual Calibration: Full Results}
\label{sec:appendix-calibration-full}

Table~\ref{tab:calibration} lists the full calibration metrics of the production judge against the blinded consensus reference of Section~\ref{sec:cross-vlm-calibration}; the paragraphs below give the protocol and per-class results.

\paragraph{Protocol.} All four models (the production judge, the two annotators, and the adjudicator) receive the same evidence: the calibration golden frame, the live frame, the task stage, a short description, and the rubric's \emph{must}/\emph{must not}/\emph{ignore} fields. Two zero-model qualification executions first find all 61 tasks eligible; 4 base states per application are then drawn under a frozen hash for 40 bases, and each is expanded into the 5 sample classes below. The pre-registered gates require A/B three-way Cohen's $\kappa \geq 0.70$, clean acceptance $\geq 90\%$, combined rejection of the two pure-visual-fault classes $\geq 90\%$, and pure-fault false-accept rate $\leq 5\%$; intervals use a 10{,}000-resample bootstrap clustered by base task under a frozen seed, paired changes use exact McNemar, and every sample is judged once with the result cache disabled. The production path short-circuits on programmatic failure and accepts a frame that is pixel-identical to its golden without a call, but every verdict it does ask for goes to \texttt{qwen3-vl-flash} with the verdict cache disabled, so these are the deployed evaluator's error rates. Of the 80 clean controls, 78 are accepted (97.5\%, 95\% CI $[93.75\%, 100\%]$).

\paragraph{Results.} The five sample classes named there are registered in the artifact as \texttt{before\_clean}, \texttt{after\_clean}, \texttt{same\_profile\_wrong\_state}, \texttt{blank\_render}, and \texttt{critical\_occlusion}, and the bootstrap resamples base tasks 10{,}000 times under seed \texttt{20260830}. The single GLM/Doubao disagreement is a jynew \texttt{same\_profile\_wrong\_state} sample. Qwen's confusion counts are 78 correct/4 incorrect where the reference is correct, and 4 correct/114 incorrect where the reference is incorrect. The 80 fault samples (blank render and distributed striped occlusion, 40 each) leave the programmatic projection, persisted fields, and runtime health unchanged, so the programmatic-only oracle accepts all 80 by construction; the visual channel and, by conjunction, the Hybrid AND channel reject blank render 40/40 and distributed striped occlusion 37/40 (Table~\ref{tab:pure-visual}). Paired against the same base's clean-after frame, exact McNemar discordance is 40/0 on blank ($p = 1.82\times10^{-12}$) and 37/0 on occlusion ($p = 1.46\times10^{-11}$). The same-profile wrong-state class is excluded since its programmatic state does change.

\begin{table}[!htbp]
\caption{Cross-VLM calibration of the visual channel, 200 formal samples.}
\label{tab:calibration}
\begin{center}
\floatsize
\begin{tabular}{lr}
\hline
Calibration metric & Formal result \\
\hline
Samples & 200 \\
A/B raw agreement & 99.5\% \\
A/B Cohen's $\kappa$ & 0.9896 \\
A/B disagreements & 1 \\
Uncertain & 0 \\
Qwen strict accuracy & 96.0\% \\
Qwen-vs-reference $\kappa$ & 0.9173 \\
Overall false accept / false reject & 3.39\% / 4.88\% \\
Selective coverage & 100\% \\
\hline
\end{tabular}
\end{center}
\end{table}

\begin{table}[!htbp]
\caption{Acceptance under controlled visual treatments (40 base tasks each).}
\label{tab:pure-visual}
\begin{center}
\floatsize
\begin{tabular}{lrrr}
\hline
Condition & Programmatic accept & Visual accept & Hybrid accept \\
\hline
Clean control & 100\% & 97.5\% & 97.5\% \\
Blank render & 100\% & 0\% & 0\% \\
Distributed striped occlusion & 100\% & 7.5\% & 7.5\% \\
\hline
\end{tabular}
\end{center}
\end{table}

\paragraph{Qualitative case: SpaceRPG playable-flight.}
Three formal samples share the same \texttt{spacerpg-playable-flight} base and the same programmatic projection: \texttt{s035} (clean after), \texttt{s094} (blank render), and \texttt{s043} (distributed striped occlusion). All three pass the programmatic oracle unchanged; the visual judge accepts \texttt{s035} and rejects both \texttt{s094} and \texttt{s043}, so the Hybrid AND channel accepts only the clean sample (Figure~\ref{fig:spacerpg-hybrid-qualitative} shows all three frames alongside \texttt{before\_clean}). The case instantiates the property stated in Section~\ref{sec:coverage}, that the target action is delivered through a transition spec rather than pixel input while the judgment still reads the real runtime state and a real screenshot, and the AND guarantee of Section~\ref{sec:conjunctive-oracle}.

\paragraph{Failures and per-application breakdown.}
All 8 Qwen/reference mismatches are accounted for: 1 pure-fault false accept in LibreOffice, 2 in JupyterLab; 2 clean false rejects in 2048 (both \texttt{before\_clean}); the remaining 3 come from the non-causal \texttt{same\_profile\_wrong\_state} class, of which 2 are judged \emph{correct} by the VLM-consensus reference. Per-application strict accuracy is 100\% for Blender, SpaceRPG, DatingSim, OpenRA, and PAL3; 95\% for LibreOffice and Luanti; and 90\% for 2048, JupyterLab, and jynew.

\subsection{Visual-Experiment Notes}

The artifact contains auxiliary \texttt{visual-experiment} support for assessing a visual judge against headed validation samples. It is positioned as appendix calibration because the primary acceptance path is programmatic-state validation, runtime health, and bounded-transition oracle evidence. The OpenRA-only flow is supported through \texttt{python3 -m warptest visual-experiment --benchmark OpenRA --sample-count 12}; verified reports include \texttt{reports/warptest/20260611T051904Z/visual-experiment.json}. The annotation UI hides truth labels and variant names from annotators, randomizes sample order, and surfaces safe run errors without exposing blind labels.

For the formal 200-sample calibration of Section~\ref{sec:hybrid-eval-results}, the three-way macro-F1 of the production judge against the VLM-consensus reference is 0.6391. Neither the reference nor Qwen has any support in the \emph{uncertain} class, which still contributes an F1 of 0 to that macro-average by definition, so the two-class metrics reported in Table~\ref{tab:calibration} are the informative summary.

\subsection{Short-Circuiting and Production Efficiency}
\label{sec:visual-audit-cascade}

Because acceptance is a conjunction, a programmatic failure short-circuits the trial before the VLM Oracle is ever called (Section~\ref{sec:conjunctive-oracle}), which is both a cost saving and a guarantee that a trial already known to fail cannot be re-scored as a success by the visual channel. One judge model decides every visual verdict in this paper. There is no second, cheaper judge anywhere in the harness: the only two backends are the real \texttt{qwen3-vl-flash} client and a fake one used by the test suite, so a verdict is either produced by the calibrated judge or by no model at all, and the error rates of Section~\ref{sec:cross-vlm-calibration} are the deployed evaluator's error rates by construction.

Three mechanisms besides the short-circuit reduce what that judge is asked. The registered \textsc{Full-Only} policy captures one frame per stage; a \textsc{Conservative} policy, which no reported run uses, instead captures three frames at a fixed interval on non-heavy runtimes so that a transient mid-animation frame cannot decide a verdict on its own. Frames that are byte-identical are collapsed by their SHA-256, and a stage left with more than one distinct frame is assembled into a single contact sheet rather than sent as several calls. Verdicts are cached under a content-addressed key over the prompt version, the stage, the prompt, the golden and live image bytes, the backend, and the judge model identity; every formal run sets \texttt{WARPTEST\_C1\_DISABLE\_VISUAL\_CACHE=1}, so no reported verdict is served from that cache. A provider-side prompt-prefix cache may still report cached prompt tokens under either policy, since it reuses prefix computation only and never reuses a stored verdict or any label that an earlier judge call had already produced.

\subsection{Visual Audit Procedure}
\label{sec:appendix-cascade-algorithm}

Algorithm~\ref{alg:visual-cascade} gives the procedure the harness runs for one stage's visual audit, under the \textsc{Full-Only} policy every reported run registers ($k{=}1$, cache disabled).

\begin{algorithm}[!htbp]
\floatsize
\caption{Visual Audit of One Stage}
\label{alg:visual-cascade}
\begin{algorithmic}[1]
\Require Selected case $c$ with programmatic-state status $s_{\mathrm{prog}}$, visual oracle $o_v$, runtime handle $\mathcal{R}$, stage golden frame $\mathrm{golden}$
\Require Capture config: frame count $k$, interval $\Delta t$, cache policy
\Ensure Visual audit result $(v_{\mathrm{aud}}, \mathrm{source}, \mathrm{artifacts})$ for the stage
\If{$s_{\mathrm{prog}} = \textsc{Fail}$}
    \State \Return $(\textsc{Skip},\, \text{``semantic\_failed''},\, \emptyset)$ \Comment{report literal, unchanged}
\EndIf
\If{$o_v$ is missing or schema-invalid}
    \State \Return $(\textsc{Fail},\, \text{``missing\_visual\_oracle''},\, \emptyset)$ \textbf{if} policy requires audit
\EndIf
\State $F \gets \Call{CaptureFrames}{\mathcal{R}, k, \Delta t}$ \Comment{$k$ screenshots at interval $\Delta t$}
\State $F' \gets \Call{DeduplicateFrames}{F}$ \Comment{collapse byte-identical frames}
\State $I \gets \Call{BuildContactSheet}{F'}$ \textbf{if} $|F'|>1$ \textbf{else} $F'[0]$
\State Run screenshot-quality checks on $I$; \textbf{return} failure if quality is insufficient
\If{$\Call{NormalizedPixelEquality}{I, \mathrm{golden}}$}
    \State \Return $(\textsc{Pass},\, \text{``normalized\_image\_equality''},\, \mathrm{artifacts})$ \Comment{no model call}
\EndIf
\If{cache enabled \textbf{and} backend $\neq$ fake}
    \State $h \gets \Call{CacheKey}{o_v, I, \mathrm{golden}, \mathrm{model\_id}, \mathrm{prompt\_ver}}$
    \If{$\Call{CacheLookup}{h}$ hits}
        \State \Return cached verdict
    \EndIf
\EndIf
\State $v \gets \Call{Judge}{o_v, I}$ \Comment{the one judge model, \texttt{qwen3-vl-flash}}
\State Write artifacts and global cache (if policy allows and backend $\neq$ fake)
\State \Return $(v.\mathrm{verdict},\, \text{``llm\_visual\_judge''},\, \mathrm{artifacts})$
\end{algorithmic}
\end{algorithm}

\subsection{Runtime Hold and Start-Capture Timing}
\label{sec:appendix-runtime-hold}

\paragraph{Start-gate order.}
The registered protocol \texttt{c1\_start\_gate\_protocol} fixes the sequence \texttt{semantic\_start\_grade}, \texttt{before\_visual\_grade\_under\_runtime\_hold}, \texttt{phase\_b\_only\_if\_start\_passed}, \texttt{after\_semantic\_and\_visual\_grade}, and every run index must agree with the manifest copy. Concretely, the harness captures the start frame at the Phase~A boundary, then reads and grades the programmatic state while the runtime still runs freely; a programmatic failure short-circuits without acquiring a hold or issuing any visual request. Only after the programmatic start grade passes does the harness suspend the runtime, send the single before-judge request, release the runtime, and enter Phase~B. The after-state grades then run unheld, because no further execution depends on their outcome.

\paragraph{What the hold covers.}
The hold brackets the visual verdict and the transition into Phase~B, not the capture and not the programmatic read. It guarantees that the frame shown to the judge is still the live state when the verdict returns, and that Phase~B departs from exactly the state both oracles accepted; it does not make the two reads simultaneous. Simultaneity is unavailable on the native profiles, where \texttt{SIGSTOP} freezes the same process that serves the programmatic read, so that read must precede the hold. The residual exposure is the interval between capture and hold acquisition, which contains a local state read and no model request.

\paragraph{Per-profile backends.}
Browser-hosted profiles (2048, SpaceRPG, DatingSim) and JupyterLab are held through a Chromium DevTools session with \texttt{Emulation.setScriptExecutionDisabled}, recorded as backend \texttt{chromium\_cdp\_script\_pause}. Luanti, the two Unity benchmarks (PAL3 and jynew), OpenRA, Blender, and LibreOffice are held at the operating-system process boundary with \texttt{SIGSTOP} and resumed with \texttt{SIGCONT}, recorded as \texttt{sigstop}; the Unity and OpenRA sessions expose the pid of the live application process for this purpose. There is no engine-level time-scale freeze for Unity: every native profile is suspended at the process boundary.

\paragraph{Which arms are held.}
The Start-grading protocol specifies a runtime hold for \confgui{}, \confguicp{}, \method{}, and the delegated diagnostic, with the applicable configurations listed in \texttt{arms\_under\_runtime\_hold}. On the state-managed Unity path, the adapter closes before visual grading, so the captured Start is evaluated after the runtime has terminated. This case records backend \texttt{runtime\_terminated\_before\_grading} with \texttt{required} and \texttt{held} both false; the protocol permits that backend only for this configuration and benchmark pair.

\paragraph{Failure handling and audit record.}
A hold is attempted once, with no retry and no settling delay. If it cannot be acquired the visual judge is never called; if it cannot be released the error is raised rather than swallowed. Either way the trial is recorded as an incomplete infrastructure failure rather than a scored outcome, so a hold defect can neither inflate nor deflate success rates. Every trial that started visual grading carries \texttt{start\_grading\_runtime\_hold} with \texttt{required}, \texttt{held}, \texttt{released}, \texttt{backend}, and, on native profiles, \texttt{held\_pid}. Aggregation rejects a report whose recorded backend differs from the one expected for that benchmark, whose hold was not both taken and released, whose native \texttt{held\_pid} differs from the process id in the same trial's seed evidence, or which claims a hold without having run the visual start grade.

\subsection{Start Parity Between the Two Self-Synthesizing Arms}
\label{sec:appendix-start-parity}

The paired-Start protocol for \method{} and \confguicp{} specifies a shared checkpoint identity, acceptance contract, and visual rubric, the controls for a matched-Start analysis.

\paragraph{One start, not two samples.}
The checkpoint-cache key depends on task, model, and seed, independently of the consuming policy. Before an entry is written, the shared acceptance predicate checks seeding, the public executable contract, and per-benchmark executability. A paired analysis requires both configurations to consume the same accepted specification and attributed synthesis cost.

\paragraph{One rubric and one hold.}
The before-state visual rubric is selected from a shared registry by benchmark. For Luanti, it tolerates day/night lighting, weather, water animation, and transient HUD text. Both configurations use the runtime-hold protocol in Appendix~\ref{sec:appendix-runtime-hold}.

\paragraph{What the report checks.}
A matched-Start analysis requires identical seeded specifications, attributed synthesis burdens, and \texttt{before\_semantic\_ok}, \texttt{before\_visual\_ok}, and \texttt{start\_success} judgments. The pairing checks are applied before computing paired differences.

\paragraph{The one remaining difference, and how it is bounded.}
The two arms read the Start predicate through different interfaces, and this cannot be equalized. The GUI arms run without state hooks and must grade through the production surface (rendered DOM, production save keys); the state arm runs with hooks and calls the profile validator. Forcing either choice on the other arm is impossible or destructive: the GUI arms have no hook to call, and the DatingSim production grader clicks the real save control and restores \texttt{localStorage} afterwards, so running it inside a state trial would perturb the state that trial is about to transition. We therefore measure the equivalence offline instead of assuming it. For each of the two benchmarks that expose a production surface, a separate gate restores a known state in a hook-enabled session and grades it with both interfaces, over both a state that satisfies the oracle target and one that does not, and reports the agreement rate; a disagreement names the task and the case, and an ungradable task fails the gate rather than counting as agreement. This bounds the only Start difference the two arms still have.

\subsection{Pixel Action Space of the \confgui{} and \confcua{} Baselines}
\label{sec:appendix-action-space}

\confgui{} and \confcua{} emit different model-facing vocabularies -- Agent S3 returns one semantic call over a grounded element description, ScaleCUA returns one \texttt{pyautogui}-style call over coordinates normalized to the policy frame -- and neither is executed as written: the parent parses the call, maps ScaleCUA's normalized coordinates to the current viewport, validates every literal against the policy frame, and compiles it to one bounded action drawn from the single executed vocabulary in Table~\ref{tab:action-space}. Model-authored code never reaches \texttt{eval} or \texttt{exec}. The verbs that would let a policy leave the software under test or bypass the pixel channel are rejected at that boundary rather than emulated, which is what the two right-hand columns make auditable: the baselines differ in how they name an action and in how they ground it, not in what the runtime lets them do. That is the precondition for reading \confgui{} against \confcua{} in Appendix~\ref{sec:appendix-latent-defect}.

\subsection{Full Performance-Study Results}
\label{sec:appendix-performance-full}

The complete matrix contains 2,196 registered trials: 61 tasks, four configurations. All 549 \confgui{} and 549 \confdelegated{} trials complete. \confguicp{} has 547 completed trials and two terminal failures, and \method{} has 548 completed trials and one terminal failure; Table~\ref{tab:performance-terminal-failures} names the three -- two Kimi \confguicp{} trials on JupyterLab and OpenRA at seed 2, and one Gemini \method{} trial on OpenRA at seed 1 -- and all three remain in the denominators and token means, including those of Table~\ref{tab:performance-main}. Table~\ref{tab:performance-matched-seed0} repeats that table with the per-row denominators and terminal-failure counts it omits. Table~\ref{tab:performance-full-model} reports all model/configuration cells, adding the per-model Start counts, token means, and \confdelegated{} rows behind the rates quoted in Section~\ref{sec:performance}. Table~\ref{tab:performance-seed} reports the seed breakdown: \method{}'s E2E count moves by one trial across them (50, 50, and 51 of 183) against four for \confguicp{} (19, 21, and 17), so the pooled gap is not produced by one favorable seed. Table~\ref{tab:performance-application} reports all ten applications: \method{} records at least one E2E success in eight of them, against four for \confguicp{} and none for \confgui{}, and LibreOffice is the only application no configuration completes at all, JupyterLab's two \confdelegated{} successes being the only ones there. Blender and PAL3 separate the execution interface most sharply: \confguicp{} reaches Start on 27/45 and 10/45 there without converting any into an E2E success, while \method{} reaches Start on 25/45 and 12/45 and converts 24 and 11 of those into E2E successes, 35 of the 151 it records in total.

\paragraph{Cost basis.}
Every cost column reported for the \perfstudy{}, here and in Table~\ref{tab:performance-main}, prices each trial's tokens at its decision model's public catalog rate and its cached prompt tokens at the cached-input rate. Where a row pools models or seeds, each trial is priced at its own decision model's rate before the row is averaged. The ledger does not separate the auxiliary grounder's calls, so a reported cost is an upper bound, at prices we did not independently verify.

\begin{figure}[H]
\centering
\includegraphics[width=0.98\linewidth]{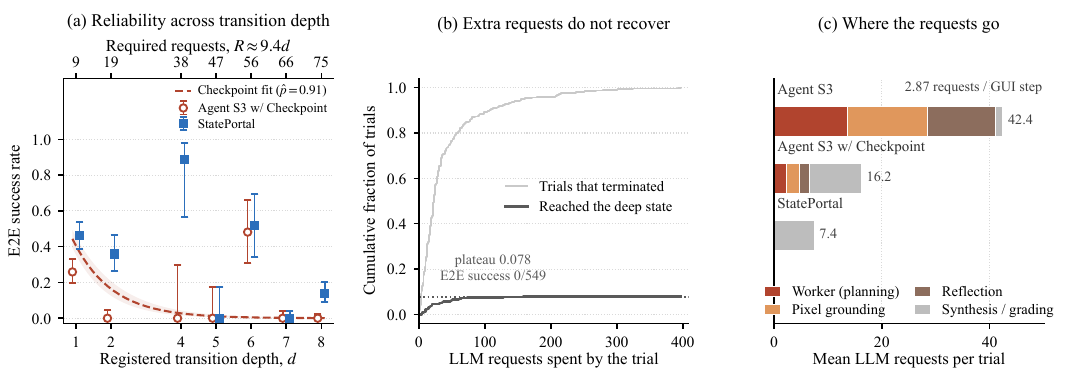}
\caption{Request-level diagnostics for the full three-seed \perfstudy{} run. (a) E2E success by registered transition depth; the upper axis maps depth to the required-request exposure it implies ($R\approx9.4d$), and the dashed \confguicp{} fit gives $\hat p=0.91$ per request (95\% CI $[0.90,0.92]$, shaded). (b) \confgui{} Start reachability plateaus at 43/549 (0.078), with 0/549 E2E outcomes at any depth. (c) Mean requests per completed trial are 42.4 for \confgui{}, 16.2 for \confguicp{}, and 7.4 for \method{}; one \confgui{} GUI step costs 2.87 requests.}
\label{fig:request-decay}
\end{figure}

Figure~\ref{fig:request-decay} gives the request-level view of the three-seed run. \confgui{} spends 42.4 requests per completed trial yet reaches Start on only 7.8\% of trials and never completes the target; checkpoint seeding reduces this to 16.2 requests, and joint checkpoint and transition synthesis uses 7.4. Wall clock follows the same ordering under all three registered models: 234--323~s per \confgui{} trial, 82--161~s per \confguicp{} trial, and 42--87~s per \method{} trial. Almost all of \confgui{}'s time is spent before Start (228--317~s of it), whereas a restore costs 1--3~s.

\begin{table}[!htbp]
\centering
\caption{The executed pixel action space and each baseline's model-facing verb.}
\label{tab:action-space}
\vspace{2pt}
\begingroup
\floatsize
\setlength{\tabcolsep}{3pt}
\renewcommand{\arraystretch}{1.05}
\begin{tabular}{@{}>{\raggedright\arraybackslash}p{0.22\linewidth}>{\raggedright\arraybackslash}p{0.22\linewidth}>{\raggedright\arraybackslash}p{0.29\linewidth}>{\raggedright\arraybackslash}p{0.22\linewidth}@{}}
\toprule
 & & \multicolumn{2}{c}{Model-facing verb} \\
\cmidrule(l){3-4}
Executed action & Description & \confgui{} & \confcua{} (ScaleCUA) \\
\midrule
\multicolumn{4}{@{}l@{}}{\textit{Pointer}} \\
\texttt{click(x, y, button, clicks, modifiers)} & Click at a pixel point & \texttt{click} & \texttt{click}, \texttt{doubleClick}, \texttt{rightClick} \\
\texttt{drag(x, y, x2, y2, button, modifiers)} & Drag between two points & \texttt{drag\_and\_drop}, \texttt{highlight\_text\_span} & \texttt{dragTo} \\
\texttt{scroll(dx, dy, x, y)} & Wheel delta at a point & \texttt{scroll} & \texttt{scroll} \\
\addlinespace[3pt]
\multicolumn{4}{@{}l@{}}{\textit{Keyboard}} \\
\texttt{key(key, modifiers)} & Press one key or chord & \texttt{hotkey}, \texttt{hold\_and\_press} & \texttt{hotkey}, \texttt{press} \\
\texttt{type(text, x, y, overwrite, enter)} & Type text after focusing & \texttt{type} & \texttt{write} \\
\addlinespace[3pt]
\multicolumn{4}{@{}l@{}}{\textit{Control}} \\
\texttt{wait(seconds)} & Wait for the UI to settle & \texttt{wait}, \texttt{save\_to\_knowledge} & \texttt{wait} \\
\texttt{done(reason)} & Declare success & \texttt{done} & \texttt{terminate}, \texttt{response} \\
\texttt{fail(reason)} & Declare failure & \texttt{fail} & \texttt{terminate}, \texttt{call\_user} \\
\bottomrule
\end{tabular}
\par\vspace{3pt}
\begin{minipage}{\linewidth}
\floatsize
\raggedright
\hypersetup{hidelinks}
\textit{Notes.} Coordinates are \texttt{frame\_pixels\_top\_left\_v1}; a batch holds at most 64 actions, text 4{,}096 characters, \texttt{clicks} 1--3, and waits and drags five seconds. Refused rather than emulated: out-of-frame coordinates, Agent S3's \texttt{open}, \texttt{switch\_applications}, \texttt{set\_cell\_values}, \texttt{call\_code\_agent}, raw \texttt{pyautogui} source, and DOM, control-tree, or task-API access. \texttt{save\_to\_knowledge} becomes a zero-second \texttt{wait}.
\end{minipage}
\endgroup
\end{table}

\begin{table}[!htbp]
\caption{Complete performance-study results by model for the four configurations of the three-seed registered run, including the \confdelegated{} diagnostic. Every row covers 183 trials. ``Goal $\mid$ Start'' conditions on trials that reached Start; the token and cost means retain terminal failures, and cost carries the pricing basis of the \emph{Cost basis} note above.}
\label{tab:performance-full-model}
\begin{center}
\floatsize
\setlength{\tabcolsep}{2.5pt}
\begin{tabular}{llrcccrrr}
\toprule
Model & Configuration & $n$ & Start & Goal $\mid$ Start & E2E & Tokens (k) & Cost (\$) & Failures \\
\midrule
gpt-5.6-terra & \confgui{} & 183 & \rate{16/183}{8.7\%} & \rate{0/16}{0.0\%} & \rate{0/183}{0.0\%} & 302.0 & 1.148 & 0 \\
gpt-5.6-terra & \confguicp{} & 183 & \rate{55/183}{30.1\%} & \rate{26/55}{47.3\%} & \rate{26/183}{14.2\%} & 96.6 & 0.367 & 0 \\
gpt-5.6-terra & \method{} & 183 & \rate{56/183}{30.6\%} & \rate{45/56}{80.4\%} & \rate{45/183}{24.6\%} & 51.6 & 0.196 & 0 \\
gpt-5.6-terra & \confdelegated{} & 183 & \rate{70/183}{38.3\%} & \rate{47/70}{67.1\%} & \rate{47/183}{25.7\%} & 63.8 & 0.242 & 0 \\
gemini-3.5-flash & \confgui{} & 183 & \rate{13/183}{7.1\%} & \rate{0/13}{0.0\%} & \rate{0/183}{0.0\%} & 554.3 & 1.580 & 0 \\
gemini-3.5-flash & \confguicp{} & 183 & \rate{86/183}{47.0\%} & \rate{21/86}{24.4\%} & \rate{21/183}{11.5\%} & 181.4 & 0.517 & 0 \\
gemini-3.5-flash & \method{} & 183 & \rate{92/183}{50.3\%} & \rate{65/92}{70.7\%} & \rate{65/183}{35.5\%} & 115.1 & 0.328 & 1 \\
gemini-3.5-flash & \confdelegated{} & 183 & \rate{101/183}{55.2\%} & \rate{54/101}{53.5\%} & \rate{54/183}{29.5\%} & 149.4 & 0.426 & 0 \\
kimi-k2.6 & \confgui{} & 183 & \rate{14/183}{7.7\%} & \rate{0/14}{0.0\%} & \rate{0/183}{0.0\%} & 368.6 & 0.552 & 0 \\
kimi-k2.6 & \confguicp{} & 183 & \rate{60/183}{32.8\%} & \rate{10/60}{16.7\%} & \rate{10/183}{5.5\%} & 213.4 & 0.320 & 2 \\
kimi-k2.6 & \method{} & 183 & \rate{63/183}{34.4\%} & \rate{41/63}{65.1\%} & \rate{41/183}{22.4\%} & 153.5 & 0.230 & 0 \\
kimi-k2.6 & \confdelegated{} & 183 & \rate{70/183}{38.3\%} & \rate{32/70}{45.7\%} & \rate{32/183}{17.5\%} & 206.2 & 0.309 & 0 \\
\bottomrule
\end{tabular}

\end{center}
\end{table}

\begin{table}[!htbp]
\caption{Performance-study results by seed for the three registered configurations, pooling the three decision models within each, so every row covers 183 trials, priced under the basis stated above.}
\label{tab:performance-seed}
\begin{center}
\floatsize
\setlength{\tabcolsep}{3.5pt}
\begin{tabular}{lrrrrrr}
\toprule
Configuration & Seed & Start & E2E & Mean tokens (k) & Mean cost (\$) & Failures \\
\midrule
\confgui{} & 0 & 16/183 & 0/183 & 465.7 & 1.219 & 0 \\
\confgui{} & 1 & 17/183 & 0/183 & 393.2 & 1.100 & 0 \\
\confgui{} & 2 & 10/183 & 0/183 & 366.0 & 0.961 & 0 \\
\confguicp{} & 0 & 66/183 & 19/183 & 145.6 & 0.360 & 0 \\
\confguicp{} & 1 & 71/183 & 21/183 & 158.9 & 0.401 & 0 \\
\confguicp{} & 2 & 64/183 & 17/183 & 186.9 & 0.443 & 2 \\
\method{} & 0 & 68/183 & 50/183 & 107.6 & 0.253 & 0 \\
\method{} & 1 & 73/183 & 50/183 & 105.6 & 0.250 & 1 \\
\method{} & 2 & 70/183 & 51/183 & 106.9 & 0.251 & 0 \\
\confdelegated{} & 0 & 78/183 & 45/183 & 140.5 & 0.331 & 0 \\
\confdelegated{} & 1 & 78/183 & 40/183 & 138.4 & 0.319 & 0 \\
\confdelegated{} & 2 & 85/183 & 48/183 & 140.5 & 0.328 & 0 \\
\bottomrule
\end{tabular}

\end{center}
\end{table}

\begin{table}[!htbp]
\caption{Performance-study results by application, pooling three models and three seeds, so each configuration/application cell contains nine trials per task, priced under the basis stated above.}
\label{tab:performance-application}
\begin{center}
\floatsize
\setlength{\tabcolsep}{3pt}
\begin{tabular}{llrrrr}
\toprule
Application & Configuration & Start & E2E & Mean tokens (k) & Mean cost (\$) \\
\midrule
2048 & \confgui{} & 2/72 & 0/72 & 418.1 & 1.247 \\
blender & \confgui{} & 14/45 & 0/45 & 288.7 & 1.047 \\
datingsim & \confgui{} & 6/63 & 0/63 & 309.8 & 1.028 \\
jupyterlab & \confgui{} & 0/45 & 0/45 & 873.7 & 1.582 \\
jynew & \confgui{} & 0/45 & 0/45 & 174.9 & 0.484 \\
libreoffice & \confgui{} & 0/45 & 0/45 & 1087.9 & 2.990 \\
luanti & \confgui{} & 5/72 & 0/72 & 686.3 & 1.770 \\
openra & \confgui{} & 16/45 & 0/45 & 92.4 & 0.261 \\
pal3 & \confgui{} & 0/45 & 0/45 & 88.2 & 0.192 \\
spacerpg & \confgui{} & 0/72 & 0/72 & 109.1 & 0.323 \\
2048 & \confguicp{} & 55/72 & 45/72 & 81.9 & 0.199 \\
blender & \confguicp{} & 27/45 & 0/45 & 273.2 & 0.827 \\
datingsim & \confguicp{} & 21/63 & 5/63 & 207.0 & 0.504 \\
jupyterlab & \confguicp{} & 8/45 & 0/45 & 301.5 & 0.591 \\
jynew & \confguicp{} & 10/45 & 0/45 & 117.1 & 0.280 \\
libreoffice & \confguicp{} & 2/45 & 0/45 & 79.9 & 0.187 \\
luanti & \confguicp{} & 9/72 & 0/72 & 112.2 & 0.268 \\
openra & \confguicp{} & 20/45 & 6/45 & 168.3 & 0.417 \\
pal3 & \confguicp{} & 10/45 & 0/45 & 122.3 & 0.275 \\
spacerpg & \confguicp{} & 39/72 & 1/72 & 209.7 & 0.541 \\
2048 & \method{} & 57/72 & 44/72 & 48.8 & 0.108 \\
blender & \method{} & 25/45 & 24/45 & 66.8 & 0.144 \\
datingsim & \method{} & 22/63 & 9/63 & 118.8 & 0.283 \\
jupyterlab & \method{} & 8/45 & 0/45 & 159.1 & 0.358 \\
jynew & \method{} & 19/45 & 12/45 & 99.2 & 0.230 \\
libreoffice & \method{} & 3/45 & 0/45 & 71.6 & 0.163 \\
luanti & \method{} & 7/72 & 2/72 & 103.8 & 0.253 \\
openra & \method{} & 18/45 & 17/45 & 119.2 & 0.273 \\
pal3 & \method{} & 12/45 & 11/45 & 119.4 & 0.263 \\
spacerpg & \method{} & 40/72 & 32/72 & 160.2 & 0.415 \\
2048 & \confdelegated{} & 60/72 & 46/72 & 66.3 & 0.146 \\
blender & \confdelegated{} & 27/45 & 16/45 & 98.3 & 0.222 \\
datingsim & \confdelegated{} & 34/63 & 11/63 & 165.6 & 0.390 \\
jupyterlab & \confdelegated{} & 14/45 & 2/45 & 220.8 & 0.494 \\
jynew & \confdelegated{} & 26/45 & 11/45 & 138.4 & 0.319 \\
libreoffice & \confdelegated{} & 0/45 & 0/45 & 71.5 & 0.164 \\
luanti & \confdelegated{} & 10/72 & 2/72 & 117.9 & 0.281 \\
openra & \confdelegated{} & 19/45 & 16/45 & 144.2 & 0.345 \\
pal3 & \confdelegated{} & 12/45 & 4/45 & 145.3 & 0.323 \\
spacerpg & \confdelegated{} & 39/72 & 25/72 & 225.2 & 0.548 \\
\bottomrule
\end{tabular}

\end{center}
\end{table}

\begin{table}[!htbp]
\caption{The three registered \perfstudy{} trials that exhausted their retry budget. They are retained as unsuccessful trials and their measured token use remains in the cost means.}
\label{tab:performance-terminal-failures}
\begin{center}
\floatsize
\begin{tabular}{llr>{\raggedright\arraybackslash}p{0.30\linewidth}r}
\toprule
Configuration & Model & Seed & Task & Attempt \\
\midrule
\confguicp{} & kimi-\allowbreak k2.6 & 2 & \texttt{jupyterlab-\allowbreak reproducible-\allowbreak experiment-\allowbreak workspace} & 2 \\
\confguicp{} & kimi-\allowbreak k2.6 & 2 & \texttt{openra-\allowbreak disable-\allowbreak edge-\allowbreak scroll} & 2 \\
\method{} & gemini-\allowbreak 3.5-\allowbreak flash & 1 & \texttt{openra-\allowbreak allied-\allowbreak combined-\allowbreak arms-\allowbreak rally} & 2 \\
\bottomrule
\end{tabular}

\end{center}
\end{table}

\paragraph{Three-seed extension on a matched population.}
A separate three-seed extension adds the second pixel-policy family and a fourth decision model, so that the \perfstudy{} carries the same configuration set as the \latentstudy{} (Section~\ref{sec:latent-defect-results}): \confcua{} and \confcuacp{} run ScaleCUA \citep{liu2026scalecua} cold and from a restored checkpoint respectively, under the three registered models, and all five configurations run under \texttt{gpt-6-astra}. Every extension row covers all 61 tasks at seeds 0, 1, and 2 ($n=183$ per model/configuration cell), is reported alongside the registered rows, and is never pooled with them. Under the three registered models, the \confgui{}, \confguicp{}, and \method{} rows of Table~\ref{tab:performance-main} come from the original three-seed run; under \texttt{gpt-6-astra}, all five rows come from the extension. Table~\ref{tab:performance-matched-seed0} restricts every configuration to that same 61-task, three-seed grid so cross-configuration comparisons share one denominator. All 2{,}013 extension trials completed; none exhausted its retry budget. The extension synthesized its own checkpoints, so within a model its three checkpoint-restoring configurations share one cache, but they do not reuse the artifacts behind the three-seed rows. The extension rows quoted in Section~\ref{sec:performance} are descriptive marginal outcomes: they establish neither a model ranking nor a controlled causal comparison between execution interfaces on this extension cohort alone.

\begin{table}[!htbp]
\caption{All five \perfstudy{} configurations on the matched population (61 tasks $\times$ 3 seeds), which is every row of Table~\ref{tab:performance-main} with the per-row denominators and terminal-failure counts that table omits for space. Every row has $n=183$, so E2E values are directly comparable across configurations and models. ``Goal $\mid$ Start'' conditions on trials that reached Start; the token, cost, and time means include terminal failures rather than dropping them. TiPS is total time per E2E success, including unsuccessful attempts; ``\textemdash{}'' denotes no observed success. Cost uses the catalog pricing basis of the \emph{Cost basis} note above, which prices each trial at its own decision model's rate.}
\label{tab:performance-matched-seed0}
\begin{center}
\floatsize
\setlength{\tabcolsep}{1.5pt}
\begin{tabular}{@{}>{\columncolor{white}[0pt][\tabcolsep]}llrS[table-format=2.1]S[table-format=2.1]S[table-format=2.1]@{\hspace{3pt}}rrrr>{\columncolor{white}[\tabcolsep][0pt]}r@{}}
\toprule
Model & Configuration & $n$ & \multicolumn{3}{c}{Success (\%) $\uparrow$} & \multicolumn{3}{c}{Mean $\downarrow$} & TiPS $\downarrow$ & Terminal \\
\cmidrule(lr){4-6}\cmidrule(lr){7-9}
 & & & {Start} & {Goal $\mid$ Start} & {E2E} & tokens (k) & cost (\$) & time (s) & (s) & failures \\
\midrule
gpt-5.6-terra & \confgui{} & 183 & 8.7 & 0.0 & 0.0 & 302.0 & 1.148 & 234 & \textemdash & 0 \\
 & \confcua{} & 183 & 10.9 & 0.0 & 0.0 & 193.2 & 0.734 & 126 & \textemdash & 0 \\
 & \confguicp{} & 183 & 30.1 & 47.3 & 14.2 & 96.6 & 0.367 & 81 & 569.0 & 0 \\
 & \confcuacp{} & 183 & 33.9 & 37.1 & 12.6 & 56.1 & 0.213 & 38 & 303.4 & 0 \\
\rowcolor{blue!4}
 & \textbf{\method{}} & 183 & 30.6 & 80.4 & \bfseries 24.6 & 51.6 & 0.196 & 41 & 166.6 & 0 \\
\addlinespace[3pt]
gemini-3.5-flash & \confgui{} & 183 & 7.1 & 0.0 & 0.0 & 554.3 & 1.580 & 323 & \textemdash & 0 \\
 & \confcua{} & 183 & 12.6 & 13.0 & 1.6 & 444.0 & 1.265 & 241 & 14694.4 & 0 \\
 & \confguicp{} & 183 & 47.0 & 24.4 & 11.5 & 181.4 & 0.517 & 159 & 1386.9 & 0 \\
 & \confcuacp{} & 183 & 51.9 & 21.1 & 10.9 & 119.6 & 0.341 & 100 & 917.6 & 0 \\
\rowcolor{blue!4}
 & \textbf{\method{}} & 183 & 50.3 & 70.7 & \bfseries 35.5 & 115.1 & 0.328 & 84 & 236.2 & 1 \\
\addlinespace[3pt]
kimi-k2.6 & \confgui{} & 183 & 7.7 & 0.0 & 0.0 & 368.6 & 0.552 & 294 & \textemdash & 0 \\
 & \confcua{} & 183 & 8.7 & 18.8 & 1.6 & 404.6 & 0.607 & 285 & 17357.2 & 0 \\
 & \confguicp{} & 183 & 32.8 & 16.7 & 5.5 & 213.4 & 0.320 & 115 & 2111.2 & 2 \\
 & \confcuacp{} & 183 & 33.9 & 16.1 & 5.5 & 175.2 & 0.263 & 88 & 1613.8 & 0 \\
\rowcolor{blue!4}
 & \textbf{\method{}} & 183 & 34.4 & 65.1 & \bfseries 22.4 & 153.5 & 0.230 & 68 & 305.4 & 0 \\
\addlinespace[3pt]
gpt-6-astra & \confgui{} & 183 & 21.3 & 0.0 & 0.0 & 248.9 & 4.280 & 205 & \textemdash & 0 \\
 & \confcua{} & 183 & 18.0 & 21.2 & 3.8 & 233.7 & 4.020 & 164 & 4296.4 & 0 \\
 & \confguicp{} & 183 & 48.1 & 29.5 & 14.2 & 162.6 & 2.797 & 128 & 903.4 & 0 \\
 & \confcuacp{} & 183 & 46.4 & 30.6 & 14.2 & 146.7 & 2.524 & 94 & 660.0 & 0 \\
\rowcolor{blue!4}
 & \textbf{\method{}} & 183 & 48.1 & 63.6 & \bfseries 30.6 & 135.7 & 2.333 & 87 & 285.8 & 0 \\
\bottomrule
\end{tabular}

\end{center}
\end{table}

\subsection{Execution-Channel Strata: Derivation and Uncertainty}
\label{sec:appendix-channel-strata}

The strata of Table~\ref{tab:performance-channel} re-split the same 2{,}196 three-seed trials by the channel each task's registered transition spec uses, derived in the parent repository from each task's registered action type and checked to sum back to the per-configuration rows of Table~\ref{tab:performance-main}. Intervals resample tasks with replacement 10{,}000 times under seed \texttt{20260918}, clustering the three seeds and three models of one task; the same procedure reproduces the pooled intervals of Section~\ref{sec:performance} to within 0.2 points, which the collector asserts before writing a table. The paired interface difference resamples both configurations on the same drawn tasks, so it retains the per-task pairing the pooled difference also has. Unseeded \confgui{} records no E2E success in any of the three strata. The native-input column is close to a per-application row of Table~\ref{tab:performance-application}: eight of its nine tasks are 2048's single-move tasks.

\begin{table}[!htbp]
\caption{The \perfstudy{} split by the execution channel of each task's registered transition spec, at three seeds and three decision models. Start and E2E are over the $n$ trials of the row; ``Goal $\mid$ Start'' conditions on the trials that reached Start; the interval is a 10{,}000-resample bootstrap clustered by task. Every configuration runs all 61 tasks, so the channel describes each task's spec and not which configuration ran it: \confgui{} and \confguicp{} execute the target action by pixel input throughout, and \confdelegated{} is the delegated-transition diagnostic of Appendix~\ref{sec:appendix-e-prime}. An em dash marks a conditional ratio with no reached Start.}
\label{tab:performance-channel}
\centering
\fontfamily{lmr}\selectfont
\floatsize
\setlength{\tabcolsep}{2.5pt}
\renewcommand{\arraystretch}{1.12}
\begin{tabular}{llrcccr}
\toprule
Channel (tasks) & Configuration & $n$ & Start & Goal $\mid$ Start & E2E & E2E 95\% CI \\
\midrule
Native input (9) & \confgui{} & 81 & \rate{2/81}{2.5\%} & \rate{0/2}{0.0\%} & \rate{0/81}{0.0\%} & $[0.0,\,0.0]$ \\
 & \confguicp{} & 81 & \rate{64/81}{79.0\%} & \rate{45/64}{70.3\%} & \rate{45/81}{55.6\%} & $[33.3,\,74.1]$ \\
 & \method{} & 81 & \rate{66/81}{81.5\%} & \rate{53/66}{80.3\%} & \rate{53/81}{65.4\%} & $[44.4,\,82.7]$ \\
 & \confdelegated{} & 81 & \rate{69/81}{85.2\%} & \rate{53/69}{76.8\%} & \rate{53/81}{65.4\%} & $[45.7,\,80.2]$ \\
\addlinespace[3pt]
UI event (2) & \confgui{} & 18 & \rate{0/18}{0.0\%} & \textemdash{} & \rate{0/18}{0.0\%} & $[0.0,\,0.0]$ \\
 & \confguicp{} & 18 & \rate{6/18}{33.3\%} & \rate{0/6}{0.0\%} & \rate{0/18}{0.0\%} & $[0.0,\,0.0]$ \\
 & \method{} & 18 & \rate{5/18}{27.8\%} & \rate{0/5}{0.0\%} & \rate{0/18}{0.0\%} & $[0.0,\,0.0]$ \\
 & \confdelegated{} & 18 & \rate{11/18}{61.1\%} & \rate{1/11}{9.1\%} & \rate{1/18}{5.6\%} & $[0.0,\,11.1]$ \\
\addlinespace[3pt]
Source API (50) & \confgui{} & 450 & \rate{41/450}{9.1\%} & \rate{0/41}{0.0\%} & \rate{0/450}{0.0\%} & $[0.0,\,0.0]$ \\
 & \confguicp{} & 450 & \rate{131/450}{29.1\%} & \rate{12/131}{9.2\%} & \rate{12/450}{2.7\%} & $[0.7,\,5.1]$ \\
 & \method{} & 450 & \rate{140/450}{31.1\%} & \rate{98/140}{70.0\%} & \rate{98/450}{21.8\%} & $[14.0,\,30.2]$ \\
 & \confdelegated{} & 450 & \rate{161/450}{35.8\%} & \rate{79/161}{49.1\%} & \rate{79/450}{17.6\%} & $[11.1,\,24.7]$ \\
\bottomrule
\end{tabular}

\end{table}

\subsection{Qualitative Failure Modes of the ScaleCUA-Driven \confcua{}}
\label{sec:appendix-cua-failures}

We inspected the per-step execution record of every \confcua{} trial on the matched population of Table~\ref{tab:performance-matched-seed0} (four models $\times$ 61 tasks $\times$ 3 seeds, 732 trials). After the harness maps ScaleCUA's normalized coordinates to the policy frame and compiles each response to one bounded action, every trial runs to the shared 100-step Phase~A budget or to a graded Start/Goal outcome; none terminate on action-decoding rejection. The resulting Start and E2E rates in Tables~\ref{tab:performance-main} and~\ref{tab:performance-matched-seed0} are therefore directly comparable to the other configurations on that population.

The dominant failure is deep-state navigation, not interface rejection. Table~\ref{tab:cua-error-types} groups Phase~A outcomes. 383/732 trials (52.3\%) exhaust the step budget without reaching Start; 257/732 (35.1\%) execute syntactically valid actions that still fail the Start gate, typically by operating on the wrong surface or leaving the application context; 92/732 (12.6\%) reach Start, of which 79 then fail Goal and 13 record E2E success. Thus, 383/640 Start failures (59.8\%) exhaust the step budget.

\begin{table}[!htbp]
\caption{Phase~A outcome breakdown for \confcua{} on the matched population of Table~\ref{tab:performance-matched-seed0} (732 trials: \texttt{gpt-5.6-terra}, \texttt{gemini-3.5-flash}, \texttt{kimi-k2.6}, \texttt{gpt-6-astra}). A count is the number of trials whose Phase~A ends in that outcome.}
\label{tab:cua-error-types}
\begin{center}
\floatsize
\begin{tabular}{p{0.82\linewidth}r}
\hline
Phase~A outcome & Trials \\
\hline
Step budget exhausted before Start & 383 \\
Valid actions fail Start gate (wrong navigation or surface) & 257 \\
Reached Start; Goal failed & 79 \\
E2E success & 13 \\
\hline
\end{tabular}

\end{center}
\end{table}

\begin{figure}[t]
\centering
\includegraphics[width=0.7\linewidth]{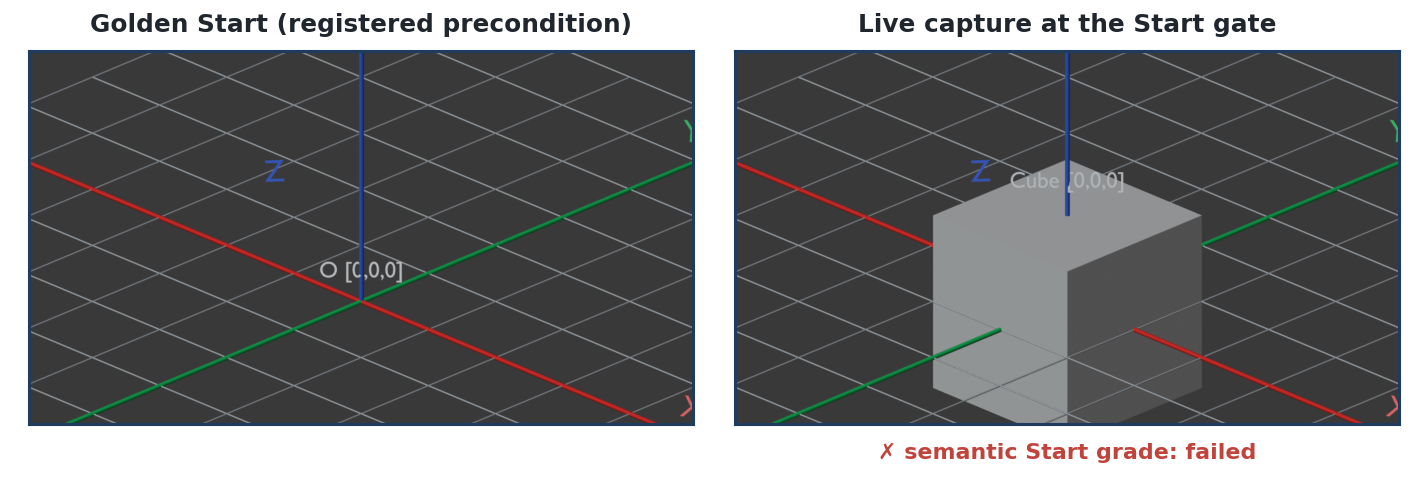}
\caption{The \texttt{blender-product-model-pipeline} case from Table~\ref{tab:cua-error-types}'s wrong-navigation row. \confcua{} executes valid pointer actions but never clears Blender's default cube; the Start-gate visual judge compares this frame against the registered empty-scene precondition and fails it.}
\label{fig:cua-failure-qualitative}
\end{figure}

Two trials illustrate the two largest rows. On \texttt{2048-cascade-256} (\texttt{gemini-3.5-flash}), the worker spends its first dozen steps toggling the browser's developer console with keyboard shortcuts, never touching the board, and exhausts the budget at 228.6s without reaching Start. On \texttt{blender-product-model-pipeline} (\texttt{gpt-5.6-terra}), the worker issues a sequence of pointer moves and menu clicks that leave Blender at its default startup cube (Figure~\ref{fig:cua-failure-qualitative}); after 41 steps and 132.4~s, the Start-gate visual judge compares this frame against the registered empty-scene precondition and rejects the trial without recording Start.

Read together with Table~\ref{tab:performance-matched-seed0}, \confcua{} spends more tokens and wall clock than \method{} on every model while recording lower Start and E2E rates; its TiPS is 15--62$\times$ higher wherever it records any E2E success. The gap is therefore a policy and reachability limitation under pixel execution, not a side effect of how the harness translates ScaleCUA's model-facing calls into bounded runtime actions.

\subsection{Diagnostic Comparison: Delegated Transition Execution (\method{} vs.\ \confdelegated{})}
\label{sec:appendix-e-prime}

\paragraph{Configuration and observed population.}
\method{} jointly emits a checkpoint and a transition spec in one source exploration. \confdelegated{} instead delegates the target action to a separate bounded semantic-action planner using the functional test case text. Both configurations use transition-spec execution. The diagnostic runs outside the \perfstudy{} rotation and covers 61 tasks, three models, and three seeds: 549 records for each configuration. Table~\ref{tab:delegated-diagnostic} reports each model separately, with all 183 task--seed trials retained in every denominator.

\begin{table}[!htbp]
\caption{Descriptive results for joint and delegated transition execution over 61 tasks and three seeds. E2E percentages, mean total tokens, and mean catalog-price cost retain failed trials.}
\label{tab:delegated-diagnostic}
\begin{center}
\floatsize
\setlength{\tabcolsep}{3pt}
\begin{tabular}{llrcccrrr}
\toprule
Model & Configuration & $n$ & Start & Goal $\mid$ Start & E2E & Tokens (k) & Cost (\$) & Failures \\
\midrule
gpt-5.6-terra & \method{} & 183 & \rate{56/183}{30.6\%} & \rate{45/56}{80.4\%} & \rate{45/183}{24.6\%} & 51.6 & 0.196 & 0 \\
gpt-5.6-terra & \confdelegated{} & 183 & \rate{70/183}{38.3\%} & \rate{47/70}{67.1\%} & \rate{47/183}{25.7\%} & 63.8 & 0.242 & 0 \\
gemini-3.5-flash & \method{} & 183 & \rate{92/183}{50.3\%} & \rate{65/92}{70.7\%} & \rate{65/183}{35.5\%} & 115.1 & 0.328 & 1 \\
gemini-3.5-flash & \confdelegated{} & 183 & \rate{101/183}{55.2\%} & \rate{54/101}{53.5\%} & \rate{54/183}{29.5\%} & 149.4 & 0.426 & 0 \\
kimi-k2.6 & \method{} & 183 & \rate{63/183}{34.4\%} & \rate{41/63}{65.1\%} & \rate{41/183}{22.4\%} & 153.5 & 0.230 & 0 \\
kimi-k2.6 & \confdelegated{} & 183 & \rate{70/183}{38.3\%} & \rate{32/70}{45.7\%} & \rate{32/183}{17.5\%} & 206.2 & 0.309 & 0 \\
\bottomrule
\end{tabular}

\end{center}
\end{table}

\paragraph{Observed outcomes.}
Across all models and seeds, \method{} reaches 151/549 E2E outcomes (27.5\%) at 106.7k mean tokens, while \confdelegated{} reaches 133/549 (24.2\%) at 139.8k. By model, E2E counts are 45 versus 47 for \texttt{gpt-5.6-terra}, 65 versus 54 for \texttt{gemini-3.5-flash}, and 41 versus 32 for \texttt{kimi-k2.6}, each out of 183 trials. Conditioned on Start, the two convert 151/211 and 133/241 of their restored trials into a Goal success, the contrast Section~\ref{sec:performance} discusses. These are marginal outcomes of the two configurations run separately, so the table reports what each policy achieved rather than a paired causal estimate of joint generation.

\subsection{Fail-to-Pass and Pass-to-Pass Tests}
\label{sec:appendix-f2p-p2p}

\paragraph{Definitions.}
Benchmarks built from real repairs label a repository's registered tests by how their verdicts change across the landed fix \citep{jimenez2024swe}.  A \emph{fail-to-pass} (F2P) test fails on the pre-fix tree and passes on the post-fix tree: it is the executable statement of what the repair was supposed to change, and a candidate patch is graded correct when it turns every F2P test green.  A \emph{pass-to-pass} (P2P) test passes on both trees: it states what the repair must leave alone, and it is the regression guard that stops a patch from buying its F2P verdict by breaking something else.  Both labels are properties of a test relative to a specific pair of trees rather than of the test's source, and both are established by executing the suite on each tree rather than by reading the patch.

\paragraph{Why a P2P-only population is the treatment target.}
Patch benchmarks begin from at least one F2P test and ask a method to produce the patch that makes it pass.  The \latentstudy{} does not ask for a patch --- the repair has already landed and both trees are real --- and instead studies the complementary blind spot: repairs for which no F2P test exists upstream, so that the failure signal a patch grader would read simply is not there.  Recovering that signal from the issue text is itself unreliable, succeeding on roughly one third of real issues or fewer \citep{swtbench2024,otter2025}, and it is least available exactly where interactive software fails, in defects confined to the rendered result \citep{visiondroid2025}.  Such a defect is visible in the real rendered or interactive output --- analogous to a game object visibly crossing a wall --- while the software and its existing tests continue to run.  F2P absence is therefore an admission requirement of this population, not a gap in our evidence: it is what makes an admitted repair a case the registered suite cannot decide.

\paragraph{The two roles P2P plays here.}
As an admission criterion, a non-empty P2P set that is green on the fixed and on the reversed tree makes the absence of a failure signal a measured property of the registered suite rather than an untested assumption: the tests run, and they do not notice the defect.  As a per-trial control, the same set is executed on the clone the trial actually drives, so that a rendered difference is attributable to the defect rather than to a broken build; the buggy tree is obtained by reversing real production hunks, and a reversal that disturbed unrelated behavior would surface as a P2P failure.  A P2P failure consequently blocks the trial instead of counting as defect evidence, and any test discovered to be F2P disqualifies the task.

\subsection{Latent-Defect Study: P2P-Only Population and Protocol}
\label{sec:appendix-latent-defect}

The \latentstudy{} asks whether a checkpoint-driven deep-GUI test exposes a real upstream defect for which the registered unit tests provide no failure signal, in the sense fixed in Appendix~\ref{sec:appendix-f2p-p2p}.  The paper population is frozen to the ordered eleven \texttt{lzp-*} rows in \texttt{benchmarks/p5/c1\_lazarus\_p2p\_only\_tasks.yaml}, listed against their upstream repairs in Table~\ref{tab:latent-defect-tasks}: 6 JupyterLab, 3 Luanti, and 2 OpenRA tasks.

\paragraph{Fail-closed admission and construction gates.}
\begin{itemize}
\item Confirm a mechanically traceable public Issue--landed-repair relation, production patch, source revision, and dependency closure.
\item Materialize the fixed tree, then obtain the buggy tree only by exact reverse application of the real production hunks; synthetic mutation, fuzzy/three-way merge, and test-patch injection are forbidden.
\item Require \texttt{fail\_to\_pass=[]}, a non-empty frozen P2P set, and a non-empty per-trial P2P set.  Any discovered F2P test disqualifies the row from this population.
\item Execute the same P2P control on fixed and buggy in isolated APFS copy-on-write clones.  Skip, invalid discovery, flake, provenance drift, or a P2P failure blocks the trial rather than becoming defect evidence.
\item Cold-start each candidate build and require fixed and buggy to restore a visually equivalent common-before Start before the target action.
\item Execute the same public target action and capture both builds' after-references, so that each run's own render can be placed against its own version's reference and against the other's, while shared invariants and per-trial P2P remain satisfied.
\end{itemize}

\begin{table}[!htbp]
\caption{The only latent-defect-study paper population: the frozen ordered task IDs of the eleven P2P-only, \texttt{at\_goal} tasks (6 JupyterLab, 3 Luanti, 2 OpenRA).}
\label{tab:latent-defect-tasks}
\begin{center}
\floatsize
\begin{tabular}{ll}
\hline
Repository/change & Task ID \\
\hline
JupyterLab 17153 & \texttt{lzp-jupyterlab-17153-shortcut-chip-contrast} \\
JupyterLab 17222 & \texttt{lzp-jupyterlab-17222-toolbar-button-pressed-toggle} \\
JupyterLab 17618 & \texttt{lzp-jupyterlab-17618-high-contrast-selection-layer} \\
JupyterLab 18464 & \texttt{lzp-jupyterlab-18464-debugger-zero-variable-detail} \\
JupyterLab 18996 & \texttt{lzp-jupyterlab-18996-htmlselect-caret-offset} \\
JupyterLab 19091 & \texttt{lzp-jupyterlab-19091-themed-button-glyph-font} \\
Luanti 16679 & \texttt{lzp-luanti-16679-unavailable-settings} \\
Luanti 16799 & \texttt{lzp-luanti-16799-world-mods} \\
Luanti 17152 & \texttt{lzp-luanti-17152-server-list-search-tooltip} \\
OpenRA 22283 & \texttt{lzp-openra-22283-tileset-template-order} \\
OpenRA 22476 & \texttt{lzp-openra-22476-load-game-faction-filter} \\
\hline
\end{tabular}
\end{center}
\end{table}

\paragraph{Execution schedule and reporting.}
The schedule contains eleven tasks, four decision models (\texttt{gpt-5.6-terra}, \texttt{gemini-3.5-flash}, \texttt{kimi-k2.6}, and \texttt{gpt-6-astra}), seed zero, two versions, and \confgui{}, \confcua{}, \confguicp{}, \confcuacp{}, and \method{}, for 440 cells in 44 ten-cell groups, of which 430 are analyzed. The ten cells of one group (\texttt{gemini-3.5-flash} on the OpenRA \texttt{load-game-faction-filter} task) are withheld because the task cannot fit its registered budget rather than because the configurations failed it: Phase~B steps average 67.1~s on OpenRA, so the registered 100-step allowance needs roughly 108 minutes against a 2400~s wall-clock limit, and three consecutive attempts were cut off at step 36 on a different configuration each time. Withheld cells contribute no analyzed evidence but remain in the registered 22-cell and 11-task denominators. This single-seed experiment is descriptive and separate from the \perfstudy{} and the visual-calibration population. The decision models request disabled thinking: GPT routes use \texttt{reasoning\_effort=none}, while Gemini/Kimi routes explicitly disable thinking. \confgui{} and \confcua{} construct Start from the real cold entry with pixel actions, the latter driven by ScaleCUA, and share a 100-step budget between construction and transition; both compile to the executed action space of Table~\ref{tab:action-space}. \confguicp{}, \confcuacp{} and \method{} share one synthesized checkpoint per task/model/seed across both versions; the first two use a 100-step pixel transition and the last at most 100 generated semantic actions. Fixed versions reuse the artifact with zero additional synthesis calls. Results are descriptive per model. In this P2P-only runtime, semantic differential assertions are not evaluated: defect evidence comes from the two after-image oracles, while the nonvisual checks establish Start reachability, runtime health, and P2P preservation. Catalog fields naming \texttt{goal\_checks} or \texttt{defect\_signature} must not be read as independent semantic evidence for the paid outcomes.

\paragraph{Scoring version discrimination.}
\label{sec:appendix-differential}
Reaching the defect state is not the same as producing a test that tells the two versions apart, and the second question is scored separately here. Each task carries an after-reference captured from the real fixed build and one from the real buggy build, so the pixels where those two disagree by more than 8 per channel -- the threshold below which every task's rubric ignores antialiasing -- are the rendered footprint of the landed fix. On those pixels alone, a run's after-frame is assigned to whichever reference it deviates less from, requiring a 10\% margin between the two deviations, and a task counts as separated when the fixed run lands on the fixed reference and the buggy run on the buggy one. Nothing in this procedure calls a model. The after-stage, including its P2P subset, is skipped when Start fails, so such a cell supplies no evidence either way. Because the study's runtime directories were pruned after publication, the frames come from the run's own log-viewer image store and are used only when a file's SHA-256 equals the hash the record stored for that frame; 141 of 141 referenced frames were recovered this way, and the collector re-derives the Goal counts of Table~\ref{tab:latent-defect-results} from the same records before scoring anything.

\begin{figure}[!htbp]
\centering
\includegraphics[width=0.92\linewidth]{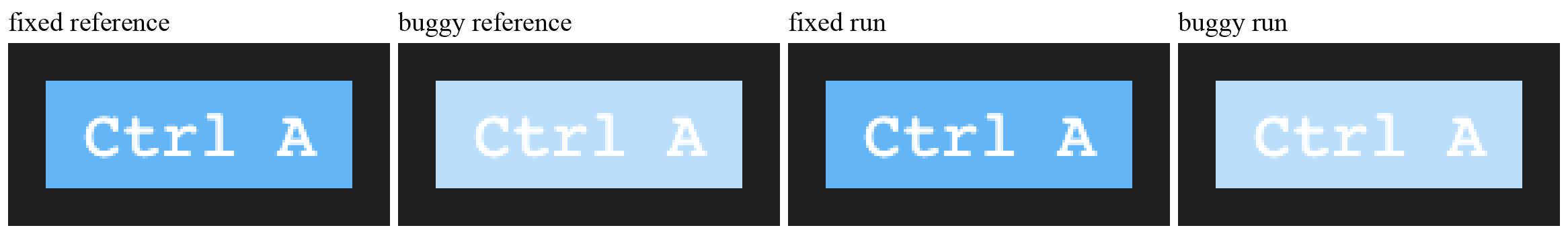}
\caption{The \texttt{jupyterlab-17153} shortcut-chip repair under \method{} with \texttt{gpt-6-astra}, cropped to the pixels where the two references disagree. The fixed run reproduces its own reference's medium blue chip and the buggy run the near-white chip that makes the white label unreadable, which is the defect the landed fix removed, so each frame sits closer to its own version's reference.}
\label{fig:latent-defect-differential-case}
\end{figure}

Table~\ref{tab:latent-defect-differential} gives the result. \method{} separates 7 of the 8 tasks on which it has a comparable pair, against 4 and 2 for the two checkpoint-seeded pixel configurations and 2 and 1 for the cold ones; under \texttt{gemini-3.5-flash} the split is 6 against 0 and 0. Two falsification controls bound it. The two versions must cold-start into a visually equivalent common-before state, and on those before-frames the criterion separates 0 of 85 pairs. Pairing two \emph{fixed}-version frames of one task from different decision models, which cannot be separated by construction either, it reports 4 of 26 -- the measured false-separation rate, concentrated on the two Luanti tasks and one JupyterLab task. Counts are identical at margins 0.05 and 0.10 and lose one \method{} task at 0.20. Figure~\ref{fig:latent-defect-differential-case} shows what one separated task looks like: the fixed run renders the medium blue shortcut chip of its own reference and the buggy run the near-white chip of its own reference.

\begin{table}[!htbp]
\caption{Deterministic version discrimination over the frozen latent-defect frames, with no model calls. Each cell counts the tasks whose fixed and buggy runs land on their own version's reference, out of the tasks for which that model and configuration have a geometrically comparable pair; ``\textemdash{}'' marks a cell with none. Nine of the eleven tasks are comparable at all, the two OpenRA tasks being excluded because their runs capture a 1280$\times$720 viewport against 1600$\times$1200 references, and no pixel comparison can absorb that change of window aspect.}
\label{tab:latent-defect-differential}
\begin{center}
\floatsize
\begin{tabular}{lccccc}
\toprule
Configuration & \texttt{terra} & \texttt{gemini} & \texttt{kimi} & \texttt{astra} & Pooled \\
\midrule
\confgui{} & 1/3 & \textemdash{} & \textemdash{} & 1/2 & 2/3 \\
\confcua{} & 1/3 & \textemdash{} & \textemdash{} & 0/2 & 1/2 \\
\confguicp{} & 2/4 & 0/7 & 0/5 & 4/7 & 4/8 \\
\confcuacp{} & 0/6 & 0/7 & 0/5 & 2/8 & 2/8 \\
\method{} & 3/6 & 6/7 & 2/5 & 6/8 & 7/8 \\
\bottomrule
\end{tabular}

\end{center}
\end{table}

Three limits apply. The mask is grader-side and defect-aware: it encodes where the landed fix shows, which is what a regression test written for a known repair always encodes, and it is never exposed to the method, but it does mean the criterion asks whether a reached state renders its own version rather than whether a defect can be found without knowing it. Geometry has to line up: two JupyterLab tasks ship a tight crop as their reference (344$\times$132 and 340$\times$104) while their runs capture the full 760$\times$420 viewport, which the criterion handles by locating the crop inside the frame first, whereas the two OpenRA tasks cannot be scored at all. And the population is the one the rest of this study carries, 9 comparable tasks of 11 under a single seed, so these counts describe how the criterion behaves on this population rather than estimating a rate that generalizes.

\paragraph{Treatment cost and reuse.}
Table~\ref{tab:latent-defect-cost} sums synthesis, Start restoration, and transition execution from complete per-phase token ledgers over 22 cells per model/configuration, and repeats each row's E2E outcome so that a cost is never read without the result it produced. Shared synthesis is attributed once to each consuming method for method-level comparison; such attributed rows must not be summed as actual provider expenditure. Shared judging and acquisition overhead are excluded, and time is summed per-cell duration rather than batch wall-clock time. Actual acquisition expenditure, including shared costs and overhead, is 27.9M tokens. All 129 analyzed fixed-version cells of the three checkpoint-restoring configurations reuse their buggy-side checkpoint with zero new synthesis calls. Artifact reuse does not imply successful Start restoration.

\begin{table}[!htbp]
\caption{Method-attributed latent-defect-study treatment cost, reported next to the outcome that cost bought. Denominators are 22 cells per row; E2E repeats the E2E column of Table~\ref{tab:latent-defect-results} as counts over the same cells. Shared judging and acquisition overhead are excluded; time is summed per-cell duration, with the measurement qualification stated in the text.}
\label{tab:latent-defect-cost}
\begin{center}
\floatsize
\setlength{\tabcolsep}{3.5pt}
\begin{tabular}{llrrr}
\hline
Model & Configuration & E2E (Goal) & Attributed tokens & Attributed time (min) \\
\hline
gpt-5.6-terra & \confgui{} & 1/22 (4.5\%) & 7{,}556{,}400 & 97.6 \\
gpt-5.6-terra & \confcua{} & 1/22 (4.5\%) & 3{,}449{,}600 & 63.8 \\
gpt-5.6-terra & \confguicp{} & 1/22 (4.5\%) & 1{,}775{,}070 & 28.3 \\
gpt-5.6-terra & \confcuacp{} & 0/22 (0.0\%) & 587{,}558 & 16.1 \\
gpt-5.6-terra & \method{} & 5/22 (22.7\%) & 273{,}471 & 3.5 \\
gemini-3.5-flash & \confgui{} & 1/22 (4.5\%) & 4{,}771{,}180 & 117.0 \\
gemini-3.5-flash & \confcua{} & 0/22 (0.0\%) & 2{,}934{,}800 & 71.9 \\
gemini-3.5-flash & \confguicp{} & 1/22 (4.5\%) & 2{,}404{,}711 & 38.4 \\
gemini-3.5-flash & \confcuacp{} & 0/22 (0.0\%) & 1{,}519{,}582 & 16.2 \\
gemini-3.5-flash & \method{} & 9/22 (40.9\%) & 1{,}449{,}541 & 11.7 \\
kimi-k2.6 & \confgui{} & 0/22 (0.0\%) & 1{,}902{,}707 & 82.4 \\
kimi-k2.6 & \confcua{} & 0/22 (0.0\%) & 2{,}523{,}400 & 43.3 \\
kimi-k2.6 & \confguicp{} & 0/22 (0.0\%) & 3{,}250{,}773 & 68.3 \\
kimi-k2.6 & \confcuacp{} & 0/22 (0.0\%) & 2{,}008{,}919 & 13.0 \\
kimi-k2.6 & \method{} & 3/22 (13.6\%) & 1{,}971{,}858 & 10.6 \\
gpt-6-astra & \confgui{} & 1/22 (4.5\%) & 2{,}544{,}052 & 78.3 \\
gpt-6-astra & \confcua{} & 1/22 (4.5\%) & 2{,}008{,}600 & 43.6 \\
gpt-6-astra & \confguicp{} & 5/22 (22.7\%) & 2{,}281{,}067 & 42.9 \\
gpt-6-astra & \confcuacp{} & 3/22 (13.6\%) & 1{,}384{,}096 & 17.8 \\
gpt-6-astra & \method{} & 10/22 (45.5\%) & 1{,}324{,}297 & 13.7 \\
\hline
\end{tabular}

\end{center}
\end{table}

\paragraph{Reproducible evidence.}
The supplementary \texttt{scripts/\allowbreak collect\_\allowbreak experiment\_\allowbreak evidence.py}, invoked with \texttt{-{}-lazarus-only}, reads the 430 analyzed cells under the frozen single-seed specification and recomputes Start, Goal, and method-attributed costs without model calls, while \texttt{scripts/\allowbreak collect\_\allowbreak latent\_\allowbreak defect\_\allowbreak differential.py} recomputes the version-discrimination counts from the frozen frames. The experiment remains descriptive and non-confirmatory: the pooled column of Table~\ref{tab:latent-defect-differential} counts each distinct task once, separated under at least one model, rather than averaging the per-model rates, no across-seed variability is estimated, and per-cell outcomes, request settings, and cost attribution accompany the supplementary evidence.

\subsection{Fuzzing Study: Mechanism and Campaign}
\label{sec:appendix-fuzzing}

\paragraph{Mechanism.}
A validated deep state is also a cheap seed for exploring the states around it, which from-scratch replay or scripted GUI testing rarely reaches. Seeded with a validated checkpoint, an LLM-driven evolutionary loop proposes neighborhood states with one of four mutation operators (field substitution, unexpected-key insertion, subtree regeneration, and type flip), chosen by a Thompson-sampling bandit over each operator's yield. Each mutated state is injected through the same warm-session restore path as an ordinary checkpoint, and a robustness oracle labels the result \textsc{defect} (the software misbehaves on a well-formed state), \textsc{robust} (it handles the mutation gracefully), or \textsc{rejected} (it correctly refuses an invalid state). Only \textsc{defect} outcomes are counted, so an operator cannot inflate its yield by proposing states that are merely invalid. Defects are deduplicated by a deterministic fingerprint over the failure signature, and each carries its mutated state as a model-free reproduction. The search is economical only because restoring one more state inside a warm session is cheap (Section~\ref{sec:replay}); otherwise every candidate would cost a full restart.

\paragraph{Campaign on \texttt{SpaceRPG}.}
We ran the \fuzzstudy{} as one unattended campaign against the \texttt{SpaceRPG} profile, seeded from its checkpoints validated in the \perfstudy{}; it surfaced the 67 deduplicated defects in Table~\ref{tab:spacerpg-fuzz-defects}. Mutations entered through five injection modes: three write the game's own persisted save keys, and two write the live store through the DEV-gated \texttt{\_\_SPACERPG\_E2E\_\_} hook that production builds compile out. Reachability is therefore a mechanical label rather than a judgment call. The 55 \emph{player-reachable} defects arrive through the production save loader, exactly as a hand-edited save or one written before a schema change would; the 12 \emph{hook-only} defects are resilience gaps rather than player-facing bugs. The dominant player-reachable class is one a programmatic oracle would miss: the crafted save loads without error and every state assertion holds, so the failure appears only in the rendered frame, the separation Section~\ref{sec:hybrid-eval} is built around. Each triggering state doubles as a model-free regression fixture.

\begin{table}[!htbp]
\caption{The 67 deduplicated defects surfaced by the \texttt{SpaceRPG} state-fuzzing campaign, by failure class and by which surface reaches them. Reachability follows mechanically from the injection mode: the three save-key modes write state the game itself reads on load, while the two store modes require the DEV-gated hook that production builds compile out.}
\label{tab:spacerpg-fuzz-defects}
\begin{center}
\floatsize
\begin{tabular}{@{}llrrr@{}}
\toprule
Failure class & Captured signature & \shortstack[r]{Player-\\reachable} & \shortstack[r]{Hook-\\only} & Total \\
\midrule
\texttt{blank\_render} & non-informative frame after restore & 45 & 0 & 45 \\
\texttt{console\_error} & unhandled error; NaN into three.js geometry & 10 & 7 & 17 \\
\texttt{soft\_lock} & no actionable UI and no open dialogue & 0 & 5 & 5 \\
\midrule
Total & & 55 & 12 & 67 \\
\bottomrule
\end{tabular}
\end{center}
\end{table}

\paragraph{Scope.}
The campaign is descriptive: no mutation budget was fixed in advance and the loop was not re-run under a frozen seed, so the 67 defects are what one campaign surfaced rather than a defect-per-mutation yield, and they are not pooled with the performance, replay, or visual-calibration populations. The failure classes are also the browser profile's vocabulary; a native-engine profile reports crashes and invariant violations instead, so these counts do not transfer across profile families.

\subsection{Evaluation Efficiency}
\label{sec:evaluation-efficiency}

\begin{figure}[t]
\centering
\includegraphics[width=0.98\linewidth]{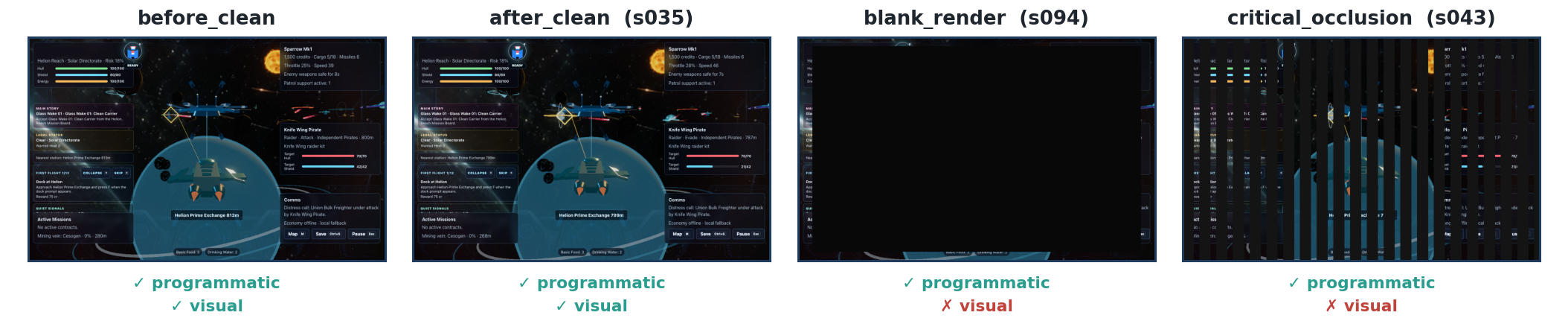}
\caption{The \texttt{spacerpg-playable-flight} qualitative case of Section~\ref{sec:hybrid-eval-results}. All four frames share the same base task and programmatic projection; \texttt{before\_clean} and \texttt{after\_clean} bracket the target action, and \texttt{blank\_render}/\texttt{critical\_occlusion} are the two fault variants applied to the \texttt{after} frame (\texttt{critical\_occlusion} is called \emph{distributed striped occlusion} in the main text). The programmatic oracle accepts all four; the visual judge accepts only the two clean frames, so the Hybrid AND channel accepts \texttt{s035} and rejects \texttt{s094}/\texttt{s043}.}
\label{fig:spacerpg-hybrid-qualitative}
\end{figure}

This section states why evaluation itself is cheaper under the design of Section~\ref{sec:hybrid-eval}, and keeps that claim cleanly separate from the 200-sample calibration protocol of Section~\ref{sec:hybrid-eval-results}: the production path issues a visual request only for a trial the Programmatic State Oracle has already passed, so the calls it avoids are exactly its programmatic failures, while Section~\ref{sec:hybrid-eval-results} judges every one of its 200 samples whatever their programmatic status. Both paths run the single judge of Section~\ref{sec:visual-audit-cascade} with the verdict cache disabled, so there is no escalation or cache-hit rate to report. A provider-side prompt-prefix cache may report cached prompt tokens under either path; it reuses prefix computation only, never a stored verdict, so it must not be read as a verdict-cache hit rate. The two disagreement directions are what motivate the conjunction in the first place, and they are not symmetric under this design. A state whose semantic assertion passes can still render wrongly, and Section~\ref{sec:hybrid-eval-results} measures that direction: the 80 fault samples pass the programmatic oracle by construction and the visual channel rejects 77 of them. The converse direction, a programmatic failure the visual judge would have accepted, is short-circuited before any call is issued and so is never measured; the monotonicity invariant is what makes discarding it free, since such a trial cannot be accepted either way. Finally, the same formal case from Section~\ref{sec:hybrid-eval-results} instantiates the delivery-channel property of Section~\ref{sec:coverage} directly: how the target action is delivered changes across the three formal configurations, but how the result is judged does not; the restored Start passes through the same visual gate under the same rubric (whose \emph{must\_not} field explicitly includes crashes, blank frames, large opaque overlays, and distributed banding) regardless of which configuration produced it, and \confdelegated{} reuses the identical rubric. A Start that is actually blank or badly corrupted therefore cannot escape judgment merely because the target action is executed as a transition spec instead of a pixel click; the \texttt{s035}/\texttt{s094}/\texttt{s043} samples in Section~\ref{sec:hybrid-eval-results} are a concrete instance of this property (Figure~\ref{fig:spacerpg-hybrid-qualitative}).

\subsection{Reference Controls and Ablations}
\label{sec:ablations}

\textbf{Reference coverage on the current 61-task population.} Every registered task in the \perfstudy{} population has a reference specification. The qualification stage of Section~\ref{sec:hybrid-eval-results} executes each reference twice, before selecting the 40 visual-calibration bases. All 61 references pass both executions, giving 122/122 successful programmatic qualification runs across all ten profiles (Table~\ref{tab:reference-coverage}). All entries are backed by reference execution; none uses a registered-golden fallback.

\begin{table}[t]
\caption{Reference qualification on the complete current performance-study catalog. Each reference is executed twice without synthesis or VLM judging; success is the programmatic qualification criterion with real runtime capture. These are feasibility controls on the catalog itself, not model-synthesis scores or full hybrid-oracle acceptance rates.}
\label{tab:reference-coverage}
\begin{center}
\floatsize
\begin{tabular}{lrrr}
\hline
Profile & Reference specs & Qualified tasks & Successful executions \\
\hline
2048 & 8 & 8/8 & 16/16 \\
SpaceRPG & 8 & 8/8 & 16/16 \\
DatingSim & 7 & 7/7 & 14/14 \\
Luanti & 8 & 8/8 & 16/16 \\
OpenRA & 5 & 5/5 & 10/10 \\
JupyterLab & 5 & 5/5 & 10/10 \\
jynew & 5 & 5/5 & 10/10 \\
PAL3 & 5 & 5/5 & 10/10 \\
LibreOffice & 5 & 5/5 & 10/10 \\
Blender & 5 & 5/5 & 10/10 \\
\hline
Total & 61 & 61/61 & 122/122 \\
\hline
\end{tabular}

\end{center}
\end{table}

\textbf{What this control establishes.} A reference-execution pass establishes that the registered state and transition can be exercised by the qualification path. It does not establish that a model can synthesize that reference, that \confgui{} can execute its pixel equivalent, or that the full hybrid grader used in the \perfstudy{} is free of errors. The reference population is therefore reported separately from the model-generated checkpoint scores of Section~\ref{sec:performance}.

\textbf{Replay-study design notes.} The \replaystudy{} of Section~\ref{sec:replay} replays each frozen checkpoint through the \emph{unmodified} \perfstudy{} execution and grading path, so only the artifact's provenance changes between a freshly synthesized checkpoint and a frozen one on disk, and a disagreement between the reference verdict and the replay verdict is attributable to the artifact rather than to two different harnesses. Three design choices follow. The population retains checkpoints the \perfstudy{} judged as both accepted and rejected, since the measurement is verdict reproduction rather than pass rate; a guard counts and rejects any model call, which makes the zero-call figure structural rather than a promise; and the replay's normalized pixel-equality check against the frozen reference frame is stricter than the Section~\ref{sec:hybrid-eval} rubric, which tolerates benign non-determinism such as 2048's randomly placed new tile, so it is reported as a separate \emph{render-identity} diagnostic and is never allowed to gate the Start that admits Phase~B. The two replay environments are two sessions on one host with separate temporary directories and ports. Determinism and reference mismatch are different claims: exactly one of the 112 checkpoints (\texttt{luanti-build-column}) is internally inconsistent across its own replays, 2 of its 13 disagreeing with each other in the engine's cold-start path, so 1 checkpoint in 112 flakes (0.0089); a second (\texttt{jupyterlab-reproducible-experiment-workspace}) is consistent across all 13 of its replays yet consistently disagrees with the original reference verdict. The two figures are counted over different units. Self-consistency (replay-to-replay agreement) is 0.9911 over checkpoints, while reproduction (agreement with the original reference, which additionally requires an unchanged harness and host) is counted over replays and gives the 0.9890 headline of Section~\ref{sec:replay}; the two environments' reproduction rates differ by 0.3 points (0.9869 and 0.9896), and the restore alone accounts for 206~ms of the 422~ms median replay time.

\textbf{Other component comparisons.} Appendix~\ref{sec:appendix-e-prime} reports unpaired descriptive results for joint and delegated transition execution. The \confgui{}-vs-\confguicp{} comparison assesses checkpoint-seeded pixel execution as a complete configuration; these outcomes do not measure the individual contribution of each semantic, persistence, runtime-health, or visual check.

\subsubsection{Component Ablations of Checkpoint Synthesis}
\label{sec:appendix-synthesis-ablation}

\textbf{Population and controlled interventions.}
We evaluate \texttt{gpt-5.6-terra} on all 61 performance-study tasks with seed 0 under
four configurations, giving 244 task--configuration trials. The full method and
each ablation share the task, model route, generation budget, restoration
interface, and hidden Start/Goal graders. Exploration permits 16 response turns
with an 8,192-token response cap; one public format repair is allowed, and
hidden feedback never triggers candidate resampling. Four rounds rotate the
configuration order by task index. Each configuration therefore covers the
same task population, including construction failures and unsuccessful runs.

The \emph{no-source} intervention disables repository queries and requires empty
source evidence, while retaining the public task and interface contract.
The \emph{schema-only} intervention removes descriptive contract prose but
retains operation names, argument types, enumerations, and bounds; execution
uses the same validators. The \emph{no-assertion-prediction} intervention asks
the model to emit $A=\emptyset$ during synthesis and any format repair.
Its public admission rules permit empty auxiliary assertions, while retaining
the action requirements and all hidden checks. This measures whether predicting
the after-state helps joint synthesis; it does not remove assertions after
generation. Strategy-selection ablation is inapplicable because every task
specifies a single legal restoration strategy.

\begin{table}[t]
\centering
\caption{Synthesis-component ablations with \texttt{gpt-5.6-terra}, seed 0. Each row
contains all 61 tasks. Tokens and active time are means including unsuccessful
trials. ``Goal $\mid$ Start'' is descriptive because the successful Start subsets
differ across configurations.}
\label{tab:synthesis-ablation-main}
\floatsize
\begin{tabular}{lrrS[table-format=2.1]rr}
\toprule
Configuration & Start & E2E & {Goal $\mid$ Start (\%)} & Tokens (k) & Time (s) \\
\midrule
\method{} (full) & 22/61 & 19/61 & 86.4 & 52.7 & 52.9 \\
w/o source exploration & 22/61 & 14/61 & 63.6 & 7.4 & 32.5 \\
Schema-only contract & 21/61 & 17/61 & 81.0 & 52.0 & 47.8 \\
w/o assertion prediction & 21/61 & 13/61 & 61.9 & 50.1 & 46.0 \\
\bottomrule
\end{tabular}

\end{table}

\textbf{Success and computational cost.}
Table~\ref{tab:synthesis-ablation-main} reports 19/61 E2E successes for the full
method, versus 14/61 without exploration, 17/61 with the schema-only contract,
and 13/61 without assertion prediction. The corresponding Start counts are
22, 22, 21, and 21. Removing exploration reduces mean tokens by 85.9\%, from
52.7k to 7.4k, alongside five fewer E2E successes. The observations support a
cost--success tradeoff on this population; neither a universal benefit nor a
significant improvement follows from the point estimates for 61 tasks under a
single generation seed.

\begin{table}[t]
\centering
\caption{Paired E2E differences: full method minus each ablation. Intervals use
10,000 bootstrap samples of tasks, stratified by application and preserving all
four configurations of each sampled task. Wins/losses count discordant pairs
in favor of the full/ablated method. Exact McNemar tests are corrected across
the three planned comparisons using Holm's procedure.}
\label{tab:synthesis-ablation-paired}
\floatsize
\begin{tabular}{lrrrrr}
\toprule
Compared configuration & $\Delta$ (pp) & 95\% CI & Wins/losses & Exact $p$ & Holm $p$ \\
\midrule
w/o source exploration & +8.2 & [0.0, 16.4] & 7/2 & 0.180 & 0.359 \\
Schema-only contract & +3.3 & [-4.9, 11.5] & 5/3 & 0.727 & 0.727 \\
w/o assertion prediction & +9.8 & [1.6, 19.7] & 8/2 & 0.109 & 0.328 \\
\bottomrule
\end{tabular}

\end{table}

\textbf{Paired evidence and limits.}
The E2E differences are $+8.2$, $+3.3$, and $+9.8$ percentage points
(Table~\ref{tab:synthesis-ablation-paired}). None reaches the 0.05 threshold
under the prespecified exact paired tests with Holm correction. Bootstrap
intervals summarize task variation within the sampled applications, rather
than variation across generation seeds or models; their endpoints do not
replace the exact-test decisions. The results are consistent with a benefit
from predicting assertions during synthesis, but do not establish that benefit
statistically or measure the assertions' independent defect-detection accuracy.
They also do not prove that every admissible checkpoint avoids a semantic
shortcut: that boundary remains the registered hidden Start checks of
Table~\ref{tab:synthesis-check-boundaries}, which assess the preconditions
before execution.

\begin{table}[t]
\centering
\caption{Application-level component-ablation results. Each entry is
Start successes / E2E successes / tasks. Every configuration covers the same
tasks within each application; these small strata are descriptive and do not
establish effects for an individual application.}
\label{tab:synthesis-ablation-applications}
\floatsize
\begin{tabular}{lrrrr}
\toprule
Application & Full & w/o exploration & Schema-only & w/o assertions \\
\midrule
2048 & 8/7/8 & 8/6/8 & 8/7/8 & 8/4/8 \\
Blender & 2/2/5 & 3/3/5 & 2/2/5 & 2/2/5 \\
SpaceRPG & 5/5/8 & 2/2/8 & 4/3/8 & 5/4/8 \\
JupyterLab & 0/0/5 & 3/0/5 & 0/0/5 & 0/0/5 \\
jynew & 2/1/5 & 3/1/5 & 2/2/5 & 1/1/5 \\
LibreOffice & 0/0/5 & 0/0/5 & 0/0/5 & 0/0/5 \\
DatingSim & 1/0/7 & 1/0/7 & 2/1/7 & 2/0/7 \\
Luanti & 0/0/8 & 0/0/8 & 1/0/8 & 1/0/8 \\
OpenRA & 2/2/5 & 2/2/5 & 1/1/5 & 2/2/5 \\
PAL3 & 2/2/5 & 0/0/5 & 1/1/5 & 0/0/5 \\
\bottomrule
\end{tabular}

\end{table}

\begin{table}[t]
\centering
\caption{Terminal stages for the 61 trials in each configuration. Pre-Start
denotes construction, format, or public-contract failure before Start grading;
Start and Goal denote failure at the corresponding hidden gate. Rows sum to 61,
including all construction failures and failed executions.}
\label{tab:synthesis-ablation-failures}
\floatsize
\begin{tabular}{lrrrr}
\toprule
Configuration & Pre-Start & Start & Goal & Success \\
\midrule
\method{} (full) & 30 & 9 & 3 & 19 \\
w/o source exploration & 25 & 14 & 8 & 14 \\
Schema-only contract & 32 & 8 & 4 & 17 \\
w/o assertion prediction & 31 & 9 & 8 & 13 \\
\bottomrule
\end{tabular}

\end{table}

\textbf{Complete accounting.}
Tables~\ref{tab:synthesis-ablation-applications} and
\ref{tab:synthesis-ablation-failures} retain the full population by application
and terminal stage. The 244 trials use 9,895,186 tokens across 1,739 model calls,
including visual judging; 20 separate preflight calls use 1,488 tokens and are
excluded from the per-trial means. Four Unity candidates in the no-assertion
configuration were reconstructed from their byte-identical recorded model
responses for subsequent execution and grading, with no new synthesis calls.
Their costs count original synthesis once plus subsequent execution and grading;
their active times sum the two execution segments, excluding the intervening
delay. Consequently, that row does not support an uninterrupted wall-time or
speedup claim. The evidence artifact retains both report hashes and each
candidate hash, while the remaining 240 trial records retain their original
scores and computational-cost provenance.

\subsection{Supplementary Provenance Tables}

Two tables record the benchmark harness on the grader's side of the hook; neither measures the hook or $\Sigma$, which the onboarding agent produces (Appendix~\ref{sec:appendix-hook-onboarding}). Table~\ref{tab:effort} reports source size per component, and Table~\ref{tab:surfaces} names, per application, the surface class the inspector recorded and example semantic fields the profile reads through the hook when validating a checkpoint. Both are engineering provenance for the benchmark harness, not results and not method input.

\begin{table}[H]
\caption{Grader-side harness effort. Semantic profiles are the oracle's runtime-state readers, invisible to the method; the hook and $\Sigma$ are agent-generated, and their patch sizes are measured in Table~\ref{tab:onboarding-replication-apps}. LOC are non-empty, non-comment source or configuration lines under the measurement script's definition; counts include the JupyterLab V1 profile and bundled specs. The \confgui{} safe execution bridge (Section~\ref{sec:agent-s}; \texttt{warptest/agent\_s\_bridge.py} and \texttt{warptest/agent\_s\_worker.py}, 3{,}013 raw lines together) is not wired into this measurement script, so it is omitted from the counts rather than mixed in under a different methodology.}
\label{tab:effort}
\begin{center}
\floatsize
\begin{tabular}{llrr}
\hline
Component & Scope & LOC & Count \\
\hline
Reusable core and adapters & all & 13236 & 21 \\
Base profile helpers & all & 145 & 1 \\
Bundled specs & all & 8449 & 140 \\
Semantic profile & 2048 & 207 & 1 \\
Semantic profile & SpaceRPG & 503 & 1 \\
Semantic profile & DatingSim & 337 & 1 \\
Semantic profile & OpenRA & 96 & 1 \\
Semantic profile & JupyterLab & 105 & 1 \\
Semantic profile & LibreOffice & 113 & 1 \\
Semantic profile & Blender & 108 & 1 \\
Semantic profile & jynew & 93 & 1 \\
Semantic profile & pal3 & 94 & 1 \\
Semantic profile & luanti & 99 & 1 \\
\hline
\end{tabular}
\end{center}
\end{table}

\begin{table}[H]
\caption{Examples of discovered semantic surfaces. The inspector records source-guided evidence for the surfaces that grader-side profiles read, through the agent-generated hook, to observe or validate checkpoints; the writable fields $\Sigma$ exposes are declared in each application's onboarding contract.}
\label{tab:surfaces}
\begin{center}
\floatsize
\setlength{\tabcolsep}{4pt}
\begin{tabular}{@{}>{\raggedright\arraybackslash}p{0.10\linewidth}>{\raggedright\arraybackslash}p{0.34\linewidth}>{\raggedright\arraybackslash}p{0.515\linewidth}@{}}
\hline
Benchmark & Surface class & Example semantic fields \\
\hline
2048 & game manager and storage & board, score, win/loss flags \\
SpaceRPG & store, save namespace, WebGL readiness & screen, station, cargo, missions \\
DatingSim & store, save key, mocked-AI gate & resources, relationships, dialogue \\
OpenRA & utility/probe reports and map YAML & mod, map, players, actors, objectives, production \\
JupyterLab & server, contents/session APIs, workspace metadata & path, text, notebook cells, outputs, workspace tabs, kernel session \\
jynew & ES3/\texttt{GameRuntimeData} save & party, attributes, money, items, skills, flags \\
PAL3 & in-engine console-command save & story vars, scene, party, money, items, regions, favor \\
Luanti & worldmod state and sqlite world DB & nodes, player pos/hp, inventories, mod storage \\
LibreOffice & UNO/\texttt{pyuno} document model & Writer text/table/bookmark, Calc cell/formula/chart, Impress slide text \\
Blender & \texttt{bpy} scene graph, \texttt{.blend} file & object location/material, modifier stack, animation keyframes \\
\hline
\end{tabular}
\end{center}
\end{table}

\end{document}